\documentclass[a4paper,12pt]{iopart}
\usepackage[utf8]{inputenc}
\usepackage[english]{babel}
\usepackage{epsfig}
\usepackage{color}% color

\usepackage{iopams,amsthm,amscd,bm}
\usepackage{arydshln}

\usepackage{a4wide}
\eqnobysec

\def\k{\kappa}
\newcommand{\dem} {\mbox{$\frac{1}{2}$}}
\newcommand{\ve} [1] {\mbox{\boldmath $#1$}}

\newcommand{\cE}{{\cal E}}
\newcommand{\diag}{\mbox{\rm diag\,}}
\newcommand{\Imag}{\mbox{\rm Im\,}}
\newcommand{\Real}{\mbox{\rm Re\,}}

\newcommand{\Am}{L}
\newcommand{\Ap}{L^\dagger}

\newcommand{\cU}{{\cal U}}

\def\tfrac#1#2{{\textstyle {#1 \over #2}}}%
\def\dfrac#1#2{{\displaystyle {#1 \over #2}}}%

\renewcommand{\ge}{\geqslant}
\renewcommand{\le}{\leqslant}

\begin{document}

\title{Single- and coupled-channel radial inverse scattering with
supersymmetric transformations}

\author{Daniel Baye$^1$, Jean-Marc Sparenberg$^1$,
Andrey M Pupasov-Maksimov$^{2}$ and Boris F Samsonov$^3$\footnote{deceased}}

\address{$^1$ Physique Nucléaire et Physique Quantique,
Université libre de Bruxelles, C.P.\ 229, 
B 1050 Bruxelles, Belgium}

\address{$^2$ Departamento de Matemática, Universidade Federal de Juiz de Fora,
Juiz de Fora, MG, Brazil}

\address{$^3$ Physics Department, Tomsk State University, 36 Lenin Avenue,
634050 Tomsk, Russia}

\eads{\mailto{dbaye@ulb.ac.be}, \mailto{jmspar@ulb.ac.be}, \mailto{pupasov@phys.tsu.ru}}

\begin{abstract}
The present status of the three-dimensional inverse-scattering method
with supersymmetric transformations is reviewed for the coupled-channel case.
We first revisit in a pedagogical way the single-channel case,
where the supersymmetric approach is shown to provide a complete, efficient and elegant solution
to the inverse-scattering problem for the radial Schrödinger equation with short-range interactions.
A special emphasis is put on the differences
between conservative and non-conservative transformations,
i.e.\ transformations that do or do not conserve the behaviour
of solutions of the radial Schrödinger equation at the origin.
In particular, we show that for the zero initial potential,
a non-conservative transformation is always equivalent
to a pair of conservative transformations.
These single-channel results are illustrated
on the inversion of the neutron-proton triplet eigenphase shifts
for the $S$ and $D$ waves.

We then summarize and extend our previous works on the coupled-channel case,
i.e.\ on systems of coupled radial Schrödinger equations,
and stress remaining difficulties and open questions of this problem
by putting it in perspective with the single-channel case.
We mostly concentrate on two-channel examples to illustrate general principles
while keeping mathematics as simple as possible.
In particular, we discuss the important difference between the equal-threshold
and different-threshold problems.
For equal thresholds, conservative transformations
can provide non-diagonal Jost and scattering matrices.
Iterations of such transformations in the two-channel case are studied
and shown to lead to practical algorithms for inversion.
A convenient particular technique where the mixing parameter can be fitted
without modifying the eigenphases is developed with iterations of pairs
of conjugate transformations.
This technique is applied to the neutron-proton triplet $S$-$D$ scattering matrix,
for which exactly-solvable matrix potential models are constructed.
For different thresholds,
conservative transformations do not seem
to be able to provide a non-trivial coupling between channels.
In contrast, a single non-conservative transformation can generate
coupled-channel potentials starting from the zero potential
and is a promising first step towards a full solution to the coupled-channel inverse problem
with threshold differences.
\end{abstract}


\centerline{\today}

\tableofcontents
\section{Introduction}
Low-energy collisions of particles with an internal structure
(i.e., atom-atom, nucleus-nucleus, etc.) generally include
inelastic processes such as excitations of internal degrees of
freedom of the colliding particles or processes with rearrangements of
their constituent parts.
For three-dimensional rotationally-invariant systems,
these processes can be approximately described by a
matrix (more precisely multichannel) radial Schrödinger equation with a local
matrix potential \cite{taylor:72,newton:82,joachain:83} in the framework of
the coupled-channel scattering theory.
The main idea of scattering theory is that the colliding particles
are supposed to move freely at large distances
(in the present work, we do not consider the Coulomb interaction).
This asymptotic behaviour is encoded in the incoming and outgoing states.
Roughly speaking, to describe the collision process one should find
the operator which transforms incoming states into outgoing states.
This operator is nothing but the scattering matrix $S$.

In principle, the scattering matrix can be extracted from collision experiments.
Subsequently, one can raise the inverse-scattering problem
about the determination of the interacting potential from the scattering matrix~\cite{chadan:89,zakhariev:90,zakhariev:07}.
A part of the problem was solved in works of Gel'fand, Levitan, Marchenko,
Newton and Jost.
They formulated prescriptions for both the single- and coupled-channel cases of
how to construct an integral equation which allows one to find the
potential from the scattering matrix or from the related Jost matrix
\cite{levitan:book,gelfand:51a,gelfand:51b,marchenko:55,newton:55b,newton:57}.
They also found some exact solutions of the integral equation, in particular,
for single-channel problems in the case of a separable kernel,
a result which was later generalized to coupled-channel problems
\cite{geramb:93,kohlhoff:93}.

Coupled-channel scattering problems can be divided into different categories.
First, the interacting particles are either charged or neutral.
In the present work, we only consider the second case,
which is simpler from the mathematical point of view.
Second, one can distinguish the cases of different and equal thresholds.
%These two situations lead to
%topologically different energy Riemann surfaces for the
%corresponding $S$-matrices. In the case of the equal thresholds
%the Riemann surface is a sphere, whereas in the case of $N$
%different thresholds..
These two situations require significantly different approaches
to the inversion.
%An example of coupled-channel scattering with different channel thresholds
%is the scattering of alkali-metal atoms in a background magnetic field.
%The presence of the magnetic field results in different energies
%for the different possible hyperfine states of the atoms.
%Thus channels with different threshold energies appear.
Multichannel scattering with different thresholds appears in all reactions.
It requires a separate treatment above and below a given threshold.
The low-energy neutron-proton scattering gives an example of two-channel
scattering with equal thresholds, because one should take into
account uncoupled channels $^1\!S_0$, $^1\!P_1$, $\ldots$, and
coupled channels $^3\!S_1- ^3\!D_1$,  $^3\!P_2- ^3\!F_2$, $\ldots$
Both the equal- and different-threshold cases will be considered in detail.

It is known that supersymmetric transformations are a powerful
tool to manipulate the properties of one-dimensional
(single-channel) potentials in quantum mechanics.
For instance, supersymmetric quantum mechanics allows
the construction of potentials that are exactly solvable or
that display interesting symmetry properties like shape invariance,
or the manipulation of the discrete spectra of these potentials.
These classical applications of supersymmetry are extremely vast
and are the subject of several textbooks \cite{junker:96,bagchi:00,cooper:01,gangopadhyaya:11}.
In the present review, we rather concentrate on the use of supersymmetric transformations to manipulate
scattering properties of one-dimensional potentials defined on the half line
that appear in the radial Schrödinger equation.
Briefly speaking, a supersymmetric transformation of this equation for a given partial wave
is a powerful tool to address its scattering properties
because under such a transormation the scattering matrix is simply multiplied
by a first-order rational function of the momentum \cite{sukumar:85c}.
The supersymmetric approach is basically equivalent 
to the Darboux transformation method (see, e.g., reference \cite{Bagrov}). 
Therefore one can use supersymmetric and Darboux transformations as synonyms.

Starting from a zero potential,
which corresponds to a unit scattering matrix,
the iteration of supersymmetric transformations may then be used to solve
the inverse-scattering problem with good accuracy:
the resulting scattering matrix reads as a rational function of arbitrary order.
The effectiveness of this approach to the inversion of scattering data
is demonstrated in \cite{sparenberg:97a,samsonov:03,baye:04}.
The potentials obtained by supersymmetric inversion are equivalent
to the potentials obtained from the Gel'fand-Levitan and Marchenko integral
equations in the case of separable kernels \cite{nieto:84,sukumar:85c}.
However, the supersymmetric approach is probably simpler to implement
because of the differential character of the transformation.
Moreover, it presents the advantage of being an iterative procedure
and of leading to compact expressions for the obtained potentials.

In the single-channel case, several types of supersymmetric
transformations exist,
that will be reviewed below.
They are obtained from solutions of the Schrödinger equation,
not necessarily physical,
at negative energies called factorization energies.
{\em Conservative} transformations map the solutions of the supersymmetric
partner Hamiltonians to each other,
while keeping their boundary behaviours at the origin unchanged
(e.g., regular solutions at the origin remain regular after transformation).
{\em Non-conservative} transformations, on the other hand,
modify the boundary behaviours of the solutions and are thus more complicated.
Below, we shall study the link between both types of transformations
and show that conservative transformations are probably sufficient
from an inverse-scattering-problem perspective.

There are several papers devoted to the generalization of these supersymmetric transformations
to multichannel three-dimensional scattering problems,
i.e.\ to systems of coupled radial Schrödinger equations
\cite{andrianov:86,andrianov:87,amado:88a,amado:88b,amado:90,cannata:92,cannata:93,andrianov:95,andrianov:97,sparenberg:97b,leeb:00,sparenberg:06}.
In the coupled-channel case, there is much freedom in the form of
supersymmetric transformations, therefore a full analysis of all types of
transformations does not exist up to now.
However, it is clear that, as in the single-channel case,
coupled-channel transformations provide a very useful tool from the point of view of scattering properties,
since under such a transformation the scattering matrix is also modified by a rational (matrix) function of the momentum.
% Arbitrary chains of first-order supersymmetric transformations
% in the case of the matrix Schrödinger equation are studied in
% \cite{samsonov:04}.
% There, compact expressions for both the transformed matrix potential and
% solutions were obtained.
This led for instance to the discovery of the phase-equivalent supersymmetric
transformations, which are based on two-fold, or second-order, differential operators.
These are described in
\cite{baye:87a,baye:93a,sparenberg:96,samsonov:02} for the
single-channel case and are generalized in \cite{sparenberg:97b,leeb:00} for the
coupled-channel case.
Such transformations keep the scattering matrix unchanged and simultaneously
allow one to reproduce given bound-state properties.

However these encouraging results are still far
from an effective supersymmetry-based inversion in the coupled-channel case,
in particular because methods based on a direct generalization
of the supersymmetry technique to the multichannel case are not able to
provide an easy control of the scattering properties for all channels simultaneously.
In the case of equal thresholds, the eigenphase shifts and the
mixing parameters are modified in a complicated way, which makes
their individual control difficult \cite{pupasov:09}.
In the case of different thresholds,
it is even impossible to modify the coupling between channels
by using standard {\it conservative} supersymmetric transformations.
This fact was established by Amado, Cannata and Dedonder \cite{amado:88b}.
We believe that these are reasons why supersymmetric transformations did not
find a wide application to multichannel scattering inversion.
Non-conservative transformations, however, can solve this problem.

In the present work, we revisit the supersymmetric approach to the single-
and multichannel inverse problems.
We summarize, unify and extend our previous works on this topic.
We study more general supersymmetric transformations
of the coupled-channel Schrödinger equation.
We establish constraints on the free parameters of supersymmetric transformations
determined by physical requirements.
In this way we solve some of the problems mentioned above.
We mostly focus on the case of equal thresholds and arbitrary partial waves,
for the two-channel case.
In this case, we present two algorithmic supersymmetry-based approaches
to the inversion of scattering data.
One of these, based on complex factorization energies appears to be
more practical.
This method consists in the inversion of the eigenphase shifts with the help
of single-channel techniques, followed by the inversion of the mixing parameter
with the help of eigenphase-preserving coupling transformations \cite{pupasov:10}.
When applied to the inversion of n-p triplet scattering data, our approach gives
a realistic potential similar to the well-known phenomenological models.
This important application will also be briefly reviewed below.

Unfortunately, this approach cannot be used in the case of different thresholds.
In this case, only a preliminary analysis of the problem is available
in the literature \cite{sparenberg:06,samsonov:07,pupasov:08}
and reviewed here.
We show that, in the different-threshold coupled-channel case,
only {\it non-conservative} supersymmetric transformations allow one to circumvent
the impossibility argument of reference \cite{amado:88b}
and modify the coupling between channels.
Though this coupling modification is not easily exploitable, it is possible to
generate several simple exactly-solvable models starting from the zero potential.
In the two-channel case,
this provides for instance exactly-solvable schematic models of atom-atom interactions
for the interplay of a magnetically-induced Feshbach resonance with a bound state or a
virtual state close to the elastic-scattering threshold \cite{pupasov:08}.
In the $N$-channel case, a general discussion of the number of bound, virtual and resonance states
of such a potential could even be made,
based on a geometrical analysis of the Jost-matrix-determinant zeros \cite{pupasov_08_jpa}.
Here we only revisit the 2-channel case in detail
and provide analytical expressions for the potential, eigenphase shifts and wave functions,
which may be useful to test numerical methods.

The structure of the paper reads as follows.
Scattering theory definitions (channels, partial waves, thresholds,
regular solutions, Jost and scattering matrices, effective range expansion, etc.)
are recalled in section \ref{sec:sumscat}.
Single-channel supersymmetric quantum mechanics and inversion are summarized
in section \ref{sec:sc}.
Conservative and non-conservative transformations are defined and
pairs of transformations with real and complex factorization
energies are discussed.
In section \ref{sec:sci}, the Bargmann potentials are revisited as basic tools
for iterative constructions of solutions of the inverse-scattering problem.
Coupled-channel supersymmetric quantum mechanics is summarized in section
\ref{sec:cc}.
The most general transformation as well as conservative and
non-conservative transformations are discussed.
In section \ref{sec:ipnot},
the inverse two-channel problem with equal thresholds is analyzed.
In particular, eigenphase-preserving supersymmetric transformations
are presented as a practical tool and applied to the neutron-proton scattering
by fitting modern data.
Non-conservative supersymmetric transformations are derived
in the two-channel case with different thresholds in section \ref{sec:ipwt}.
Concluding remarks are presented in section \ref{sec:conc}
\section{Summary of three-dimensional scattering theory}
\label{sec:sumscat}
The quantum theory of scattering in three dimensions is well described in many textbooks and
monographs, see e.g.\ \cite{taylor:72,newton:82,joachain:83,levitan:book,chadan:89}.
This section introduces necessary notions and fixes notations.
\subsection{Single-channel scattering}
\label{sec:scscat}
Before discussing complications related to the existence of several channels,
let us summarize the basic properties of single-channel scattering between two particles
interacting through a central potential.
Let us consider the real energy $E$.
In units $\hbar=2\mu=1$ where $\mu$ is the reduced mass of the particles,
this energy is the square of the wave number $k$,
\begin{equation}
E = k^2\,.
\label{sc.1}
\end{equation}
We only consider wave functions that are factorized in spherical coordinates $\ve{r} = (r,\Omega)$
as $\psi(\ve{r}) = r^{-1} \psi_l (k,r) Y_l^m (\Omega)$.
The variable $\ve{r}$ represents the relative coordinate between the two particles.
The spherical harmonics $Y_l^m (\Omega)$ depend on the orbital
and magnetic quantum
numbers $l$ and $m$, and on the angles $\Omega = (\theta,\varphi)$.
We are interested in properties of the radial wave function $\psi_l (k,r)$
for a given partial wave $l$.
The subscript $l$ is in general understood below.

All physical properties arise from the one-dimensional 
stationary radial Schrödinger equation 
\begin{equation}
H \psi(k,r) = k^2 \psi(k,r)\,,
\label{sc.2}
\end{equation}
where $H$ is the Hamiltonian operator 
\begin{equation}
H = -\frac{\rmd^2}{\rmd r^2} + V(r)\,.
\label{sc.3}
\end{equation}
We consider that the particles interact through
a central effective potential $V(r)$.
This effective potential includes the centrifugal term $l(l+1)/r^2$.
Since the distance $r$ varies on the interval $[0,\infty[$,
physical wave functions must be regular over this interval and satisfy 
\begin{equation}
\psi(k,0) = 0\,.
\label{sc.2a}
\end{equation}

The orbital momentum $l$ characterizes the asymptotic behaviour of the potential
at large distances.
Let us write the potential as 
\begin{equation}
V(r) = \frac{l(l+1)}{r^2} + \bar{V}(r)\,.
\label{sc.4}
\end{equation}
We assume that $\bar{V}$ is short-ranged,
i.e.\ there exist $\varepsilon > 0$ and $a>0$ such that 
\begin{equation}
\int_a^\infty \rme^{\varepsilon r} \bar{V}(r) < \infty\,.
\label{sc.4a}
\end{equation}
The inverse of the upper bound of the $\varepsilon$ values is the range
of the potential.
The Coulomb asymptotic behaviour is excluded here.
We assume that the potential is continuous with
a single singularity located at the origin.
An integer $\nu$ determines the singularity of the potential at the origin 
\begin{equation}
V(r\to 0)=\frac{\nu(\nu+1)}{r^2} + V_0 + \Or(r)\,.
\label{sc.5}
\end{equation}
Note that, here also, $V$ does not contain a Coulomb-like $r^{-1}$ singularity.
For usual effective potentials, $\nu$ is equal to the orbital quantum number $l$.
Since this singularity can change when supersymmetric transformations
are performed, we consider a more general case where $\nu$ may differ from $l$.

The wave functions satisfying the Schrödinger equation \eref{sc.2} display two important properties.
First, the logarithmic derivative of such a wave function satisfies a Riccati equation 
\begin{equation}
\left[\frac{\psi'(k,r)}{\psi(k,r)}\right]'
= V(r) - k^2 - \left[\frac{\psi'(k,r)}{\psi(k,r)}\right]^2\,,
\label{sc.27}
\end{equation}
where prime means derivation with respect to $r$.
Second, the Wronskian of two solutions $\psi(k,r)$ and $\chi(\tilde{k},r)$ at energies $k^2$ and $\tilde{k}^2$,
as defined by
\begin{equation}
 \mathrm{W}[\psi(k,r),\chi(\tilde{k},r)] \equiv \psi(k,r) \chi'(\tilde{k},r) - \psi'(k,r) \chi(\tilde{k},r), 
\label{sc.9}
\end{equation}
has the property 
\begin{equation}
\frac{\rmd}{\rmd r} \mathrm{W}[\psi(k,r),\chi(\tilde{k},r)]
= (k^2-\tilde{k}^2) \psi(k,r) \chi(\tilde{k},r)\,.
\label{sc.28}
\end{equation}
This Wronskian is thus constant for equal energies.

The regular solution $\varphi(k,r)$ of \eref{sc.2} behaves at the origin as 
\begin{equation}
\varphi(k,r\to 0) = \frac{r^{\nu+1}}{(2\nu+1)!!}
+ {\textstyle \frac{1}{2}} (V_0-k^2) \frac{r^{\nu+3}}{(2\nu+3)!!}
+ {\rm o}(r^{\nu+3})\,.
\label{sc.6}
\end{equation}
When extending the wave number $k$ to the whole complex plane,
this solution has interesting analyticity properties.
As a function of the energy, it is multivalued because of \eref{sc.1};
it is at least analytical on the physical energy Riemann sheet,
which corresponds to the upper complex $k$-plane.
An infinity of irregular solutions exist.
An irregular solution is given by 
\begin{equation}
\chi(k,r) = \varphi(k,r) \int_r^a \frac{1}{[\varphi(k,t)]^2}\, dt
\label{sc.7}
\end{equation}
where $a$ is an arbitrary positive constant.
Up to an arbitrary amount of regular function $\varphi(k,r)$,
this function has the behaviour at the origin 
\begin{equation}
\chi(k,r\to 0) = \frac{(2\nu-1)!!}{r^\nu}
- {\textstyle \frac{1}{2}} (V_0-k^2) \frac{(2\nu-3)!!}{r^{\nu-2}}
+ {\rm o}(r^{-\nu+2})\,.
\label{sc.8}
\end{equation}
The Wronskian of these solutions is a constant, as shown by \eref{sc.28}.
The value of this constant is given by \eref{sc.6} and \eref{sc.8} as 
\begin{equation}
\mathrm{W} [\varphi(k,r),\chi(k,r)] = -1\,.
\label{sc.10}
\end{equation}

The regular and irregular solutions $\varphi(k,r)$ and $\chi(k,r)$ form
a basis of solutions, chosen with respect to the singularity at the origin.
For a scattering problem, the behaviour at infinity plays a crucial role.
Another important basis involves the Jost solutions $f(k,r)$ and $f(-k,r)$
which behave asymptotically as outgoing and incoming solutions, respectively.
At the origin, they are proportional to $r^{-\nu}$ according to \eref{sc.8}.
The Jost solution $f(k,r)$ is a solution of the Schrödinger equation with the
asymptotic behaviour of a free particle, 
\begin{equation}
f(k,r\to 0) \to h_l(kr)\,,
\label{sc.11}
\end{equation}
with 
\begin{equation}
h_l(z)=\rmi^{l+1}\left(\frac{\pi z}{2}\right)^{\frac{1}{2}}
H_{l+\frac{1}{2}}^{(1)}(z)\,,
\label{sc.12}
\end{equation}
where $H_{l+\frac{1}{2}}^{(1)}(z)$ is a first Hankel function
\cite{abramowitz:65}.
The behaviour of $h_l(z)$ at large distances is 
\begin{equation}
h_l(z\to\infty) = \rme^{\rmi z}\Bigl(1+\frac{\rmi l(l+1)}{2z} +
\Or(z^{-2})\Bigr)\,.
\label{sc.13}
\end{equation}
With \eref{sc.11} and \eref{sc.13},
the Wronskian of the Jost solutions is given by 
\begin{equation}
\mathrm{W} [f(k,r),f(-k,r)] = -2\,\rmi\, k\,.
\label{sc.14}
\end{equation}

The regular solution is expressed as a linear combination
of the Jost solutions as 
\begin{equation}
\varphi(k,r)=\frac{\rmi}{2k}\left[f(-k,r)F(k)-f(k,r)F(-k)\,\right]\,,
\label{sc.15}
\end{equation}
where the coefficient $F(k)$ is known as the Jost function.
From \eref{sc.14}, one obtains 
\begin{equation}
F(k) = \mathrm{W} [f(k,r),\varphi(k,r)]\,.
\label{sc.16}
\end{equation}
From \eref{sc.6} and \eref{sc.15}, it follows that 
\begin{equation}
f(-k,r\to 0) = \frac{F(-k)}{F(k)}\,f(k,r\to 0) + \Or(r^{\nu+1})\,,
\label{sc.16a}
\end{equation}
and 
\begin{equation}
F(k)=\lim\limits_{r\to 0} \frac{f(k,r)r^\nu }{(2\nu-1)!!}\,.
\label{sc.17}
\end{equation}

A zero $k_0$ of 
\begin{equation}
F(k_0) = 0
\label{sc.18}
\end{equation}
which is purely imaginary with a positive imaginary part
corresponds to a bound state at energy $-|k_0|^2$.
Indeed, the regular solution at the origin
then decreases exponentially at infinity
as shown by the non-vanishing term of \eref{sc.15}.
An imaginary zero with a negative imaginary part
corresponds to a virtual state at energy $-|k_0|^2$.
A pair of complex symmetric zeros with opposite real parts and $\Imag k_0 <0$
corresponds to a resonance at energy $\Real k_0^2$
with width $|2\,\Imag k_0^2|$.

A physical solution, which appears in the partial-wave decomposition
of the stationary scattering state, is proportional to the regular
solution $\varphi(k,r)$ and behaves at infinity as 
\begin{equation}
\psi(k,r\to\infty) \propto \rme^{-\rmi(kr-l\frac{\pi}{2})}-
\rme^{\rmi(kr-l\frac{\pi}{2})}S_l(k)\,.
\label{sc.19}
\end{equation}
When the coefficient of the incoming wave is unity,
the coefficient $S_l(k)$ of the outgoing wave is known
as the scattering matrix or $S$ matrix
although it is a function in the present single-channel case.
By comparing \eref{sc.19} with \eref{sc.15},
the scattering matrix $S_l(k)$ is obtained in terms of the Jost function as 
\begin{equation}
S_l(k) = (-1)^l \frac{F(-k)}{F(k)}\,.
\label{sc.20}
\end{equation}
Since the modulus of the scattering matrix is one,
it is conveniently expressed with the phase shift $\delta_l$ according to 
\begin{equation}
S_l(k) = \rme^{2\rmi\delta_l(k)}\,.
\label{sc.21}
\end{equation}

For studying the inverse-scattering problem, it is convenient to introduce
the effective-range function \cite{joachain:83}
\begin{equation}
K_l(k^2) = k^{2l+1} \cot \delta_l(k) = \rmi k^{2l+1} \frac{S_l(k)+1}{S_l(k)-1}
\label{sc.22}
\end{equation}
which is meromorphic and hence can be expanded as a Padé approximant.
Equation \eref{sc.22} is inverted as 
\begin{equation}
S_l(k) =  \frac{K_l(k^2)+\rmi k^{2l+1}}{K_l(k^2)-\rmi k^{2l+1}}\,,
\label{sc.23}
\end{equation}
which allows one to relate the location of the scattering-matrix poles
to the coefficients of a Padé expansion for $K_l(k^2)$.

At low energy, both the Jost function and the effective-range function of
a potential satisfying condition \eref{sc.4a} are analytical \cite{taylor:72}.
A Taylor expansion is then sufficient,
which leads to the effective-range expansion 
\begin{equation}
K_l(k^2 \to 0) =
-\frac{1}{a_l}+\frac{r_l}{2} k^2 -P_l r_l^3 k^4 + \Or(k^6)\,,
\label{ere}
\end{equation}
where $a_l$ is the scattering length, $r_l$ is the effective range
and $P_l$ is the shape parameter (generalized to $l\ne 0$ partial waves).
These numbers characterize the low-energy properties of the phase shift.
Since the effective-range function is analytical near the origin,
this expansion is valid in some domain for negative energies.
It is valid for a bound state with wave number
$k_b = \rmi \kappa_b$ with $\kappa_b > 0$
provided the state is weakly bound,
more precisely provided $(2\kappa_b)^{-1}$ is larger than the
range of the potential,
as deduced from the analyticity region of the Jost function \cite{taylor:72}.
The effective range function then satisfies the relation 
\begin{equation}
K_l(-\kappa_b^2) = (-1)^{l+1} \kappa_b^{2l+1}
\label{sc.24}
\end{equation}
arising from \eref{sc.23}.
This relation introduces constraints on the parameters in \eref{ere}
when the first few terms provide an accurate approximation of $K_l(k^2)$.
These coefficients can also be constrained in another way.
The so-called {\it asymptotic normalization constant} $C_b$
(abbreviated below as ANC) is the coefficient in the
asymptotic expression of the radial function of a normalized bound state, 
\begin{equation}
\psi (i\kappa_b,r\to\infty) \to C_b \exp(-\kappa_b r).
\label{sc.25}
\end{equation}
This coefficient is measurable experimentally.
It is related to the residue of the $S$-matrix pole at the bound-state energy
[see \eref{sc.18} and \eref{sc.20}].
This property leads to the relation 
\begin{equation}
|C_b|^{-2} = \kappa_b^{-2l} \left. \frac{\rmd}{\rmd k^2}
\left[\rmi k^{2l+1} - K_l(k^2) \right] \right|_{k^2=-\kappa_b^2}\,,
\label{sc.26}
\end{equation}
valid when $2\kappa_b$ is smaller than the inverse of the range of the potential.
\subsection{Multichannel scattering}
\label{sec:ms}
As mentioned in the introduction, two cases must be considered, i.e.\
collisions involving different or equal thresholds.
Let us first comment on the notion of channel for which the usual vocabulary
is somewhat ambiguous.
Elastic scattering corresponds to the case where
the nature and the state of the colliding particles are not
modified during a collision.
By particles, we mean any composite system such as atoms, molecules or nuclei.
This physical channel is always open.
When no other channel is allowed by energy conservation,
only one physical channel is open.
However its description may require several components
when at least one of the particles has a non-vanishing spin.
The single-channel physical problem thus requires a coupled-channel
mathematical description in this case.
After reduction of the spin and angular parts of the Schrödinger equation,
the radial problem takes the form of a system of coupled equations.
This case will be called here a coupled-channel problem with equal thresholds.

When the energy increases, the number of open channels varies because
the colliding particles can be excited or reactions leading to
a rearrangement of the components of the particles can occur.
When energy conservation allows a new channel to open,
the corresponding energy is the threshold energy.
The description of such a case requires one or several additional
coupled radial equations.
This case will be called here a coupled-channel problem with different thresholds.
In practice, however, we will only consider for this case
that all thresholds are different.
This corresponds to a simplified description where the spins are neglected.
This simplification is reasonable when spin-dependent effects are weak
or not measured.

Since interactions are assumed to be invariant under rotation and reflection,
the good quantum numbers are the total angular momentum and parity.
For each pair of good quantum numbers, one has a specific system
of $N$ coupled radial equations, where the number $N$ of equations
depends on these quantum numbers.
We assume that we consider a given partial wave and
consider $N$ channels in the mathematical sense, labelled with $i = 1, \ldots, N$.
Any order can be selected for the channels.
Here we assume that they are ordered by increasing threshold energies.

The coupled radial equations for these $N$ channels read for $i = 1, \ldots, N$, 
\begin{equation}
-\frac{1}{2\mu_i} \frac{\rmd^2 \Psi_i}{\rmd r^2}
+ \sum_{j=1}^N {\cal V}_{ij}(r) \Psi_j = (E-E_{{\rm thr},i}) \Psi_i
\label{mc.0}
\end{equation}
where $\mu_i$ is the reduced mass 
(the units are fixed by $2\mu_1 = 1$) 
and $\Psi_i$ is the wave function component of channel $i$.
The coupling potentials ${\cal V}_{ij}$ are symmetric.
Diagonal potentials include centrifugal terms.
Although, potentials should in general be non-local,
we make here the usual approximation
to consider them as local.
Each channel is characterized by a threshold energy difference 
\begin{equation}
\Delta E_{{\rm thr},i} = E_{{\rm thr},i} - E_{{\rm thr},1}\,,
\label{mc.1}
\end{equation}
where 1 denotes the reference channel with the lowest threshold energy.
For simplicity, we choose the lowest threshold as zero of energies, 
$E_{{\rm thr},1} = 0$.
The system thus presents possible bound states for negative energies $E$
and scattering states with at least one open channel for positive energies $E$.
Channels with different reduced masses always have different threshold energies,
while channels with equal reduced masses may display either equal threshold energies
(e.g.\ for partial waves coupled by a tensor interaction)
or different threshold energies
(e.g.\ for inelastic collisions).

The wave numbers $k_i$ of the different channels are defined as 
\begin{equation}
k_i = \pm \sqrt{2\mu_i (E-E_{{\rm thr},i})}\,.
\label{mc.2}
\end{equation}
When some channels are closed, the corresponding wave numbers are imaginary.
Let us stress that the signs in \eref{mc.2} can be chosen arbitrary
and independent of each other.
The energy Riemann surface is thus $2N$-fold.
When the collision energy is larger than all thresholds $E_{{\rm thr},i}$,
all channels are open and all wave numbers can be chosen positive.
This is the main situation that we study below
in the $N$-channel inverse problem.

In order to simplify the mathematical treatment, 
let us multiply each equation \eref{mc.0} by $\sqrt{2\mu_i}$. 
After introducing the notations 
\begin{equation}
\psi_i(r) = \Psi_i(r) /\sqrt{2\mu_i}\,, \quad
V_{ij}(r) = 2\,\sqrt{\mu_i\mu_j}\,{\cal V}_{ij}(r)\,,
\end{equation}
one obtains a system of $N$ coupled radial Schrödinger equations 
that reads in matrix notation \cite{taylor:72,newton:82} 
\begin{equation}
H \psi(k,r) = k^2 \psi(k,r)\,,
\label{mc.3}
\end{equation}
where $\psi(k,r)$ is either an $N$-dimensional column vector or a matrix made of an arbitrary number
of $N$-dimensional column vectors, each of which being an independent solution of the equation system.
In the following, we shall in particular make a frequent use of square $N\times N$ matrix solutions
made of $N$ linearly-independent column vectors. 
The linearity of the coupled equations \eref{mc.3} implies that if the square matrix $\psi$ is such a solution,
matrix $\psi C$ obtained by right multiplication of $\psi$ by an invertible square matrix $C$ is also such a solution,
the columns of which are linear combinations of the column vectors appearing in $\psi$.
The Hamiltonian is defined as 
\begin{equation}
H = -\frac{\rmd^2}{\rmd r^2} I_N + V(r)
\label{mc.4}
\end{equation}
where $I_N$ is the $N\times N$ identity matrix and
$V$ is an $N\times N$ real symmetric matrix with elements $V_{ij}(r)$ 
depending on a single coordinate $r$.
By $k$ we denote a diagonal matrix with the non-vanishing entries $k_i$,
$k=\mbox{diag}(k_1,\ldots,k_N)$.
In matrix form, one can also write 
\begin{equation}
k^2 = E - \Delta\,, \quad E = \diag(E_1,\ldots,E_N)\,,
\quad \Delta = \diag(\Delta_1,\ldots,\Delta_N)\,,
\end{equation}
where 
\begin{equation}
E_i = 2\mu_i E, \quad \Delta_i = 2\mu_i E_{{\rm thr},i}
\end{equation}
When all reduced masses and all thresholds are equal, 
$k$ and $E$ are {\it scalar} matrices
and the notation $k$ or $E$ can be alternatively considered 
as representing a number. 
By scalar matrix, we mean a matrix proportional to the identity matrix.
Another comment about the notation is useful:
in the notation for matrix $\psi(k,r)$,
the symbol $k$ should generally not be taken as representing $N$ parameters.
All wave numbers $k_i$ are related to the energy by equation
\eref{mc.2} and are depending on each other.
The matrix wave function thus depends in reality on a single parameter,
the energy,
even though this dependence is in fact multivalued
(or single valued on the whole Riemann surface)
when the wave numbers are extended to their whole complex planes.

Another important comment concerns the notion of coupled channels.
When matrix $V$ is diagonal, the equations are uncoupled.
They correspond to an independent set of single-channel problems.
One can also say that this potential is trivially coupled.
Such a potential will often be the starting point 
of inversion techniques in the following.
In the case of equal masses and thresholds, 
we assume in addition that a non-diagonal matrix $V$
cannot be diagonalized with a matrix independent of $r$.
Otherwise, we would have a trivially coupled problem,
which can also be seen as an uncoupled problem in disguise
since the matrix that diagonalizes $V$ actually
diagonalizes the whole system \eref{mc.3}.
Let us stress that such a possibility of trivial coupling in
the potential matrix does not exist in the presence 
of unequal masses or unequal thresholds, 
as matrix $k^2$ is then diagonal but not scalar,
which implies it becomes non-diagonal when multiplied
by the matrix that diagonalizes $V$.

In \S \ref{sect_diff_coupl}, for the two-channel case, 
we give a practical recipe about how to distinguish trivially 
and non-trivially coupled potentials. 
Moreover, we also analyze cases where the potential matrix 
has a non-trivial coupling but the corresponding 
Jost matrix or scattering matrix is trivially coupled. 
This may happen when the Jost matrix or scattering matrix 
can be diagonalized by a $k$-independent transformation. 
Otherwise they are non-trivially coupled.

Let $l=\diag(l_1,\ldots,l_N)$ be a diagonal matrix
of orbital momentum quantum numbers $l_j$.
Let us write potential $V$ as
\begin{equation}
V = r^{-2} l(l+I_N) + \bar{V}(r).
\label{mc.5}
\end{equation}
We assume that the second term of \eref{mc.5}
is short-ranged at infinity,
i.e.\ there exists $\varepsilon>0$ and $a>0$ such that
\begin{equation}
\int_a^\infty \rme^{\varepsilon r}|\bar{V}_{ij}(r)| \rmd r < \infty\,
\label{mc.6}
\end{equation}
where $\bar{V}_{ij}$, $i,j=1,\ldots,N$ are entries of matrix $\bar{V}$.
Matrix $l$ thus defines the asymptotic behaviour of the potential
at large distances 
\begin{equation}
V(r\to \infty) \to r^{-2} l(l+I_N).
\label{mc.7}
\end{equation}
Relation \eref{mc.5} is typical of coupled channels involving
various partial waves
when the Coulomb interaction is absent.
This expression is similar to \eref{sc.4}
but the orbital momenta can differ according to the channel.

We limit ourselves to bounded potentials for $r>0$.
At the origin, the potential matrix $V$ can be singular.
Its singularity is determined by a diagonal matrix $\nu$,
with positive integer $\nu_j$, $j= 1,\ldots,N$, as diagonal elements,
according to 
\begin{equation}
V(r\to 0) = r^{-2} \nu(\nu+I_N) + \Or(1)\,.
\label{mc.8}
\end{equation}
Note that we assume that $V$ does not contain a Coulomb-like $r^{-1}$ singularity.

Let us now define the regular solution matrix $\varphi(k,r)$
of system \eref{mc.3}.
For the assumed potentials, a unique regular solution with
a well-defined normalization
is fixed by its behaviour at the origin 
\begin{eqnarray}
\varphi(k,r\to 0) & \to & 
\diag\left(\frac{r^{\nu_1+1}}{(2\nu_1+1)!!},
\ldots,\frac{r^{\nu_N+1}}{(2\nu_N+1)!!}\right)
\nonumber \\[.5em]
& \equiv & r^{\nu+I_N}[(2\nu+I_N)!!]^{-1}\,,
\label{mc.9}
\end{eqnarray}
where the double factorials act on the diagonal elements of the diagonal matrix.
The Schrödinger equation possesses an infinity of irregular solutions
differing by arbitrary amounts of regular components.
Let us define a subset behaving at the origin as
\begin{eqnarray}
\chi(k,r\to 0) & \to & 
{\diag}\left(\frac{(2\nu_1-1)!!}{r^{\nu_1}},
\ldots,\frac{(2\nu_N-1)!!}{r^{\nu_N}}\right)
\nonumber \\[.5em]
&\equiv&r^{-\nu}(2\nu-I_N)!!\,.
\label{mc.10}
\end{eqnarray}
For real energies, both solutions $\varphi$ and $\chi$
are purely real because they satisfy
a system of differential equations with real coefficients
and real boundary conditions.
They form a basis in the space of matrix solutions.
The columns of the regular and irregular solution matrices
are vector solutions of the Schrödinger equation.
They also form a basis in the vector solution space of this equation.
Let us note that the non-diagonal next-order terms of solutions $\varphi$ and $\chi$
are dependent on the particular value of the potential at the origin
and on whether some singularity parameters $\nu_i$ differ by two units or not.

Let us define the Wronskian of two matrix functions by 
\begin{equation}
\mathrm{W}[\psi_1,\psi_2] = \psi_1^T \psi_2' - \psi_1'^T \psi_2\,,
\label{mc.11}
\end{equation}
where $T$ means transposition.
For solutions of the coupled Schrödinger equation one has the generalization of \eref{sc.28}, 
\begin{equation}
 \frac{\rmd}{\rmd r} \mathrm{W}[\psi(k,r),\psi(\tilde{k},r)] 
 = (k^2-\tilde{k}^2) \psi(k,r)^T \psi(\tilde{k},r)\,,
\label{Wder}
\end{equation}
as can be shown by using the Schrödinger equation twice.
This implies that the Wronskian of two solutions
at identical energies is a constant matrix.
In particular, for functions with behaviours
\eref{mc.9} and \eref{mc.10}, one obtains
\begin{equation}
\mathrm{W}[\varphi(k,r),\chi(k,r)] = -I_N\,.
\label{mc.12}
\end{equation}

Under the same assumptions,
the Schrödinger equation has two $N\times N$
matrix-valued solutions $f(\pm k,r)$
called Jost solutions.
The Jost solutions $f(k,r)$ have the exponential asymptotic behaviour
at large distances
\begin{equation}
f(k,r\to\infty) \to \exp(\rmi k r)
= \diag[\exp(\rmi k_1 r),\ldots,\exp(\rmi k_N r)]\,,
\label{mc.13}
\end{equation}
where we use the exponential of a diagonal matrix.
In general, these solutions are complex and satisfy the symmetry property
$f(k,r)=f^*(-k^*,r)$, where an asterisk denotes complex conjugation.
For real energies below all thresholds, $k=-k^*$ and the Jost solutions are real.
The Jost solutions $f(k,r)$ and $f(-k,r)$
form a basis in the matrix solution space.
The columns of these matrices form a basis
in the $2N$-dimensional vector-solution space
of the Schrödinger equation with a given value of $k$.
The Wronskian of the Jost solutions is given by 
\begin{equation}
\mathrm{W}[f(k,r),f(-k,r)] = -2 \rmi k I_N,
\label{mc.14}
\end{equation}
where $k$ can be either a matrix or a number depending
on the case and on the choice of convention.

The regular solution is expressed in terms of the Jost solutions as 
\begin{equation}
\varphi(k,r) = \frac{\rmi}{2} \left[f(-k,r)k^{-1}F(k) - f(k,r)k^{-1}F(-k)\right],
\label{mc.15}
\end{equation}
where $F(k)$ is known as the Jost matrix.
From \eref{mc.14}, it follows for any $r$ that 
\begin{equation}
F(k) = \mathrm{W}[f(k,r),\varphi(k,r)]\,.
\label{mc.16}
\end{equation}
Equations \eref{mc.9} and \eref{mc.15} provide 
\begin{equation}
f(-k,r\to 0) = f(k,r\to 0) k^{-1} F(-k) F^{-1}(k) k [I_N + {\rm o}(r^{2\nu})]
\label{mc.16a}
\end{equation}
and
\begin{equation}
F(k) = \lim\limits_{r\to 0} f^T(k,r) r^\nu [(2\nu-I_N)!!]^{-1}\,.
\label{mc.17}
\end{equation}

Before using the Jost matrix for scattering,
let us consider it for bound-state properties.
Bound states are obtained when the result
of the multiplication of $\varphi(k,r)$
by some column vector \ve{v} leads to a square-integrable vector solution.
This requires that some wave number matrix $k_0$ exists for which
the first term of $\varphi$ in \eref{mc.15}
multiplied by \ve{v} vanishes, i.e.\ 
\begin{equation}
F(k_0) \ve{v} = 0\,,
\label{mc.18}
\end{equation}
and the remaining Jost solution $f(k_0,r)$ is exponentially decreasing,
i.e.\ the diagonal elements of $k_0$ are purely imaginary with
positive imaginary parts.
The bound-state energies thus correspond to zeros of
the determinant of the Jost function, 
\begin{equation}
\det F(k_0) = 0\,,
\label{mc.19}
\end{equation}
with $\Real k_{0,i} = 0$ and $\Imag k_{0,i} > 0$, $i=1,\ldots,N$.
The corresponding energies $2\mu_i E_0 = k_{0,i}^2+\Delta_i$
are located below all thresholds.
For potentials satisfying the above assumptions,
the number $M$ of bound states is finite.
More generally, if the rank of $F(k_0)$ is $N-m$ ($m>1$),
the problem possesses $m$ degenerate
bound states at the same energy.
While this case does not seem to occur in physical situations,
it may be encountered in exactly solvable potentials generated
by supersymmetric transformations.
Other zeros in \eref{mc.19} correspond to virtual states or resonances.

A physical solution can be obtained from the regular solution
by right multiplication with an invertible matrix.
It has diagonal incoming waves in all channels (assumed open), 
\begin{equation}
\psi(k,r\to\infty) \propto k^{-1/2}
\left[ \rme^{-\rmi kr} \rme^{\rmi l\frac{\pi}{2}}
- \rme^{\rmi kr} \rme^{-\rmi l\frac{\pi}{2}} S(k) \right].
\label{mc.20}
\end{equation}
Each column of matrix $\psi$ describes a collision starting
from a different entrance channel.
In general, not all these entrance channels are accessible to experiment.
The knowledge of the scattering matrix may then be incomplete
which makes the inversion of purely experimental data ambiguous.
A less ambiguous inverse problem is the construction of local coupled-channel potentials
by inversion of theoretical scattering matrices,
calculated e.g.\ with a more elaborate microscopic model 
which takes account of the internal structure of the colliding particles.

From \eref{mc.15}, \eref{mc.13} and \eref{mc.20},
the complex scattering matrix $S(k)$ is expressed in terms of the Jost matrix as 
\begin{equation}
S(k)=\rme^{\rmi l\frac{\pi}{2}} %
k^{-1/2}F(-k)F^{-1}(k)k^{1/2} \rme^{\rmi l\frac{\pi}{2}}.
\label{mc.21}
\end{equation}
The scattering matrix $S(k)$ is unitary and symmetric
and depends thus on $\dem N(N+1)$ independent real parameters.
It can be diagonalized with a real orthogonal matrix $T(k)$ as 
\begin{equation}
S(k) = T(k)\, \rme^{2\rmi \delta(k)} T^T(k)\,,
\label{mc.22}
\end{equation}
where the diagonal elements of matrix
$\delta = \diag(\delta_1, \ldots, \delta_N)$
are the eigenphases \cite{blatt:1952a,blatt:1952b}.
In the multichannel inverse problem, the goal is to derive potentials providing
a given collision matrix or given eigenphases
$\delta_j(k)$ and orthogonal matrix $T(k)$.

We shall sometimes consider the particular case where $l = 0$ and $\nu = 0$.
In this case, the regular solution satisfies \eref{mc.9} under the form 
\begin{equation}
\varphi(k,0)=0, \qquad \varphi'(k,0)=I_N\,.
\label{mc.23}
\end{equation}
The Jost matrix $F(k)$ is given by \eref{mc.17} as 
\begin{equation}
F(k)=f^T(k,0)\,.
\label{mc.24}
\end{equation}
To single out one of the irregular solutions,
we define the irregular solution $\eta(k,r)$ by its behaviour at the origin
\begin{equation}
\eta(k,0)=I_N, \qquad \eta'(k,0)=0\,.
\label{mc.25}
\end{equation}
In terms of the Jost solutions, this irregular solution can be written as
\begin{equation}
\eta(k,r) = \frac{\rmi}{2} \left[f(-k,r)k^{-1}G(k)-f(k,r)k^{-1}G(-k)\right].
\label{mc.26}
\end{equation}
With \eref{mc.14}, matrix $G(k)$ is given by 
\begin{equation}
G(k)=-[f'(k,0)]^T\,.
\label{mc.27}
\end{equation}

The effective range expansion can be generalized to several channels
\cite{delves}.
Let us consider the opening of channels
$i = c, c+1, \ldots, N$ with equal thresholds
and thus equal wave numbers $k_i = k_c = \dots = k_N$.
For $k_i > 0$, one has for the new eigenphases due to the opening channels, 
\begin{equation}
k_i^{2l_i+1} \cot \delta_i = \Delta_{i0} + \Delta_{i1} k_i^2 +
\ldots\,.
\label{mc.28}
\end{equation}
This behaviour is the same as for a single channel although
the eigenphase does not correspond to a specific channel.
The eigenphases of the pre-existing channels with $j<c$ verify
\begin{equation}
\cot \delta_j = \Delta_{j0} + \Delta_{j1} k_i^2 + \ldots
\label{mc.29}
\end{equation}
The new elements of the orthogonal matrix $T$ verify 
\begin{eqnarray}
k_i^{-|l_i-l_{i'}|}T_{ii'} &=&
 T_{ii'0} + T_{ii'1} k_i^2 + \ldots, \qquad i,i' \ge c\,,
\label{mc.30} \\[.5em]
k_i^{-(l_i+1/2)}T_{ij} &=& T_{ij0} + T_{ij1} k_i^2 + \ldots,
\qquad i \ge c, j < c\,,
\label{mc.31} \\[.5em]
k_i^{-(l_i+1/2)}T_{ji} &=& T_{ji0} + T_{ji1} k_i^2 + \ldots,
\qquad i \ge c\,, j < c\,,
\label{mc.32}
\end{eqnarray}
while the old ones satisfy 
\begin{equation}
T_{jj'} = T_{jj'0} + T_{jj'1} k_i^2 + \ldots, \qquad
\qquad \ \ j,j{\,'} < c\,.
\label{mc.33}
\end{equation}
%
%%%%%%%%%%%%%%%%%%%%%%%%%%%%%%%%%%%%%%%%%%%%%%%%%%%%%%%%%%%%%%%%%%%%%%%%%%%%
\section{Single-channel supersymmetric quantum mechanics}
\label{sec:sc}
The general field of supersymmetric transformations of one-dimensional quantum-mechanical systems
is well described in several textbooks \cite{junker:96,bagchi:00,cooper:01,gangopadhyaya:11}.
Here, we focus on the particular application of supersymmetric quantum mechanics to the radial Schrödinger equation
that appears in three-dimensional scattering with central potentials.

\subsection{General properties of single-channel transformations}
A supersymmetric transformation of the radial Schrödinger
equation is an algebraic transformation
of the initial Hamiltonian $H_0$ into a new Hamiltonian $H_1$,
with all properties of $H_1$ being directly expressed
in terms of those of $H_0$ \cite{sukumar:85c}.
The transformation is based on a factorization
of the initial Hamiltonian $H_0$ under the form 
\begin{equation}
H_0 = \Ap \Am + \cE\,, \qquad \cE = -\kappa^2\,,
\label{H0fact}
\end{equation}
where $\cE$ is the so-called factorization energy,
corresponding to a complex wave number $\rmi  \kappa$. 
Below, the corresponding Schrödinger equation is called the $H_0$ equation.
In the following, we shall mostly use real negative factorization energies,
which correspond to imaginary wave numbers,
 i.e.\ real $\kappa$'s chosen positive.
The operators $\Am$ and $\Ap$ are mutually
adjoint first-order differential operators 
\begin{equation}
\Am = - \frac{\rmd}{\rmd r} + w(r)\,, \qquad \Ap = \frac{\rmd}{\rmd r} + w(r)\,,
\label{A0pm}
\end{equation}
where the {\it superpotential} 
\begin{equation}
w(r)=u'(r) u^{-1}(r)
\label{w0}
\end{equation} 
is expressed in terms of the so-called factorization solution $u(r)$,
which satisfies the initial Schrödinger equation at energy $\cE$ 
\begin{equation}
H_0 u(r) = \cE u(r)\,.
\end{equation}
Solution $u(r)$ may be either a normalizable bound-state wave function
or an unbound mathematical solution.
In both cases, its asymptotic behaviour is exponential,
which implies that the superpotential tends to a constant
\begin{equation}
w_\infty = \lim\limits_{r\to \infty} w(r).
\end{equation}
To avoid singularities in $w$, $u$ has to be nodeless.
%which is only possible for an energy
%$\cE$ lower or equal to the ground-state energy
%of $H_0$. (why suppress this sentence?)
This restriction is however lifted when
supersymmetric transformations are iterated
(see \S \ref{sec:pairs} below).

The supersymmetric partner of $H_0$ is 
\begin{equation}
H_1=\Am \Ap +\cE.
\label{H1fact}
\end{equation}
It is Hermitian provided $\cE$ is real.
Again, iterations of transformations allow one to lift that restriction
(see \S \ref{sec:conjE} below).
The corresponding Schrödinger equation is called the $H_1$ equation.
Using \eref{A0pm} and \eref{w0},
it can be checked that $H_1$ has the same form as $H_0$,
except for a different potential 
\begin{equation}
 V_1(r)=V_0(r)-2 w'(r)
\label{V1V0}
\end{equation}
with the useful property \eref{sc.27},
\begin{equation}
w'(r) = V_0(r) - \cE - w^2(r)\,.
\label{wp}
\end{equation}
The Hamiltonians $H_0$ and $H_1$ are related by the intertwining relation 
\begin{equation}
L H_0 = H_1 L\,.
\label{LHHL}
\end{equation}
Moreover, a solution $\psi_0(E,r)$ of the initial
Schrödinger equation at energy $E$
gives rise to a solution $\psi_1(E,r)$ of the transformed
Schrödinger equation at the same energy, which reads 
\begin{equation}
\psi_1(E,r) \propto  \Am \psi_0(E,r) = \mathrm{W}[\psi_0(E,r)\,,
u(r)] u^{-1}(r)\,,
\label{psi1}
\end{equation}
except if $u(r)$ is proportional to $\psi_0(E,r)$.
With two independent solutions of the initial equation
$H_0 \psi_0 = E\psi_0$ at a fixed value of $E$,
one obtains in general two independent solutions of the transformed equation
$H_1 \psi_1 = E\psi_1$. 
In particular, this equation can be used to relate physical solutions
of both Hamiltonians, as well as their Jost solutions: 
the asymptotic behaviour of \eref{psi1}, 
together with the definition \eref{sc.11}, 
implies that the precise relation between the 
Jost solutions of $H_1$ and $H_0$ is 
\begin{eqnarray}
 f_1(k,r) & = & \Am f_0(k,r)(w_\infty-\rmi k)^{-1}.
%& = & \Amf_0(k,r) (\pm \kappa -\rmi k)^{-1}
%\qquad
%(u(\cE) \mathop{\sim}_{r\to\infty} \rme^{\pm \kappa r}).
\label{f1f0}
\end{eqnarray}
At the factorization energy $\cE$, the Wronskian in \eref{psi1} is constant
[see \eref{sc.28}]
and equation $H_1 \psi_1 = \cE \psi_1$ has the solution 
\begin{equation}
 \psi_1(\cE,r) \propto u^{-1}(r).
\label{psi1E0}
\end{equation}
The linearly independent solutions have then to be calculated 
with an equation similar to \eref{sc.7}.
These results allow us to relate all the physical properties 
of $H_1$ to those of $H_0$,
in particular their bound spectra and scattering matrices.
\subsection{Conservative and non-conservative transformations}
Six types of transformations have to be distinguished,
depending on the behaviour of $u$ both at the origin 
and at infinity (see table \ref{transf_prop}).
\begin{table}
%\begin{indented}
%\item[]
\begin{tabular}{@{}cccccccc}
\br
Notation & $\displaystyle\lim_{r \to 0} u$ &
$\nu_0$ & $\nu_1$ &
$\displaystyle\lim_{r \to \infty} u$ &
$F_1(k)$ & mod.\ & $\delta_1(k) - \delta_0(k)$  \\[.5em]
\mr

$T_\mathrm{rem}(E_0)$ & $r^{\nu_0+1}$ &
$\ge 0$ & $\nu_0+1$ &
$\rme^{-\kappa r}$ & $-\displaystyle\frac{F_0(k)}{\k+\rmi k}$ &
rem & $\displaystyle\arctan\frac{k}{\kappa}$ \\[.9em]

$T_\mathrm{l}(\cE)$ & $r^{\nu_0+1}$ &
$\ge 0$ & $\nu_0+1$ &
$\rme^{\kappa r}$ & $\displaystyle\frac{F_0(k)}{\k-\rmi k}$ &
none & $-\displaystyle\arctan\frac{k}{\kappa}$   \\[.5em]

$T_\mathrm{add}(\cE, \alpha) $ &  $r^{-\nu_0} $ &
$>0$ & $\nu_0-1$ &
$\rme^{\kappa r}$  & $-(\k+\rmi k)F_0(k)$ &add &
$-\displaystyle\arctan\frac{k}{\kappa}$  \\[.5em]

$T_\mathrm{r}(\cE)$ &  $r^{-\nu_0}$  &
$>0$ & $\nu_0-1$ &
$\rme^{-\kappa r}$ & $(\k-\rmi k)F_0(k)$ &
none & $\displaystyle\arctan\frac{k}{\kappa}$ \\[.5em]

$T_\mathrm{nc}(\cE,w(0))$ & 1 &
0 & 0 &
$\rme^{\kappa r}$ &
$\displaystyle\frac{G_0(k)+w(0)F_0(k)}{\kappa -\rmi k}$ & all & ? \\[.9em]

$T_\mathrm{rnc}(\cE)$ & 1 &
0 & 0 &
$\rme^{-\kappa r}$ &
$\displaystyle\frac{G_0(k)+w(0)F_0(k)}{-\kappa -\rmi k}$ & all & ?

\end{tabular}
%\end{indented}
\caption{Summary of the properties of single-channel supersymmetric transformations: 
singularity parameter, Jost function, bound-spectrum modification 
(mod.) and phase shift (up to a multiple of $\pi$). 
``rem'' stands for removal, ``add'' for addition, 
``l'' for left-regular factorization solution, 
``r'' for right-regular factorization function 
and ``nc'' for non-conservative transformation. 
}
\label{transf_prop}
\end{table}
Whereas \eref{f1f0} implies that
all transformations conserve the behaviour of solutions at infinity, 
e.g.\ transform an exponentially decreasing
solution $\psi_0$ into an exponentially decreasing solution $\psi_1$,
only four transformation types conserve the behaviour of $\psi_0$ at the origin,
i.e.\ transform a regular solution into a regular solution.
Such transformations are called {\em conservative} and denoted by $T_\mathrm{rem}$,
$T_\mathrm{l}$, $T_\mathrm{add}$ and $T_\mathrm{r}$.
These notations summarize the main feature of each transformation:
$T_\mathrm{rem}$ (resp.\ $T_\mathrm{add}$) modifies the bound spectrum
by removing (resp.\ adding) a bound state
while $T_\mathrm{l}$  (resp.\ $T_\mathrm{r}$) does not modify the bound spectrum
but corresponds to a factorization solution regular ``on the left'' (resp.\ ``on the right'') only,
i.e.\ regular at the origin but not at infinity (resp.\ regular at infinity but not at the origin).

For conservative transformations,
the singularity parameter of the new potential reads $\nu_1=\nu_0\pm 1$,
as obtained by expanding \eref{V1V0} at the origin
and by defining the singularity parameters $\nu_0$ and $\nu_1$
for potentials $V_0$ and $V_1$ as in \eref{sc.5}.
The bound spectrum of $H_1$ is identical to that of $H_0$
(transformations $T_\mathrm{l}$ and $T_\mathrm{r}$),
with the possible exception of $\cE$,
which may be either added to (transformation $T_\mathrm{add}$)
or removed from (transformation $T_\mathrm{rem}$) the bound spectrum.
Finally, the transformed Jost function and scattering matrix 
must be determined. 

With \eref{sc.6} and \eref{sc.8},
for $u(r \to 0) \to r^{-\nu_0}$ ($\nu_0 > 0$, $T_\mathrm{r}$ and $T_\mathrm{add}$),
the superpotential behaves at the origin as 
\begin{equation}
w(r \to 0) = - \frac{\nu_0}{r}
\label{sct.12a}
\end{equation}
and the regular solution transforms according to 
\begin{equation}
L \varphi_0^{(\nu_0)}(k,r) = - \varphi_1^{(\nu_0-1)}(k,r)\,,
\label{sct.12}
\end{equation}
where the superscript recalls the behaviour at the origin.
For $u(r \to 0) \to r^{\nu_0+1}$ ($T_\mathrm{l}$ and $T_\mathrm{rem}$),
the superpotential behaves at the origin as 
\begin{equation}
w(r \to 0) = \frac{\nu_0 + 1}{r}
\label{sct.13a}
\end{equation}
and the regular solution transforms according to
\begin{equation}
L \varphi_0^{(\nu_0)}(k,r) = (k^2 + \kappa^2)\, \varphi_1^{(\nu_0+1)}(k,r)\,.
\label{sct.13}
\end{equation}
These relations combined with \eref{sc.15} and \eref{f1f0}
lead to the Jost function.
The transformed Jost function and scattering matrix
are obtained by the multiplication of the initial quantities
by a first-order rational function
of the wave number,
which corresponds to the phase-shift modification
\begin{equation}
 \delta_1(k) = \delta_0(k)-\arctan \frac{k}{w_\infty}
\equiv \delta_0(k)-\epsilon \arctan \frac{k}{\kappa} \,,
\end{equation}
up to a multiple of $\pi$.
The parameter $\epsilon$ is defined according to the asymptotic behaviour of the factorization solution as 
\begin{equation} 
\epsilon =\left\{
 \begin{array}{ll}
+1 & (T_\mathrm{l}, T_\mathrm{add}), \\
-1 & (T_\mathrm{r}, T_\mathrm{rem})\,.
\end{array}
\right.
\label{eps}
\end{equation}
Indeed, the asymptotic value of the superpotential,
$w_\infty$, takes the value $\pm\kappa$,
depending on the exponentially increasing or
decreasing asymptotic behaviour of the factorization solution.
These properties are established in \cite{sukumar:85b,sukumar:85c}
and summarized in the first four lines of table \ref{transf_prop}.
For the $T_\mathrm{rem}$ transformation,
the factorization energy $\cE$ equals the ground-state energy
$E_0$ and this ground state is removed,
while for the $T_\mathrm{add}$ transformation,
the factorization solution being irregular
at the origin depends on an arbitrary parameter
(called $\alpha$ in table \ref{transf_prop})
in addition to the factorization energy and a new state is added at this energy.

The fifth and sixth types of transformation are less studied in the literature;
they are more complicated
as they do not transform the regular solution into a regular solution.
They are thus called {\em non conservative}.
They occur when the initial potential is regular at the origin,
i.e., $\nu_0=0$,
and the factorization solution is chosen singular, i.e., 
\begin{equation}
u(r\to 0) \sim 1 + w(0) r + \Or(r^2)\,,
\end{equation}
where the arbitrary parameter $w(0)$
is the value of the superpotential at the origin.
For non-conservative transformations, the superpotential
is regular at the origin.
Equation \eref{V1V0} then implies that
the transformed potential is also regular at the origin,
hence $\nu_1=\nu_0=0$.
The modifications of the bound spectrum and of the phase shifts
through these transformations
are determined by the modification of the Jost function.
A regular solution of the transformed potential is obtained with 
\begin{equation}
L [\eta_0(k,r) + w(0) \varphi_0(k,r)] = (k^2 + \kappa^2) \varphi_1(k,r)
\label{sct.15}
\end{equation}
where $\eta_0(k,r)$ is defined by \eref{mc.25}
for $N = 1$ and \eref{wp} has been used.
Using this expression or definition \eref{sc.17}
with $\nu=0$ or \eref{mc.24} for $N=1$,
together with the Jost solution transformation \eref{f1f0},
one gets the transformed Jost function
\begin{equation}
 F_1(k) = f_1(k,0) = \frac{G_0(k)+w(0) F_0(k)}{w_\infty - \rmi k}\,,
%\qquad
%(u(\cE) \mathop{\sim}_{r\to\infty} \rme^{\pm \kappa r}).
\label{nonconsF1}
\end{equation}
where $G_0$ is defined by \eref{mc.27} for $N=1$.
This transformation thus introduces a pole to the Jost function
in $k = -\rmi w_\infty = \mp \rmi \kappa$.
Moreover, it may introduce one or several zeros,
the wave numbers of which satisfy
\begin{equation}
 G_0(k)=-w(0) F_0(k)
\label{nonconsBS}
\end{equation}
and are determined by the slope of the factorization
solution at the origin, $w(0)$.
Equation \eref{nonconsBS} shows that in the present case
there is no simple connection between the bound states of $V_0$ and $V_1$,
in contrast with conservative transformations.
Similarly, the scattering matrix and phase shifts of
$V_1$ do not have simple expressions,
except in particular cases where the function $G_0(k)$ is simple
(see examples in \S \ref{sec:Eckart} and \S \ref{sec:l1pot} below).
The properties of these non-conservative transformations are summarized
in the fifth and sixth lines of table \ref{transf_prop}.
Note that, as for a $T_\mathrm{add}$ transformation,
the factorization solution of a $T_\mathrm{nc}$ transformation
depends on an arbitrary parameter $w(0)$,
in addition to the factorization energy.
\subsection{Normalization of solutions through supersymmetric transformations}
\label{sec:norm}
Let us now specify the normalization in \eref{psi1} for different
types of solutions $\psi_0(E, r)$.

When $\psi_0(E_i, r)$ is a normalizable bound-state wave function,
it is normalized as 
\begin{equation}
\int_0^\infty \psi_0^2(E_i,r) \rmd r = 1\,.
\end{equation}
Solution $\psi_1(E_i,r)$ is also normalizable.
Since the factorization energy $\cE$ must be lower or equal to the ground state
of $H_0$ in order to avoid singularities in potential $V_1(r)$,
one has 
\begin{equation}
\cE<E_i.
\label{E0<E}
\end{equation}
It is then well known
that the appropriate normalization in \eref{psi1} must read 
\begin{eqnarray}
\psi_1(E_i,r) & = & \frac{1}{\sqrt{E_i-\cE}}\, \Am \psi_0(E_i,r) \\
& = & \frac{1}{\sqrt{E_i-\cE}}\, \mathrm{W}[\psi_0(E_i,r), u(r)] u^{-1}(r)
\label{psi1psi0norm}
\end{eqnarray}
in order to have $\psi_1(E_i,r)$ normalized \cite{sukumar:85a}.

Let us now consider solutions $\psi_0(E,r)$ of the initial Schrödinger equation
with a normalizable inverse and normalize them as 
\begin{equation}
\int_0^\infty \frac{1}{\psi_0^2(E,r)}\, \rmd r = 1\,.
\label{invnorm}
\end{equation}
Such solutions have to be singular both at the origin and at infinity;
hence they are always non physical.
Moreover, they have to be nodeless, which means $E$ has to be lower than the
ground-state energy.
Let us now consider the corresponding solutions $\psi_1(E,r)$
of the transformed equation, as defined by \eref{psi1}.
For conservative transformations,
they are also singular at the origin and at infinity.
For $\psi_1(E,r)$ to be nodeless with $\psi_0(E,r)$ nodeless,
\eref{E0<E} needs be satisfied, as in the previous case.

When all these conditions are satisfied,
$\psi_1(E,r)$ is normalized as in \eref{invnorm}, 
\begin{equation}
\int_0^\infty \frac{1}{\psi_1^2(E,r)}\, \rmd r = 1\,,
\label{psi1norm}
\end{equation}
as can be directly verified by replacing
$\psi_1(E,r)$ in \eref{psi1norm} by its explicit expression
\eref{psi1psi0norm} and integrating by parts using \eref{sc.28}: 
\begin{eqnarray}
\int_0^\infty \frac{1}{\psi_1^2(E,r)}\, \rmd r
& = & - \int_0^\infty \frac{u(r)}{\psi_0(E,r)}
\frac{\rmd \mathrm{W}^{-1}[\,\psi_0(E,r), u(r)\,]}{\rmd r}\, \rmd r 
\nonumber \\[.5em]
& = & \left[ \frac{1}{\sqrt{E-\cE}\psi_0(E,r)\psi_1(E,r)} 
\right]^0_\infty + \int_0^\infty \frac{1}{\psi_0^2(E,r)}\, \rmd r\,,
\label{norminv}
\end{eqnarray}
where the last equality follows from
$(u/\psi_0)'=\mathrm{W}[\psi_0,u]/\psi_0^2$\,.
The first term of \eref{norminv} vanishes since both $\psi_0(E,r)$
and $\psi_1(E,r)$ are singular at the origin and at infinity,
hence the announced property.
\subsection{Pairs of transformations}
\label{sec:pairs}
Let us now iterate two supersymmetric transformations
\begin{equation}
H_0 \mathop{\longrightarrow}_{u (\cE_0, r)}
H_1 \mathop{\longrightarrow}_{v (\cE_1, r)} H_2\,,
\end{equation}
where $u (\cE_0, r)$ is a solution of the $H_0$ equation at energy $\cE_0$
and $v (\cE_1, r)$ is a solution of the $H_1$ equation at energy $\cE_1$.
Solution $v (\cE_1, r)$ can be expressed as
$Lu (\cE_1, r)$ if $\cE_1 \neq \cE_0$.
The new potential $V_2(r)$ reads, by iteration of \eref{V1V0},
\begin{eqnarray}
V_2(r) & = & V_1(r)-2 \frac{\rmd^2}{\rmd r^2} \ln v (\cE_1, r) \\[.5em]
& = & V_0(r) - 2 \frac{\rmd^2}{\rmd r^2} \ln[v (\cE_1, r) u (\cE_0, r)]
\equiv V_0(r) - 2 W'(r)\,,
\label{V2V1V0}
\end{eqnarray}
where $W(r)$, the pair superpotential, has been defined.
This superpotential $W(.)$ should not be confused
with the Wronskian $\mathrm{W}[.,.]$.
Since the solutions of the $H_0$ equation are in general simpler
than the solutions of the $H_1$ equation,
it is interesting to express $W(r)$ in terms of solutions of
the $H_0$ equation only.
The obtained expressions depend on whether $\cE_0$ and $\cE_1$
are equal or not;
these different cases will be discussed separately below
(\S \ref{sect_equal_fact} and \S \ref{subsubsec:pairdiff}).
Let us also note that, for conservative transformations,
the transformed phase shift has the simple expression,
up to a multiple of $\pi$,
\begin{equation}
 \delta_2(k)=\delta_0(k) - \epsilon_0 \arctan\frac{k}{\kappa_0}
 - \epsilon_1 \arctan\frac{k}{\kappa_1}\,,
\label{d2}
\end{equation}
where $\epsilon_{0,1}$ are defined as in \eref{eps}.
Iterating more conservative transformations simply adds more
arctangent terms to \eref{d2},
which is the basis of the supersymmetric inversion algorithm of
\cite{sparenberg:97a} (see \S \ref{sec:chatde}).
\subsubsection{Equal factorization energies and phase-equivalent potentials}
\label{sect_equal_fact}
Equation \eref{d2} shows that phase-equivalent potentials,
i.e.\ potentials sharing the same phase shifts,
can be obtained when the two successive transformations
have the same factorization energies,
with factorization solutions displaying different asymptotic behaviours.
Three such transformation pairs have been extensively studied
in the literature \cite{baye:87a,baye:87b,baye:94}:
$(T_\mathrm{rem},T_\mathrm{l})$ for a phase-equivalent bound-state removal,
$(T_\mathrm{r},T_\mathrm{add})$ for a phase-equivalent bound-state addition,
and $(T_\mathrm{rem},T_\mathrm{add})$ for a phase-equivalent arbitrary change
of the bound-state ANC,
which can be linked to the arbitrary parameter $\alpha$
appearing in $T_\mathrm{add}$.
Such transformation pairs with equal factorization
energies are also sometimes called ``confluent''
\cite{mielnik:00,fernandez:03b,fernandez:05}.

As an example, let us detail the expressions for
the addition of a bound state at energy $\cE$.
The singularity of the initial potential must be larger 
than 2 in this case: $\nu_0 \ge 2$.
The first factorization solution, corresponding to the
$T_\mathrm{r}$ transformation,
is a solution vanishing at infinity and diverging at the origin,
i.e.\ proportional to the Jost solution,
$u(r) \propto f(\rmi \kappa,r)$.
The second factorization solution, corresponding
to the $T_\mathrm{add}$ transformation,
diverges both at the origin and at infinity;
it reads
\begin{equation}
v(r) \propto u^{-1}(r) \int_r^\alpha u^2(t)\, dt\,.
\end{equation}
Hence, the pair superpotential is
\begin{equation}
 W(r) = \frac{\rmd}{\rmd r} %
 \ln \int_r^\alpha f^{\,2}(\rmi \kappa, t)\, dt\,.
\label{W_PEadd}
\end{equation}
Remarkably, this superpotential and $V_2$ have no singularity,
even for a factorization energy larger than the ground-state energy of potential $V_0$,
whereas $V_1$ is singular in this case;
the second transformation removes the singularities
of the intermediate potential.
This feature is general for phase-equivalent transformation pairs \cite{baye:93a}
and is a particular case of irreducible second-order supersymmetric transformations \cite{andrianov:12}.
Moreover, the function $f(\rmi \kappa, t)$ may belong to
the continuous spectrum of $V_0$ thus producing a
{\em continuum bound state},
i.e.\ a bound state embedded in the continuum of scattering states at $\cE>0$ \cite{pappademos:93,Weber:1994}.
\subsubsection{Different factorization energies}
\label{subsubsec:pairdiff}
In this case, a compact expression for the superpotential
is well known and has even been established for an
arbitrary number of transformations with different factorization energies
\cite{deift:79,Crum,Krein,sparenberg:95} (see \S \ref{sec:chatde}).
For two transformations, it reads
\begin{equation}
W(r) = \frac{\rmd}{\rmd r}\ln \mathrm{W}[u (\cE_0, r), u (\cE_1, r)\,]
\qquad (\cE_1 \ne \cE_0)\,,
\label{W}
\end{equation}
where $u (\cE_1, r)$ is a transformation
 solution of the $H_0$ equation at energy
$\cE_1$, related to $v (\cE_1, r)$ through \eref{psi1},
the precise normalization having no influence on the
result because of the logarithmic derivative.

Though this result is very compact and elegant,
it is not very convenient for applications because
of the successive derivative calculations.
An alternative formula for $W(r)$ in terms of solutions of
the initial equation and of their first derivative only reads,
using \eref{sc.9} and \eref{sc.28},
\begin{equation}
W(r) =
\frac{(\cE_0-\cE_1)\,u(\cE_0, r)\, u(\cE_1, r)}
{\mathrm{W}[u(\cE_0, r), u(\cE_1, r)]}
= \frac{\cE_0-\cE_1}
{w_1(r)-w_0(r)}\,,
\label{W2}
\end{equation}
with
\begin{equation}
w_i(r) = \frac{u'(\cE_i, r)}{u(\cE_i, r)}\,, \quad i=0,1\,.
\end{equation}
Introducing this result into \eref{V2V1V0}
and taking \eref{sc.27}
into account, one gets
\begin{equation}
V_2(r) = V_0(r) + 2 (\cE_1-\cE_0)\,
\frac{\cE_1-\cE_0+w_1^2(r) -w_0^2(r)} {[w_1(r)-w_0(r)]^2}\,.
\label{V2expl}
\end{equation}

Let us now discuss the conditions under which
the above formulas lead to physically acceptable potentials,
i.e.\ potentials without singularity.
Equation \eref{V2expl} is somewhat misleading in this respect:
it seems to imply that neither $u(\cE_0,r)$ nor $u(\cE_1,r)$
may vanish in order for the potential to be regular.
Actually, \eref{W} shows that physical potentials
are obtained as soon as the Wronskian does not vanish
and is either real or purely imaginary for $r\in\left(0,\infty\right)$. 
These conditions guarantee the absence of singularities 
and the reality of the potential.

\subsubsection{Mutually conjugate factorization energies}
\label{sec:conjE}

For single-channel inverse-scattering applications, 
real factorization energies turn out to be sufficient in general 
\cite{samsonov:02}. 
One noticeable exception, for which complex energies are required, 
is the fit of resonances. 
Let us detail this situation 
as it will also be encountered in the coupled-channel case. 
A similar discussion, for the full-line Schrödinger equation rather than 
for the radial one, can be found in \cite{andrianov:95,fernandez:03}. 

We consider a pair of $T_\mathrm{r}$ transformations with mutually 
conjugate factorization energies 
$\cE \equiv \cE_R + \rmi \cE_I \equiv -\alpha^2$ and $\cE^*$,
and with mutually conjugate factorization solutions $u(\cE,r)$ and $u(\cE^*, r) \equiv u^*(\cE, r)$
exponentially decreasing at infinity.
Defining $\alpha \equiv \alpha_R + \rmi  \alpha_I$ with $\alpha_R > 0$,
these solutions thus behave asymptotically like $\rme^{-\alpha r}$.
To fix ideas, we make the sign choice $\alpha_I>0$, 
which corresponds to $\cE_R>0$ and $\cE_I<0$. 
Equation \eref{d2} shows that this pair adds a resonant term 
to the phase shift, 
\begin{eqnarray}
 \delta_2(k)-\delta_0(k) && = \arctan \frac{k}{\alpha}
 + \arctan \frac{k}{\alpha^*}
= \arctan \frac{2 \alpha_R k}{|\alpha|^2 - k^2}
\label{d2res} \\[.5em]
&& \mathop{\approx}_{k\approx \alpha_I \gg \alpha_R}%
 \arctan \frac{|\cE_I|}{\cE_R-E}\,.
\end{eqnarray}
For a narrow resonance, i.e.\ when
$\cE_R \gg |\cE_I|$ or $\alpha_I \gg \alpha_R$,
this expression reduces to a Breit-Wigner
term when $k$ approaches the resonance wave number \cite{Roman}.

Equation \eref{W2} also shows that the pair superpotential is real in this case,
as $u(\cE^*, r)=u^*(\cE, r)$.
Indeed, defining $w(r) \equiv u'(\cE, r)/u(\cE, r)$, one has 
\begin{equation}
 W(r)=-\frac{\cE_I}{\Imag w(r)}\,.
\label{sct.35}
\end{equation}
This pair of transformations thus relates two real potentials with each other,
whereas the intermediate potential obtained after one transformations is complex.
This possibility, first explored in \cite{AIN2,AIN1} (see also \cite{BS})
is a particular case of second-order irreducible supersymmetry \cite{irred,andrianov:12}.
Let us finally remark that the $T_\mathrm{r}$ transformation functions are singular at the origin 
and can thus only be used for a potential $V_0$ with $\nu_0 \ge 2$. 
This drawback can be eliminated in chains of transformations 
as shown in \S \ref{sec:chatsp}.
\subsection{Chains of transformations}
\label{sec:chat}
\subsubsection{Chains of transformations with different factorization energies}
\label{sec:chatde}
Supersymmetric transformations can be iterated 
to form a chain of transformations \cite{sparenberg:95}. 
With some restrictions discussed below, the six types of transformations
displayed in table \ref{transf_prop} can appear
and can be useful in such chains. 
The iteration transforms Hamiltonian $H_0$ into $H_n$,
\begin{equation}
H_0 \mathop{\longrightarrow}_{v_0 (\cE_0, r)}
H_1 \mathop{\longrightarrow}_{v_1 (\cE_1, r)} H_2\ 
\dots\  
H_{n-1} \mathop{\longrightarrow}_{v_{n-1} (\cE_{n-1}, r)} H_n\,,
\label{chatde.0}
\end{equation}
where $v_i (\cE_i, r)$ is a solution of $H_i v_i (\cE, r) = \cE v_i (\cE, r)$ 
at energy $\cE_i = -\kappa_i^2$ ($\Real{\kappa_i} > 0$) 
and all energies are different and may be complex. 
The transformed potential is given by 
\begin{equation}
V_n(r) = V_0(r) - 2\frac{\rmd^2}{\rmd r^2} \ln \left[ v_{n-1}(\cE_{n-1},r) \dots v_1(\cE_1,r) v_0(\cE_0,r) \right].
\label{chatde.1}
\end{equation}
The wave functions of $H_n$ are transformed from the wave functions of $H_0$ according to 
\begin{equation}
\psi_n(r) = L_{n-1} \dots L_1 L_0 \psi_0(r) 
\label{chatde.2}
\end{equation}
where 
\begin{equation}
L_i = - \frac{\rmd}{\rmd r} + \frac{v_i' (\cE_i, r)}{v_i (\cE_i, r)}.
\label{chatde.3}
\end{equation}
Equation \eref{chatde.2} can also be written with the $n$th-order differential operator
\begin{equation}
\label{Ln}
L^{(n)} = L_{n-1} \dots L_1 L_0 = (-1)^{n} \frac{\rmd^n}{\rmd r^n}+a_{n-1}(r)\frac{\rmd^{n-1}}{\rmd r^{n-1}}+\ldots+a_0(r),
\end{equation}
which defines the functions $a_i, i=0,\dots,n-1$.
These depend on the factorization solutions $v_i (\cE_i, r)$,
which can be expressed as functions of solutions $u_0(\cE_i, r)$ of $H_0 u_0 (\cE, r) = \cE u_0 (\cE, r)$ 
at energy $\cE_i$ by 
\begin{equation}
v_i (\cE_i, r) = L^{(i)} u_0(\cE_i, r), 
\quad v_0 (\cE_0, r) = u_0 (\cE_0, r).
\label{chatde.4}
\end{equation}
Hence the potential $V_n$ in $H_n$ can be expressed with a Wronskian 
of functions $u_i(r) \equiv u_0(\cE_i, r)$ as
\begin{equation}
V_n(r) = V_0(r) - 2\frac{\rmd^2}{\rmd r^2} \ln \mathrm{W}[u_0, u_1, \dots, u_{n-1}].
\label{chatde.5}
\end{equation}
In the same way, one obtains  
\begin{equation}
\psi_n(r) = \frac{\mathrm{W}[u_0, u_1, \dots, u_{n-1}, \psi_0]}
{\mathrm{W}[u_0, u_1, \dots, u_{n-1}]}.
\label{chatde.6}
\end{equation}
Such expressions are known as Crum-Krein formulas \cite{Crum,Krein,sparenberg:95}.

A consequence of \eref{chatde.5} and \eref{chatde.6} 
is that the order of the transformations in \eref{chatde.0} is irrelevant. 
Between expressions \eref{chatde.1} and \eref{chatde.5} of potential $V_n$, 
intermediate expressions may also be useful. 
For $0 \le k < n$, this potential can be written as 
\begin{equation}
V_n(r) = V_k(r) - 2\frac{\rmd^2}{\rmd r^2} \ln \mathrm{W}[v_{kk}, v_{kk+1}, \dots, v_{kn-1}].
\label{chatde.8}
\end{equation}
where $v_{ki}(r) \equiv v_k(\cE_i,r)$. 
Examples of use of this formula can be found 
in \S \ref{sec:trivtr} and \S \ref{sec:npS}. 

Potential $V_n$ must however satisfy two conditions. 
(i) It should be real. 
In practice, this restricts energies $\cE_i$ to real values 
and to pairs of mutually conjugate values. 
(ii) For conservative transformations, the singularity index $\nu_n$ of $V_n$ should be positive. 
The total number of $T_\mathrm{r}$ and $T_\mathrm{add}$ transformations 
should then be smaller than, or equal to, $\nu_0$ plus the total number 
of $T_\mathrm{l}$ and $T_\mathrm{rem}$ transformations.

For conservative transformations, the phase shifts of potential $V_n$ are given by 
\begin{equation}
\delta_n(k) = \delta_0(k) - \sum_{i=0}^{n-1} \epsilon_i \arctan\frac{k}{\kappa_i}.
\label{chatde.7}
\end{equation}
The coefficients $\epsilon_i = \pm 1$ are defined as in \eref{eps}. 
This expression is real since complex $\kappa_i$ values appear in conjugate pairs.
For non-conservative transformations,
we could not find a compact expression for the phase shift in the general case.
In section \ref{sec:sci}, explicit examples will be provided in the important particular case
of a vanishing initial potential.
\subsubsection{Phase-equivalent chains of transformations}
\label{sec:chatpe}
Chains where factorization energies are not all different 
can also be useful. 
An important example is given by phase-equivalent potentials 
\cite{ancarani:92,baye:93a,baye:94}. 
With a chain of $n$ phase-equivalent pairs of transformations 
(see \S \ref{sect_equal_fact}), 
the most general form of such potentials is given by 
\begin{equation}
V_{2n} = V_0 - 2\frac{\rmd^2}{\rmd r^2} \ln \det X_0.
\label{chatpe.1}
\end{equation}
where $X_0$ is an $n \times n$ matrix with elements 
\begin{equation}
X_{0; ij} = \beta_i \delta_{ij} 
+ \int_r^\infty u_0(\cE_i,r') u_0(\cE_j,r') \rmd r'.
\label{chatpe.2}
\end{equation}
The functions $u_0(\cE_i,r)$ are various solutions of the $H_0$ equation at energy $\cE_i$. 
See \cite{baye:94} for details and for the choices of $\cE_i$ 
and $\beta_i$. 
According to these choices, bound states are suppressed or added 
to the spectrum of $H_0$ and the ANCs of some states are modified. 
\subsubsection{Chains of transformations adding resonances}
\label{sec:chatsp}
We now consider a particular chain of transformations with different energies 
where resonances can be introduced without the drawback mentioned in \S \ref{sec:conjE}, 
i.e.\ the final potential has the same singularity at the origin as the initial potential 
\cite{samsonov:02,SPRS}. 
As we show below in \S \ref{sec:trivtr}, 
the whole class of potentials known as Bargmann-type potentials \cite{bargmann:49b,chadan:89}
can be obtained with the help of 
either pairs of usual supersymmetric transformations or their confluent forms \cite{SJPA95}. 

A special chain of $4n$ transformations is realized by using 
$n$ pairs of $T_\mathrm{r}$ transformation solutions 
at mutually conjugate energies $\cE_{2j-1} = -\alpha_j^2$ 
with $\alpha_{jR} > 0$ and $\cE_{2j} = \cE_{2j-1}^*$ for $j=1$ to $n$
and $2n$ $T_\mathrm{l}$ transformation solutions at real energies 
$\cE_{2n+j+1} = -\kappa_j^2$ with $\kappa_j > 0$ for $j = 0$ to $(2n-1)$. 
All factorization constants $\cE_j$ should be different from each other.
The functions $u_0(\cE_j,r) $ are distinguished by their behaviour at the origin. 
The $2n$ $T_\mathrm{l}$ functions are regular and 
the $2n$ $T_\mathrm{r}$ functions are irregular at the origin. 
Thanks to the $T_\mathrm{r}$ functions, $V_{4n}$ has the same singularity as $V_0$. 

Using the results of table \ref{transf_prop},
it is shown that such a chain of transformations modifies 
the initial Jost function $F_0(k)$ of Hamiltonian $H_0$ into the Jost function 
\begin{equation}
F_{4n}(k) = F_0(k) \prod_{j=1}^{n} 
\frac{(k+\rmi\alpha_j)(k+\rmi\alpha_j^*)}{(k+\rmi\kappa_{2j-2})(k+\rmi\kappa_{2j-1})}\,,
\label{F_n}
\end{equation}
corresponding to Hamiltonian $H_{4n}$. 
The Jost function $F_{4n}(k)$ differs from $F_0(k)$ by a rational function of momentum $k$. 
% Every purely imaginary $\alpha_j$ with a positive imaginary part $\Imag{\alpha_j} > 0$ 
% corresponds to a discrete level with energy $-|\alpha_j|^2$ of $V_{4n}$ given 
% by \eref{chatde.5}. 
Every pair of complex $\alpha_j$ corresponds to a resonance with 
complex energy $-\alpha_j^2$ like in \S \ref{sec:conjE}. 
% States with $\Real{\alpha_j} > |\Imag{\alpha_j}|$ 
% correspond to a visible resonance \cite{Roman} 
% provided $(\Real{\alpha}_j)^2 - (\Imag{\alpha}_j)^2$ is small enough. 
The corresponding expression for $\delta_{4n}(k)$ is a particular case of \eref{chatde.7} 
which generalizes \eref{d2res}, 
\begin{equation}
\delta_{4n}(k) = \delta_0(k) 
+ \sum_{j=1}^n \arctan \frac{2 \alpha_{jR} k}{|\alpha_j|^2-k^2} 
- \sum_{j=0}^{2n-1} \arctan \frac{k}{\kappa_j} \,.
\label{delta_n}
\end{equation}
In \S \ref{sec:trivtr} we apply this technique to obtain 
potentials with either one or two resonance states, starting from $V_0=0$. 
It is worth noting that solutions of the Schrödinger equation 
for these potentials are expressed in terms of elementary functions 
although their explicit form may be rather involved.
\section{Supersymmetric potentials for single-channel inverse problems}
\label{sec:sci}
Let us now review several examples of inverse problems,
as solved by supersymmetric quantum mechanics,
for the single-channel case.
We limit ourselves to the neutral case
(for the charged case, we refer the reader to \cite{sparenberg:97a}).
We first consider schematic problems that are the building blocks of the
iterative inversion procedure.
They provide a detailed comparison of conservative and
non-conservative transformations. 
Then we apply the results to the neutron-proton system,
both for $l=0$ and $l\ne 0$ partial waves.
For $l=0$, the constructed potentials generalize the Bargmann potentials,
i.e.\ potentials for which the Jost function and scattering matrix are rational functions
of the wave number \cite{bargmann:49a,bargmann:49b,newton:82,chadan:89,zakhariev:90}.
\subsection{$S$-wave Bargmann-type potentials with one bound state}
\subsubsection{Bargmann potentials with conservative transformations}
We construct a potential with a single bound state 
by the iterative application of two supersymmetric transformations 
on the zero potential $V_0(r)=0$:
\begin{equation}
V_0 \equiv 0 \mathop{\longrightarrow}_{T_\mathrm{l} (\cE_0)}
V_1 \mathop{\longrightarrow}_{T_\mathrm{add}(\cE_1, \alpha)} V_2. 
\label{Barg0cons}
\end{equation}
The initial potential has unity Jost function $F_0(k)=1$, unity scattering matrix $S_0(k)=1$, 
and the free-wave scattering solution $\psi_0(k,r)=\sin(kr)$.
For negative energies, the solutions of the initial
Schrödinger equation are linear combinations
of exponential functions, which leads to simple expressions for the transformed
potentials.

A first transformation, of the $T_\mathrm{l}$ type,
is performed with a nodeless factorization function $u(\cE_0, r)$
regular at the origin and singular at infinity,
\begin{equation}
u(\cE_0, r) \propto \sinh(\kappa_0 r)\,,
\label{Bargsol0}
\end{equation}
with the definition $\kappa_0=\sqrt{-\cE_0}>0$.
It leads to a purely-repulsive (and hence without bound state)
transformed potential
\begin{equation}
V_1(r)=\frac{2 \kappa_0^2}{\sinh^2(\kappa_0 r)}\,.
\label{BargV1}
\end{equation}
This potential is singular at the origin,
has a Jost function with a pole in the lower half plane
\begin{equation}
F_1(k) = \frac{1}{k+\rmi  \kappa_0}
\end{equation}
and an $S$ matrix with a pole in the upper half plane
\begin{equation}
S_1(k) = \frac{k+\rmi  \kappa_0}{-k+\rmi  \kappa_0}\,.
\label{BargS1}
\end{equation}
Let us stress that this pole is not associated to any bound state,
as the Jost function has no zero in the upper half plane;
this shows that the Jost function contains more physical information than
the scattering matrix.
Such a behaviour is not usually stressed in textbooks
where the Jost function is often supposed to be analytic on the whole plane. 
This is only true for truncated potentials 
whereas most short-range potentials used in practice display an exponential decay,
as potential $V_1(r)$, and hence there is no systematic link 
between $S$-matrix poles and bound states. 
The phase shift corresponding to $S$ matrix \eref{BargS1} reads
\begin{equation}
\delta_1(k) = -\arctan\frac{k}{\kappa_0}\,,
\label{Bargd1}
\end{equation}
which can also be extracted from the asymptotic behaviour
of the scattering wave function 
\begin{equation}
 \psi_1(k,r)=L\psi_0(k,r) =
 -k \cos(kr) + \kappa_0 \, \mathrm{cotanh}(\kappa_0 r) \sin(kr).
\end{equation}
The phase shift \eref{Bargd1} corresponds to a one-term
effective-range expansion
\eref{ere} with scattering length
\begin{equation}
a=\kappa_0^{-1}
\end{equation}
and all other terms zero.
This phase shift does not vanish at infinity,
in agreement with the generalized
Levinson theorem for an $r^{-2}$ singular potential
with a singularity parameter $\nu_1=1$ \cite{swan:63}.

A second transformation of the $T_\mathrm{add}$ type
is performed with a nodeless factorization function
$v(\cE_1, r)$ singular both at the origin and infinity.
This transformation regularizes the potential at the origin
and adds a bound state at energy $\cE_1$,
giving a one-bound-state Bargmann potential.
In the spirit of the discussion of \S \ref{subsubsec:pairdiff},
we do not directly discuss the properties of $v(\cE_1, r)$,
which is solution of the $H_1$ equation and has a rather intricate expression;
we rather concentrate on $u(\cE_1, r)$,
which is also singular at the origin and at infinity but which is solution
of the $H_0$ equation and hence has a simple expression 
\begin{equation}
u(\cE_1, r) = \frac{1}{\sqrt{2 \kappa_1 |1+\alpha|}}\,
[\,\exp(\kappa_1 r) + \alpha \exp(-\kappa_1 r)\,]
\label{Bargtau0}
\end{equation}
where $\kappa_1 = \sqrt{-\cE_1} > 0$.
When this function does not vanish ($\alpha>-1$),
its inverse is normalized in agreement 
with convention \eref{invnorm}.
The case $\alpha<-1$ will also be useful below,
provided $v({\cal E}_1,r)$ does not vanish;
in this case, \eref{Bargtau0}
implies that $v({\cal E}_1,r)$ is normalized according to \eref{psi1norm}.
The potential can then be written explicitly,
using either \eref{V2V1V0}, \eref{W} or \eref{V2expl}, as 
\begin{eqnarray}
V_2(r) & = & - 2 \frac{\rmd^2}{\rmd r^2}\ln
\mathrm{W}[\sinh(\kappa_0 r)\,, \exp(\kappa_1 r) + \alpha \exp(-\kappa_1 r)]
\label{BargV2} \\[.5em]
& = &
2 (\kappa_1^2-\kappa_0^2)\,
\dfrac{\dfrac{\kappa_0^2}{\sinh^2(\kappa_0 r)}
+\dfrac{4 \alpha \,\kappa_1^2}{[\,\exp(\kappa_1 r)+\alpha \exp(-\kappa_1 r)]^2}}{
\Bigl[\kappa_0 \coth(\kappa_0 r)-\kappa_1
\dfrac{\exp(\kappa_1 r)-\alpha \exp(-\kappa_1 r)}
{\exp(\kappa_1 r)+\alpha \exp(-\kappa_1 r)}\Bigr]^2}\,,
\label{BargV2expl}
\end{eqnarray}
with the particular value $V_2(0)=2(\kappa_1^2-\kappa_0^2)$.
Such compact expressions for the Bargmann potential were not known
to our knowledge (see reference \cite{chadan:89} for a comparison);
they are only valid for ${\cal E}_1 \ne {\cal E}_0$ (see reference \cite{zakhariev:90}
for the confluent case ${\cal E}_1 = {\cal E}_0$).

Let us now discuss the possible values of parameter $\alpha$.
In order for $v(\cE_1, r)$ to be nodeless,
it has to satisfy one of the following conditions 
\begin{eqnarray}
\alpha>-1 \quad {\rm for} \quad \kappa_0>\kappa_1\,,
\label{subtbs} \\
\alpha<-1 \quad {\rm for} \quad \kappa_0<\kappa_1\,.
\label{deepbs}
\end{eqnarray}
In the first case, $u(\cE_1, r)$ is also nodeless,
while in the second case, $u(\cE_1, r)$ has a node which disappears
during the first transformation.
The usual Bargmann potential \cite{chadan:89}
corresponds to the first case only;
it was not realized in previous works that the second case is also possible
because the potential was constructed in a more complicated way.

The main interest of this potential family is that both
its Jost function and its scattering matrix have rational
expressions in terms of the wave number.
The Jost function reads
\begin{equation}
F_2(k)=\frac{k-\rmi  \kappa_1}{k+\rmi  \kappa_0}\,,
\label{BargF2}
\end{equation}
with one pole in $-\rmi  \kappa_0$ as before and one zero in
$\rmi  \kappa_1$ which corresponds to the added bound state.
The scattering matrix reads
\begin{equation}
S_2(k)=
\frac{k+\rmi  \kappa_1}{k-\rmi  \kappa_1}\,\, %
\frac{k+\rmi  \kappa_0}{k-\rmi  \kappa_0}\,,
\label{BargS2}
\end{equation}
it has an unphysical pole in $\rmi  \kappa_0$, as before,
and a new pole corresponding to the added bound state in $\rmi  \kappa_1$.
The corresponding phase shift \eref{d2} reads
\begin{equation}
\delta_2(k)=\pi-\arctan\frac{k}{\kappa_0}-\arctan\frac{k}{\kappa_1}\,,
\label{Bargd2}
\end{equation}
which can also be extracted from the asymptotic
behaviour of the scattering state
\begin{eqnarray}
\psi_2(k,r) & = & \kappa_0 k \coth (\kappa_0 r) \cos (kr)
+ \left[ k^2 - \tfrac{\kappa_0^2}{\sinh^2 (\kappa_0 r)} \right] \sin(kr)
\nonumber \\
& - &
\left\{ \left[\kappa_1^2 -\kappa_0 \kappa_1 \coth(\kappa_0 r) + \tfrac{\kappa_0^2}{\sinh^2 (\kappa_0 r)}\right]
\exp (\kappa_1 r) \right. \nonumber \\
&& +
\left. \alpha \left[\kappa_1^2 +\kappa_0 \kappa_1 \coth(\kappa_0 r) + \tfrac{\kappa_0^2}{\sinh^2 (\kappa_0 r)}\right]
\exp (-\kappa_1 r) \right\} \label{Bargpsi2} \\
& \times &
\frac{k\cos (kr) - \kappa_0 \coth (\kappa_0 r) \sin (kr)}
{[-\kappa_1 + \kappa_0 \coth (\kappa_0 r)] \exp(\kappa_1 r)
+ \alpha  [\kappa_1 + \kappa_0 \coth (\kappa_0 r)] \exp(-\kappa_1 r)}. \nonumber
\end{eqnarray}
The phase shift \eref{Bargd2} is equivalent to a truncated
two-term effective-range expansion with
the scattering length $a$ and effective range $r_0$ 
reading \cite{babenko:05}
\begin{equation}
 a=\frac{1}{\kappa_0}+\frac{1}{\kappa_1}\,,
\quad
 r_0=\frac{2}{\kappa_0+\kappa_1}
\label{ar0}
\end{equation}
and all other terms vanishing.
It is important to stress that these expressions are independent
of the value of parameter $\alpha$.
Hence, different potentials corresponding to different values
of $\alpha$ have identical Jost function and scattering matrix,
on the whole complex plane.
Such potentials are phase equivalent.
\subsubsection{Link between bound- and scattering-state properties}
Let us now study in more detail the properties of the bound state
of this potential (see reference \cite{chadan:89} for a similar discussion).
Since the inverse of $u({\cal E}_1,r)$ is normalized to unity, the inverse of 
\begin{equation}
v({\cal E}_1, r)=\frac{1}{\sqrt{|\kappa_0^2 - \kappa_1^2|}}\, L u({\cal E}_1, r)
\end{equation}
also is, according to \S \ref{sec:norm}.
Hence, the wave function of the added bound state reads explicitly 
\begin{eqnarray}
\psi_2({\cal E}_1,r) = v^{-1}({\cal E}_1, r) \nonumber \\[.5em]
= \frac{\sqrt{2 \kappa_1 (\kappa_0^2-\kappa_1^2) (1+\alpha)}}
{[-\kappa_1+\kappa_0 \coth(\kappa_0 r)\,] \exp(\kappa_1 r)
+ \alpha\, [\,\kappa_1+\kappa_0 \coth(\kappa_0 r)\,] \exp(-\kappa_1 r)}
\label{psi2Barg}
\end{eqnarray}
and behaves asymptotically as 
\begin{equation}
\pm \sqrt{2 \kappa_1 \frac{\kappa_0+\kappa_1}%
{\kappa_0-\kappa_1}(1+\alpha)}\, \exp(-\kappa_1 r)\,,
\end{equation}
where the plus sign corresponds to case \eref{subtbs}
 and the minus sign corresponds to case \eref{deepbs}.
This behaviour defines the ANC [see \eref{sc.25}]
of the bound state 
\begin{eqnarray}
C = \sqrt{2 \kappa_1 \frac{\kappa_0+\kappa_1}{\kappa_0-\kappa_1}(1+\alpha)}\,,
\label{C}
\end{eqnarray}
chosen real and positive in all cases.

As is obvious from the above equation, the ANC is related
to the value of parameter $\alpha$ and can be chosen arbitrarily.
We have thus constructed a family of phase-equivalent potentials
with different ANC values for the bound state.
This illustrates a general result well known in the context
of the inverse-scattering problem \cite{chadan:89},
i.e.\ that there is no general link between the scattering matrix and the ANC.
%as already emphasized in Ref.\ \cite{sparenberg:04a}.
This may seem to contradict a result used in several
references \cite{mukhamedzhanov:99, blokhintsev:08},
where appears the relationship between the ANC 
and the residue of the scattering matrix pole in $\rmi \kappa_1$,
\begin{equation}
T(k) \equiv 1-S(k) \mathop{\to}_{k\to \rmi\, \kappa_1}
\frac{\rmi\, |C|^2}{k-\rmi \kappa_1}\,,
\end{equation}
for an $s$ wave without Coulomb interaction, as is the case here. 
For the potential constructed above, we can calculate
the $S$-matrix residue explicitly
from the analytical expression of $S$ given by \eref{BargS2}.
This leads to
\begin{equation}
 T_2(k) \equiv 1-S_2(k) \mathop{\to}_{k\to \rmi\, \kappa_1}
 \frac{2\,\rmi\,\kappa_1\,(\kappa_0+\kappa_1)}
 {(\kappa_0-\kappa_1)(k-\rmi \kappa_1)}\,.
 \label{Tres}
\end{equation}
Comparing this expression with \eref{C} 
shows that, for the above potentials,
this relation only holds for the case $\alpha=0$,
which implies that $\kappa_0>\kappa_1$ according to \eref{subtbs}. 

In case \eref{deepbs}, $\kappa_0<\kappa_1$, 
the pole of the Jost function in $-\rmi \kappa_0$ 
(which implies that the Jost function is analytic only
for complex wave numbers such that $\Imag k > - \kappa_0$) 
is closer to the origin of the complex plane than
the zero of the Jost function corresponding to the bound state.
Hence, there is no particular link between the ANC and the scattering matrix,
as seen in \S \ref{sec:scscat}.

The particular case $\alpha=0$ is actually quite important,
both from the mathematical and physical points of view.
Mathematically, the above expressions strongly simplify;
for instance, the potential \eref{BargV2expl} 
becomes the Eckart potential \cite{eckart:1930}
\begin{eqnarray}
V_2(r)
& = & \frac{2 \kappa_0^2 (\kappa_1^2-\kappa_0^2)}{[\kappa_0 \cosh(\kappa_0 r) - \kappa_1 \sinh(\kappa_0 r)]^2}
\label{Barg0} \\
& = & -8\, \kappa_0^2\, \frac{\beta\exp(-2 \kappa_0 r)}
{[\,1+\beta\exp(-2 \kappa_0 r)]^2}\,,\quad
\beta=\frac{\kappa_0+\kappa_1}{\kappa_0-\kappa_1}>1.
\label{Eckart0}
\end{eqnarray}
This is one of the potentials derived by Bargmann 
(see reference \cite{bargmann:49a} for $\sigma=1$,
equation (4.11) of \cite{bargmann:49b} and reference \cite{newton:57}).
With respect to the whole phase-equivalent family,
this potential has a shorter range as seen on the asymptotic
behaviour of \eref{BargV2expl} for arbitrary $\alpha$,
\begin{equation}
 V_2(r\to \infty) \to -8 \beta
 \left[ \kappa_0^2 \exp(-2 \kappa_0 r) %
 + \alpha \kappa_1^2 \exp(-2 \kappa_1 r) \right].
\end{equation}

For case \eref{subtbs} with $\alpha\ne0$, the $\exp(-2 \kappa_1 r)$
term is dominant;
the asymptotic behaviour of the potential is then related
to the bound-state energy
and to the particular value of $\alpha$ chosen.
When the bound state is added at a small binding energy,
the potential thus decreases slowly in general, though at an exponential rate.
For $\alpha=0$ or for case \eref{deepbs},
on the contrary, the $\exp(-2 \kappa_0 r)$ term is dominant
and the asymptotic behaviour of the potential only depends on
$\kappa_0$ and $\beta$.
The case $\alpha=0$ is thus the only possibility
to get a rapidly-decreasing potential with a small binding energy.
The potential for which the relationship between the $S$ matrix and the ANC holds
is thus shorter-ranged than the other potentials of the phase-equivalent family.
It is more physical as its range is independent of the binding energy.

Let us now study the case $\alpha=-1$ in \eref{Bargtau0}.
The second factorization solution is now regular at the origin,
which implies that no bound state is introduced.
The second transformation is thus of the $T_\mathrm{l}$ type,
like the first one;
both transformations play an equivalent role and there
is no condition on $\kappa_0$ and $\kappa_1$ in this case.
The potential formulae \eref{BargV2} and \eref{BargV2expl}
remain valid but $V_2(r)$ is now singular at the origin:
it behaves like
\begin{equation}
 V_2(r \to 0)\to \frac{6}{r^2}\,.
\end{equation}
At infinity, it decreases exponentially,
with a rate determined by the smallest factorization wave number.
The Jost function reads
\begin{equation}
F_2(k)=\frac{1}{(k+\rmi  \kappa_0)(k+\rmi  \kappa_1)}\,,
\end{equation}
which leads to the same $S$ matrix \eref{BargS2} as before.
The phase shift is given by \eref{Bargd2} minus $\pi$;
this is an example of phase equivalent potentials with different number
of bound states,
as first studied in \cite{baye:87a,baye:87b}.

This singular potential is an interesting intermediate
 step in a decomposition of the inverse problem
proposed in \cite{sparenberg:97a}:
it is the unique bound-state-less potential with the phase shift
\eref{Bargd2}, up to a multiple of $\pi$.
A bound state can then be added to this potential,
at an arbitrary energy ${\cal E}_2$ and with an arbitrary
asymptotic normalization constant,
with the additional phase-equivalent supersymmetric pair
$\{T_\mathrm{r}(\cE_2), T_\mathrm{add}(\cE_2, \alpha)\}$.
This generates the most-general phase-equivalent potential
with the phase shift \eref{Bargd2} and one bound state.
Since $\cE_0$ and $\cE_1$ fix the phase shifts, the potential
still depends on two parameters, $\cE_2$ and $\alpha$,
in agreement with general theorems for the
fixed-angular-momentum inverse problem \cite{chadan:89}.
The two-transformation potentials \eref{BargV2expl} can be obtained from
this general four-transformation family:
when $\cE_2=\cE_1$ (or $\cE_2=\cE_0$),
two transformations simplify and one gets
\begin{equation}
 \bigg\{T_\mathrm{l}(\cE_0), T_\mathrm{l}(\cE_1), %
 T_\mathrm{r}(\cE_1), T_\mathrm{add}(\cE_1,\alpha) \bigg\}=
\bigg\{T_\mathrm{l}(\cE_0), T_\mathrm{add}(\cE_1,\alpha) \bigg\}\,.
\end{equation}
The four-transformation potential family illustrates once again
the complete disconnection between bound- and scattering-state properties:
for these potentials, neither the bound-state ANC, determined by $\alpha$,
nor the binding energy, are related to the scattering-matrix poles.
However, when their binding energy is small, $|\cE_2| < |\cE_0|, |\cE_1|$, 
these potentials have a particular physical feature:
they decrease slowly, as their dominant term given by the asymptotic 
behaviour of \eref{W_PEadd} behaves like $\exp(-2\kappa_2 r)$. 
Hence, in the more general family of phase-equivalent potentials with phase shift \eref{Bargd2}
displaying both an arbitrary binding energy and an arbitrary normalization constant,
the two-transformation $\alpha=0$ potential has the shortest
range and is the only potential
with a range independent from its binding energy;
this potential, for which the ANC is related to the residue
of the $S$-matrix pole,
is thus unique and physically well defined \cite{blokhintsev:08}.

Let us finally mention that the limiting case $\alpha \rightarrow \infty$
can also be treated with the above formalism.
It corresponds to a pair of transformations
$\{T_\mathrm{l}(\cE_0), T_\mathrm{r}(\cE_1)\}$
which leads to a regular bound-state-less potential.
Equations \eref{BargF2}, \eref{BargS2}, \eref{Bargd2}, \eref{Barg0} and \eref{Eckart0} are still valid,
except for a change of sign for $\kappa_1$ \cite{chadan:89}.
In this case, no particular condition holds for $\kappa_0$ and $\kappa_1$.
This potential is not phase equivalent to the above ones as $\kappa_1$ has the opposite sign.
The one-bound-state potential \eref{BargV2expl} can be obtained from it by applying the pair
$\{T_\mathrm{l}(\cE_1), T_\mathrm{add}(\cE_1,\alpha) \}$.
This four-step procedure is the one followed in \cite{chadan:89}, for instance.
Once again, since 
\begin{equation}
 \bigg\{T_\mathrm{l}(\cE_0), T_\mathrm{r}(\cE_1), %
 T_\mathrm{l}(\cE_1), T_\mathrm{add}(\cE_1,\alpha) \bigg\}=
\bigg\{T_\mathrm{l}(\cE_0), T_\mathrm{add}(\cE_1,\alpha) \bigg\}\,,
\end{equation}
our method is the most direct way to generate the one-bound-state potential.

\subsubsection{Eckart potential with non-conservative transformations}
\label{sec:Eckart}

An alternative writing for the Eckart potential \eref{Barg0} is
\begin{equation}
V_2(r) = -2 \frac{\rmd^2}{\rmd r^2} \ln \left[\,\kappa_0 \cosh(\kappa_0 r)%
-\epsilon_1 \kappa_1 \sinh(\kappa_0 r)\,\right],
\end{equation}
where $\kappa_{0,1}$ are chosen positive and the sign of $\epsilon_1$
determines the presence of a bound state
($\epsilon_1=+1$, with the condition $\kappa_0>\kappa_1>0$)
or not ($\epsilon_1=-1$, no particular condition on $\kappa_{0,1}$).
This equation shows that this potential can also be obtained through 
a single supersymmetric transformation of the zero potential,
with factorization energy  $\cE_0=-\kappa_0^2$.
The corresponding transformation solution, 
\begin{equation}
v(\cE_0, r)=\kappa_0 \cosh(\kappa_0 r)-\epsilon_1 \kappa_1 \sinh(\kappa_0 r),
\label{Bargsig0}
\end{equation}
does not vanish at the origin;
hence this transformation is non-conservative.
Notation $v$ is chosen because this single transformation can actually be interpreted as the second step of
a chain of two non-conservative transformations,
\begin{equation}
V_0 \equiv 0 \mathop{\longrightarrow}_{T_\mathrm{nc}(\cE_1, \epsilon_1 \kappa_1)}
V_1 \equiv 0 \mathop{\longrightarrow}_{T_\mathrm{nc} (\cE_0, -\epsilon_1 \kappa_1)} V_2, 
\label{Barg0ncons}
\end{equation}
where the first transformation leaves the potential unchanged
as its factorization solution is just $\exp(\epsilon_1 \kappa_1 r)$ and its superpotential is constant,
while the second transformation has a non-constant superpotential,
the value at the origin of which is
\begin{equation}
w(0) = - \epsilon_1 \kappa_1\,.
\label{Bargsig2}
\end{equation}
In the following we call the first transformation ``purely exponential''.
The chain \eref{Barg0ncons} of non-conservative transformations
is thus equivalent to the chain \eref{Barg0cons} of conservative transformations in this case
but the order in \eref{Barg0ncons} strongly simplifies things:
\eref{chatde.8} can be applied
and the intermediate potential vanishes.
The factorization solution \eref{Bargsig0} of the second transformation can be obtained
by applying the operator corresponding to the purely-exponential transformation to the solution of $V_0$
regular at the origin, i.e.,
\begin{eqnarray}
 v(\cE_0, r) & = & \left(-\frac{\rmd}{\rmd r} + \epsilon_1 \kappa_1 \right) \sinh(\kappa_0 r)
\label{nctregsol}\\
& = & \left\{ \begin{array}{ll}
           \cosh\left(\kappa_0 r -\epsilon_1 \mathrm{arctanh} \frac{\kappa_1}{\kappa_0}\right) & (\kappa_1<\kappa_0), \\
	    \sinh\left(\kappa_0 r + \mathrm{arctanh} \frac{\kappa_0}{\kappa_1}\right) & (\kappa_1>\kappa_0),
          \end{array}\right.
\label{petsinh}
\end{eqnarray}
where the first case leads to a potential with or without bound state, depending on the sign of $\epsilon_1$,
while the second case necessarily leads to a bound-state-less potential.

Let us check that the Jost function obtained
with the single non-conservative transformation coincides
with the one obtained with the pair of conservative transformations.
Since the Jost solution of the zero potential reads $f_0(k,r)=\exp(\rmi  kr)$,
\eref{mc.27} implies that $G_0(k)=-\rmi  k$.
Hence, using \eref{Bargsig2} and the fact that $w(\infty)=\kappa_0$
for factorization solution \eref{Bargsig0},
\eref{nonconsF1} leads to the Jost function
\begin{equation}
 F_2(k) = \frac{k-\rmi\epsilon_1 \kappa_1}{k+\rmi\kappa_0},
\label{BargF2nc}
\end{equation}
which agrees with \eref{BargF2} and generalizes it to the bound-state-less case.

Let us finally express the regular solutions of the Eckart potential using the transformation
chain \eref{Barg0ncons}.
The purely-exponential transformation transforms the regular solution of the vanishing potential
into a solution that is not regular at the origin,
similarly to \eref{nctregsol}.
At energy $\cE_1$ and for $\epsilon_1=+1$,
this transformed solution is simply a decreasing exponential,
which can be normalized as 
$\sqrt{2\kappa_1} \exp(-\kappa_1 r)$.
Applying the second non-conservative transformation to this solution leads to
a solution which is regular at the origin again
and still decreases exponentially at infinity,
i.e.\ the bound-state wave function
\begin{eqnarray}
 \psi_2(r) & = & \frac{1}{\sqrt{\cE_1-\cE_0}}\, L %
 \left[\sqrt{2\kappa_1} \exp(-\kappa_1 r) \right] \\[.5em]
& = & \sqrt{2\kappa_1 (\kappa_0^2-\kappa_1^2)}\,%
 \frac{\sinh(\kappa_0 r)\exp(-\kappa_1 r)}
{\kappa_0 \cosh(\kappa_0 r) -\kappa_1 \sinh(\kappa_0 r)}\,.
\end{eqnarray}
This solution is consistent with \eref{psi2Barg} for $\alpha = 0$.
For positive energies,
the doubly transformed regular solution reads
\begin{eqnarray}
&& \varphi_2(k,r) = \left. \frac{1}{k^2 + \kappa_0^2}\, \right\{
 k\sin(kr) + \epsilon_1 \kappa_1\cos(kr)
 \nonumber \\[.5em]
&& + \left. \frac{\kappa_0}{k}\,%
 \frac{\kappa_0 \sinh(\kappa_0 r) -\epsilon_1 \kappa_1 \cosh(\kappa_0 r)}
{\kappa_0 \cosh(\kappa_0 r) -\epsilon_1 \kappa_1 \sinh(\kappa_0 r)}
 \left[\,k\cos(kr) - \epsilon_1 \kappa_1\sin(kr)\, \right]\, \right\},
\end{eqnarray}
which is consistent with \eref{Bargpsi2} for $\alpha=0$. 
Comparing its asymptotic expression with \eref{sc.15},
one recovers the Jost function \eref{BargF2nc}.
\subsection{Purely-exponential transformations and resonant states}
\label{sec:trivtr}
As illustrated above on the Eckart-potential example,
purely-exponential transformations lead to interesting simplifications of the analytical expressions
of the potentials constructed by supersymmetric transformations.
Of course, such transformations are only possible for a vanishing starting potential.
Nevertheless, this case is important enough for practical applications to deserve a detailed study.
In the present paragraph, we thus study their iterations and show in particular
that they can be used to construct potentials with resonances.

Let us first notice that a purely-exponential solution,
when transformed by a purely-exponential transformation,
remains purely exponential.
Indeed, one has,
\begin{equation}
 v_1(\cE_1,r) = L_0 u_1(\cE_1,r)
= \left[-\frac{\rmd}{\rmd r} + \epsilon_0 \kappa_0\right] \exp(\epsilon_1 \kappa_1 r)
 \propto u_1(\cE_1,r). 
\end{equation}
This implies that iterating purely-exponential transformations
keeps the initial vanishing potential unchanged;
in this case, as in the Eckart potential one,
\eref{chatde.8} still applies with $V_k=0$,
where $k$ is the number of exponential transformations.

As for non-purely-exponential solutions,
the chain of purely-exponential transformations modifies them
similarly to \eref{petsinh}.
For instance, a hyperbolic sine function $u_m(\cE_m, r) = \sinh (\kappa_m r)$
transforms through a chain of two purely-exponential transformations as
\begin{eqnarray}
 v_{2m}(\cE_m,r) = L_1 L_0 u_m(\cE_m,r) \nonumber \\
 \propto \left\{
\begin{array}{ll}
 \sinh\left(\kappa_m r -\epsilon_0 \mathrm{arctanh}\frac{\kappa_0}{\kappa_m}
-\epsilon_1\mathrm{arctanh}\frac{\kappa_1}{\kappa_m}\right)
& (\kappa_m > \kappa_0, \kappa_1) \\
 \sinh\left(\kappa_m r -\epsilon_0 \mathrm{arctanh}\frac{\kappa_m}{\kappa_0}
-\epsilon_1 \mathrm{arctanh}\frac{\kappa_m}{\kappa_1}\right)
& (\kappa_m < \kappa_0, \kappa_1) \\
 \cosh\left(\kappa_m r -\epsilon_0 \mathrm{arctanh}\frac{\kappa_0}{\kappa_m}
-\epsilon_1 \mathrm{arctanh}\frac{\kappa_m}{\kappa_1}\right)
& (\kappa_0 < \kappa_m < \kappa_1) \\
 \cosh\left(\kappa_m r -\epsilon_0 \mathrm{arctanh}\frac{\kappa_m}{\kappa_0}
-\epsilon_1 \mathrm{arctanh}\frac{\kappa_1}{\kappa_m}\right)
& (\kappa_1 < \kappa_m < \kappa_0)
\end{array} \right.
\label{v2mexpexp}
\end{eqnarray}

An important application is the case of two purely-exponential transformations
with complex-conjugate factorization solutions:
applying two right-regular $T_\mathrm{r}$ transformations with transformation functions 
\begin{eqnarray}
u_0(r) = \exp(-\alpha r), \quad u_1(r) = \exp(-\alpha^* r),
\label{irreg}
\end{eqnarray}
adds a resonance, the parameters of which depend on the value
of the complex parameter $\alpha$, chosen here with $\alpha_R \equiv \Real \alpha > 0$
(see discussion in \S \ref{sec:conjE}).
The above results can be generalized to this case by replacing
$\epsilon_0 \kappa_0$ by $-\alpha$ and $\epsilon_1 \kappa_1$ by $-\alpha^*$.
In particular, the generalization of \eref{v2mexpexp} shows
that a hyperbolic sine function $\sinh(\kappa_m r)$ transforms into
\begin{eqnarray}
 v_{2m}(\cE_m,r) & = & (\kappa_m^2+|\alpha|^2) \sinh(\kappa_m r) - 2 \alpha_R \kappa_m \cosh(\kappa_m r)
\nonumber \\
& \propto & \sinh \left( \kappa_m r -\zeta_{m}\right),
\label{hsinexpexp}
\end{eqnarray}
where only one option of \eref{v2mexpexp} survives and where the parameter
\begin{equation}
 \zeta_m \equiv \mathrm{arctanh}\tfrac{\alpha}{\kappa_m} + \mathrm{arctanh}\tfrac{\alpha^*}{\kappa_m}
= \mathrm{arctanh} \frac{ 2 \alpha_R \kappa_m}{\kappa_m^2 + |\alpha|^2}
\label{zetam}
\end{equation}
is real and positive
since $(\alpha-\kappa_m)(\alpha^*-\kappa_m)$ is always positive.
Purely-exponential transformations thus permit us
to enlarge the class of standard Bargmann potentials,
which typically only support a finite number of bound states,
to potentials supporting resonance states. 

Let us illustrate this by building the simplest possible potential displaying one resonance.
This potential can be obtained with two purely-exponential $T_\mathrm{r}$ transformations.
Using the technique of \S \ref{sec:chatsp}, these two transformations are compensated 
in a chain of four transformations by two left-regular transformations $T_\mathrm{l}$,
with hyperbolic sine factorization solutions with real parameters $\kappa_0$ and $\kappa_1$.
These solutions transform according to \eref{hsinexpexp},
which implies that the final potential reads
\begin{eqnarray}\fl
   V_4(r) & = & -2\frac{\rmd^2}{\rmd r^2} \ln
\mathrm{W}[\exp(-\alpha r), \exp(-\alpha^* r), \sinh(\kappa_0 r), \sinh(\kappa_1 r)]
\\ \fl
& = &-2\frac{\rmd^2}{\rmd r^2} \ln \mathrm{W}\left[
\sinh \left( \kappa_0 r - \zeta_0 \right),
\sinh \left( \kappa_1 r - \zeta_1 \right)\right]
\\ \fl
& = & \frac{2\left(\kappa_0^2-\kappa_1^2\right)\left[\kappa_1^2\sinh^2\left(\kappa_0 r-\zeta_0\right)
   -\kappa_0^2\sinh^2\left(\kappa_1 r-\zeta_1 \right)\right]}
   {\left[\kappa_1\sinh\left(\kappa_0 r-\zeta_0\right)\cosh\left(\kappa_1 r-\zeta_1 \right)-
   \kappa_0\sinh\left(\kappa_1 r-\zeta_1\right)\cosh\left(\kappa_0 r-\zeta_0\right)\right]^2},
\label{V_1}
\end{eqnarray}
where the real positive constants $\zeta_i$ are defined by \eref{zetam}.
If we assume $\kappa_1 > \kappa_0$, the potential is not singular 
if $\kappa_0 \tanh \zeta_1 > \kappa_1 \tanh \zeta_0$. 
Potential $V_4$  represents a generalization 
of a two-soliton potential defined on the positive 
semi-axis \cite{Matveev}. Instead of two discrete levels present 
in the two-soliton potential, potential \eref{V_1} has one resonance state. 

The Jost function \eref{F_n} assumes the form 
\begin{equation}
F_4(r) = \frac{(k+\rmi\alpha)(k+\rmi\alpha^*)}{(k+\rmi\kappa_0)(k+\rmi\kappa_1)}\,.
\end{equation}
The S-matrix resonance pole occurs at $k = -\rmi\alpha$ ($\Imag{\alpha} > 0$) 
with the mirror pole at $k = -\rmi\alpha^*$.
For the phase shift \eref{delta_n}, one obtains 
\begin{equation}
\delta_4(k) = \arctan \frac{2 \alpha_R k}{|\alpha|^2-k^2} 
- \sum_{j=0}^1 \arctan \frac{k}{\kappa_j}.
\label{delta_V_1}
\end{equation}

Let us finally consider a more complicated case: 
a potential with two resonances on a scattering background.  
The resonance poles occur at $k = -\rmi \alpha_1, -\rmi \alpha_1^*, -\rmi \alpha_2, -\rmi \alpha_2^*$, 
with $\alpha_{iR}, \alpha_{iI} > 0$, $i=1, 2$. 
Four $T_\mathrm{l}$ transformations with functions $\sinh \kappa_i r$, 
$\kappa_i>0$, $i=0,\dots,3$, regularize the potential and add the background. 
Since $V_4(r) = 0$, the 8-transformation potential can be written 
in a more compact form by means of a 4th-order Wronskian, 
\begin{equation}
V_8(r) = -2\frac{\rmd^2}{\rmd r^2}\ln    \mathrm{W} [v_{40}, v_{41}, v_{42}, v_{43}]\;,
\end{equation}
where
\begin{equation}
v_{4i}(r) = \sinh (\kappa_i r - \zeta_i - \varrho_i), \quad i=0,\dots,3\;,
\end{equation}
and
\begin{equation}
\tanh\zeta_i = \frac{2 \alpha_{1R} \kappa_i}{|\alpha_1|^2+\kappa_i^2}\,,
\quad \tanh\varrho_i = \frac{2 \alpha_{2R} \kappa_i}{|\alpha_2|^2+\kappa_i^2}\;.
\end{equation}
The phase shift is given by a generalization of \eref{delta_V_1}. 
\subsection{Application to the neutron-proton $^3S_1$ wave}
\label{sec:npS}
\begin{figure}
\centerline{\scalebox{0.55}{\includegraphics{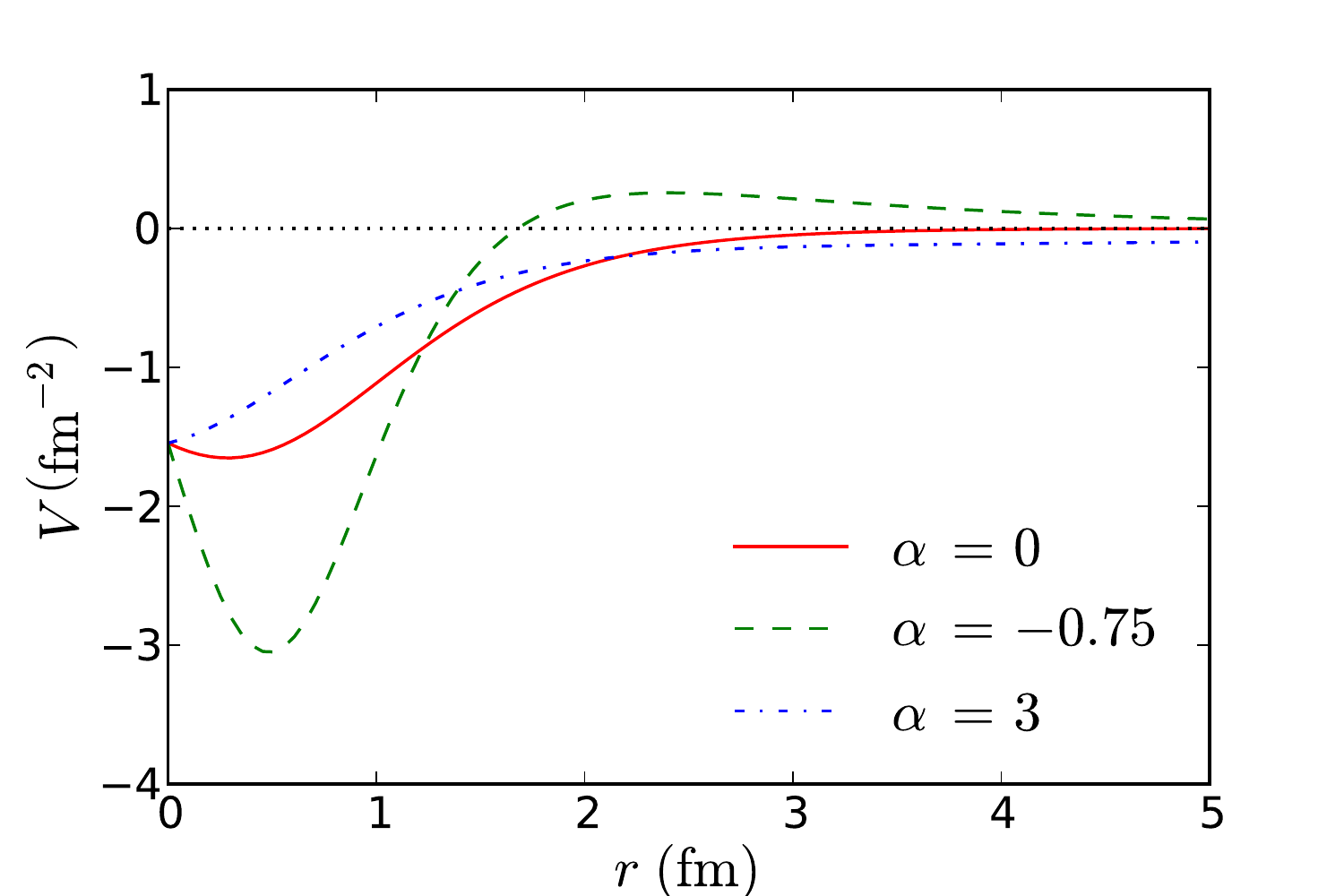}}
\hspace{-0.5cm} \scalebox{0.55}{\includegraphics{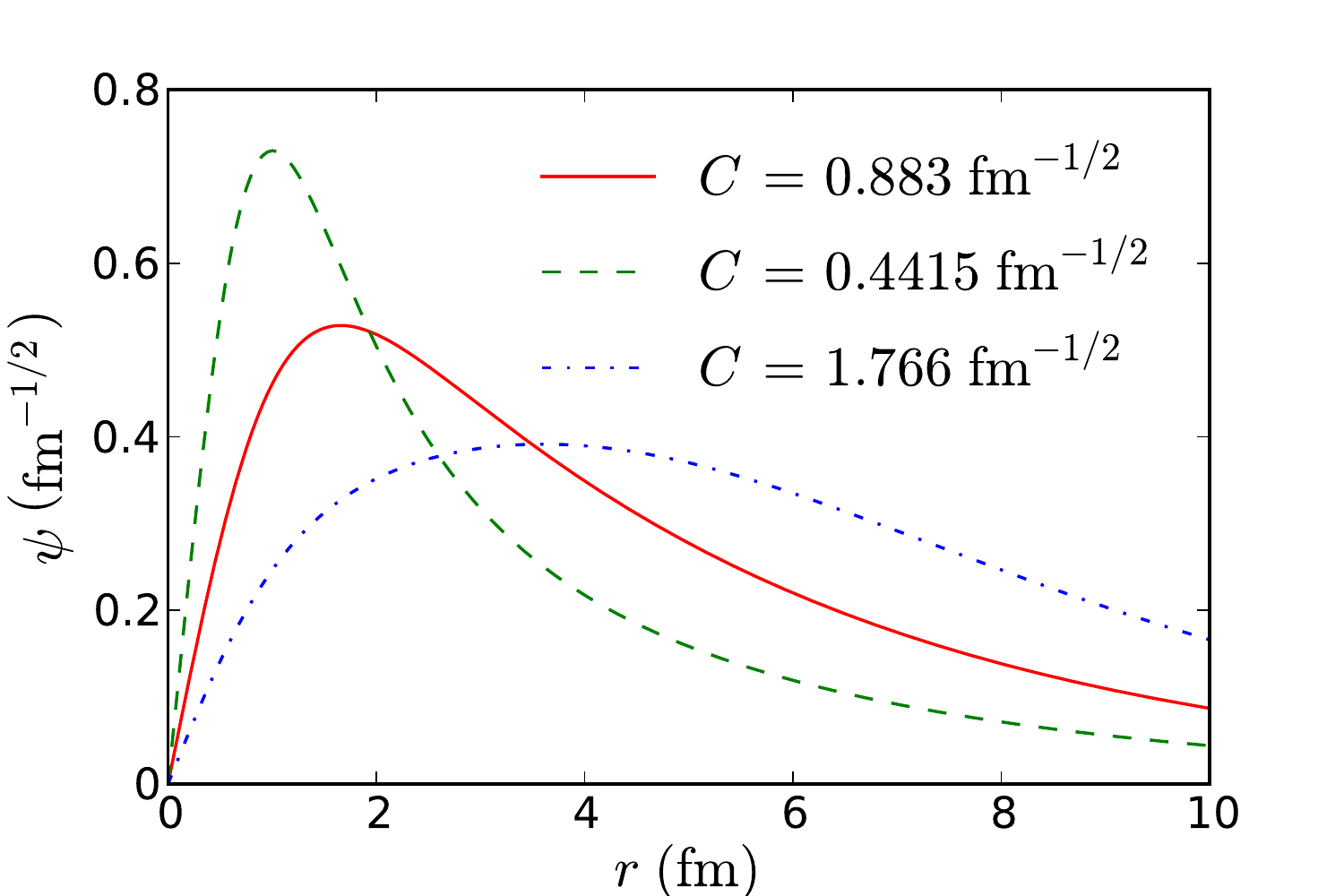}}}
 \caption{\label{fig:np_s-wave}
 Neutron-proton triplet $S$-wave potentials (left panel)
constructed with two supersymmetric transformations with factorization
wave numbers \eref{kappa0}-\eref{kappa1}
and parameter $\alpha=-0.75,$ 0, 3.
All potentials share the same effective-range parameters \eref{ar},
phase shifts \eref{Bargd2} and binding energy,
but differ by their bound-state ANC \eref{C}.
The $\alpha=0$ short-range potential has the most physical ANC \eref{Cnp},
while the longer-range potentials have the double or the half of this value,
according to \eref{C}, as seen on their bound-state wave functions (right panel).
}
\end{figure}

As an application of the above considerations,
let us now construct $S$-wave potentials for the neutron-proton system,
that fit the neutron-proton triplet scattering length
and effective range recommended in \cite{deswart:95}, 
\begin{equation}
 a = 5.4194(20) \mbox{ fm}, \quad r_0 = 1.7536(25) \mbox{ fm}.
\label{ar}
\end{equation}
From \eref{ar0}, the corresponding scattering-matrix poles have the values 
\begin{eqnarray}
 \kappa_0 & = & \frac{1}{r_0}+\sqrt{\frac{1}{r_0^2}-\frac{2}{a\,r_0}} =
 0.9090 \mbox{ fm}^{-1},
\label{kappa0}
\\[.5em]
 \kappa_1 & = & \frac{1}{r_0}-\sqrt{\frac{1}{r_0^2}-\frac{2}{a\,r_0}} =
  0.2315 \mbox{ fm}^{-1}.
\label{kappa1}
\end{eqnarray}
Taking relativistic corrections into account \cite{deswart:95},
this value of $\kappa_1$ perfectly matches
the experimental deuteron binding energy $B_d$: 
$|\cE_1| = 2\mu \kappa_1^2$ is equal to $B_d = 2.225$ MeV.
Several such potentials are displayed in figure \ref{fig:np_s-wave}.
Among them, the shortest-range potential ($\alpha=0$)
decreases asymptotically like $\exp(-2\kappa_0 r)$,
i.e.\ faster than the expected behaviour
from the one-pion exchange, $\exp(-0.7 r)/r$.
All other potentials, on the other hand, decrease like $\exp(-2\kappa_1 r)$,
with a repulsive tail when $\alpha<0$ and an attractive tail when $\alpha>0$.
These potentials thus decrease more slowly than the one-pion exchange potential;
therefore they are not physically acceptable.
The short-range potential has a bound state,
the ANC of which is related to the residue of the pole of the scattering matrix,
which in turn can be related to the scattering-matrix pole locations 
\cite{blokhintsev:08}
and the effective-range-expansion parameters \cite{babenko:05,sparenberg:10}, 
\begin{equation}
 C = \sqrt{2 \kappa_1 \frac{\kappa_0+\kappa_1}{\kappa_0-\kappa_1}}
= \sqrt{\frac{2\,\kappa_1}{{1}-\kappa_1\,r}}
= \sqrt{\frac{2 a \kappa_1^2}{2 - a\kappa_1}}
= 0.883 \mbox{ fm}^{-1/2}.
\label{Cnp}
\end{equation}
Despite the too fast asymptotic decrease of this potential,
this ANC agrees fairly well with the recommended value of reference
\cite{deswart:95}, $C= 0.8845(8)$ fm$^{-1/2}$.

In reference \cite{pupasov:11},
this unique short-range potential is generalized
to fit the scattering phase shifts at higher energies,
which corrects both its asymptotic behaviour and its ANC value.
Remarkably, a good quality fit can be obtained with only five $S$-matrix poles
$\rmi \epsilon_i \kappa_i$ lying on the imaginary wave number axis, 
where $\epsilon_i \kappa_i = 0.23154, -0.45146, 0.43654,1.6818, 2.3106$ fm$^{-1}$ for $i=0$ to 4. 
The phase shifts are thus parametrized as
\begin{equation}
\delta_4(k) = - \sum_{i=0}^4 \epsilon_i \arctan \frac{k}{\kappa_i}.
\end{equation}
The corresponding potential $ V_{S;5}$ reads 
\begin{equation}
 V_{S;5}(r)  = -2\frac{\rmd^2}{\rmd r^2} \ln \textrm{W}[u_0,\ldots,u_4]\,,
\label{np-Spot1}
\end{equation}
with the $T_\mathrm{add}$ and $T_\mathrm{r}$ solutions 
\begin{equation}
u_i(r) = \rme^{\epsilon_i \kappa_i r}\,,\quad  i=0,1\,,
\label{np-s1s2}
\end{equation}
and the three $T_\mathrm{l}$ solutions  
\begin{equation}
u_i(r) = \sinh(\kappa_i r)\,, \quad i=2,3,4\,.
\label{np-s3s4s5}
\end{equation}
Potential $V_{S;5}$ can be simplified by using expression \eref{chatde.8} 
which is interesting because $V_{S;2}(r) = 0$ with the exponential 
transformation functions \eref{np-s1s2}. 
Hence potential $V_{S;5}$ can also be seen as resulting from three non-conservative transformations
and can be rewritten as 
\begin{eqnarray}
\fl
 V_{S;5}(r) = -2\frac{\rmd^2}{\rmd r^2} \ln \textrm{W}
\left[\cosh\left(\kappa_2 r -\mathrm{arctanh}\tfrac{\kappa_0}{\kappa_2}+ \mathrm{arctanh} \tfrac{\kappa_2}{\kappa_1}\right),\right.
\nonumber \\ \fl
\left. \sinh\left(\kappa_3 r -\mathrm{arctanh}\tfrac{\kappa_0}{\kappa_3}+ \mathrm{arctanh} \tfrac{\kappa_1}{\kappa_3}\right),
\sinh\left(\kappa_4 r -\mathrm{arctanh}\tfrac{\kappa_0}{\kappa_4}+ \mathrm{arctanh} \tfrac{\kappa_1}{\kappa_4}\right)\right].
\label{np-Spot2}
\end{eqnarray}

The effective-range function corresponding to these five poles expands as a Padé approximant
of order [2/2], instead of a Taylor expansion.
The interest of effective-range Padé approximants
for the neutron-proton triplet $S$ wave
has been stressed by several authors
(see \cite{hartt:81,babenko:05} and references therein).
In particular, in \cite{babenko:05},
a [2/1] Padé expansion is found satisfactory,
which corresponds to a four-pole parametrization of the scattering matrix.
The corresponding potential, not provided in \cite{babenko:05},
can be constructed with four supersymmetric transformations.
We have checked that this potential displays a non-physical oscillating tail,
which seems to indicate that four poles
are not sufficient to build a satisfactory potential.
The oscillatory behaviour can be explained by the complex character of two poles,
a difficulty already met in \cite{sparenberg:97a} for the singlet case. 
Since they avoid oscillations, purely imaginary poles seem to be a better choice,
as illustrated by the potentials of \cite{samsonov:03,pupasov:11}. 

Let us finally note that such nucleon-nucleon potentials 
based on the Bargmann ansatz for the scattering matrix, 
though recently introduced in the context of 
the supersymmetric quantum mechanics inverse problem 
\cite{sparenberg:97b,samsonov:03}, 
had already been studied a long time ago with inversion 
techniques based on integral equations \cite{hartt:81}. 
The main advantage of the supersymmetric approach is 
that it leads to compact analytical expressions for the potential.
%
%%%%%%%%%%%%%%%%%%%%%%%%%%%%%%%%%%%%%%%%%%%%%%%%%%%%%%%%%%%%%%%%%%%%%%%%%%%%%%%%%%%%%%%%%%%%%
\subsection{$l>0$ Bargmann-type potentials}
\subsubsection{Purely centrifugal potential (free particle)}
An interesting limiting case of potential \eref{BargV1} is obtained
with a vanishing factorization energy $\cE_0=0$.
The factorization solution regular at the origin then reads
\begin{equation}
 u_0(r) \propto r\,,
\end{equation}
which corresponds to the superpotential
\begin{equation}
 w_0(r) = \frac{1}{r}\,.
\label{w0free}
\end{equation}
This can be used in a conservative transformation
to get the purely centrifugal $l=1$ potential,
\begin{equation}
 V^\mathrm{\,free}_0(r)=0\,
\mathop{\Longrightarrow}_{T_\mathrm{l}(0)}\,
V^\mathrm{\,free}_1(r)=\frac{2}{r^2}\,,
\end{equation}
with Jost function
\begin{equation}
 F^\mathrm{\,free}_1(k)=\frac{\rmi}{k}
\end{equation}
and the constant total phase shift
\begin{equation}
 \delta^\mathrm{\,free}_1(k)=-\frac{\pi}{2}\,.
\end{equation}

Iterating this transformation (with vanishing
factorization energy and vanishing factorization solution at the origin)
$l$ times leads to the purely centrifugal potential for the $l$ wave,
\begin{equation}
 V_l^\mathrm{\,free}(r) \mathop{\longrightarrow}_{T_\mathrm{l}(0)}%
  V^\mathrm{\,free}_{l+1} (r)\,,
\end{equation}
which can be seen as an illustration of the shape-invariance concept
\cite{gendenshtein:83,cooper:01}.
Indeed, the zero-energy solution of the free potential
for partial wave $l$ reads
\begin{equation}
 u_l(r) \propto u_{l-1}^{-1}(r) \int_0^{\,r} u_{l-1}^2(t)
 \, dt \propto r^{\,l+1},
\end{equation}
which corresponds to
\begin{equation}
 w_l(r) = \frac{l+1}{r}
\end{equation}
and which allows us to generate the free $l+1$ potential,
\begin{equation}
 V_{l+1}^\mathrm{\,free}(r) = V^\mathrm{\,free}_l(r) -2 w'_l(r)
 = \frac{(l+1)(l+2)}{r^2}\,.
\end{equation}
The corresponding Jost function reads
\begin{equation}
 F^\mathrm{\,free}_l(k)=\Bigl(\frac{\rmi}{k}\Bigr)^l,
\end{equation}
which corresponds to the total phase shift
\begin{equation}
 \delta^\mathrm{\,free}_l(k)=-l\ \frac{\pi}{2}.
\end{equation}

These transformations can be seen as an unnecessarily sophisticated
way of building the free-wave potential!
In the following, we will directly start from the free-wave potential
for the partial wave we are interested in,
but we will apply these transformations to construct the solutions
of this potential, used as factorization solutions.

\subsubsection{$l=1$ potentials with conservative and non-conservative
transformations}
\label{sec:l1pot}
Let us first construct a solution regular at infinity for potential 
$V_1^\mathrm{free}$, at energy $\cE_0=-\kappa_0^2$. 
Using \eref{A0pm}, \eref{psi1} with superpotential \eref{w0free}, 
this solution reads 
\begin{equation}
 u(r)=\Bigl(-\frac{\rmd}{\rmd r}+\frac{1}{r}\Bigr) \exp(-\kappa_0 r)
\propto \Bigl(1 + \frac{1}{\kappa_0 r} \Bigr) \exp(-\kappa_0 r)
\end{equation}
and is related to the modified spherical Bessel function of the second kind with $l=1$.
Using it as a factorization solution for a conservative 
$T_\mathrm{r}$ transformation of $V_1^\mathrm{free}$, 
one gets the potential 
\begin{equation}
 V_1(r)=\frac{2}{r^2}-2\frac{\rmd^2}{\rmd r^2} \ln u(r)
= \frac{2}{(r+\kappa_{\,0}^{-1})^2}
\equiv \frac{2}{(r+x_0)^2}\,,
\label{Vr0}
\end{equation}
where the parameter $x_0$ has to be positive for $V_1$ to be finite for all $r$ \cite{samsonov:03}. 

Since the transformation is conservative, the Jost function reads 
\begin{equation}
 F_1(k)=F^\mathrm{\,free}_1(k) (\kappa_0-\rmi k)\,,
\label{F1F1free}
\end{equation}
while the phase shift due to the central potential has the form 
\begin{equation}
 \delta_1(k) \equiv \delta_1^\mathrm{\,tot}(k) - \delta^\mathrm{\,free}_1(k)
= \arctan\frac{k}{\kappa_0} \equiv \arctan k x_0\,.
\label{d1Vr0}
\end{equation}
This phase shift does not satisfy the $l=1$ effective-range expansion, 
as the $k$ term does not vanish. 
This is related to the long-range behaviour of the potential, 
which reads 
\begin{equation}
 V_1(r\to\infty) \to \frac{2}{r^2} - \frac{4 x_0}{r^3}
\end{equation}
and does not satisfy the short-range hypothesis required for 
the effective-range expansion to be valid. 
To restore this expansion and make the corresponding potential short ranged, 
potential $V_1$ can in turn be transformed by additional 
conservative transformations \cite{sparenberg:97a} 
with factorization energies $\cE_i=-\kappa_i^2$. 
The final potential is short ranged and thus satisfies 
the $l=1$ effective-range expansion \eref{ere} 
provided the factorization wave numbers satisfy the condition 
\begin{equation}
\sum_{i=0}^{N-1} \frac{\epsilon_i}{\kappa_i} = 0\,,
\label{cond1}
\end{equation}
where $N$ is the total number of transformations and 
$\epsilon_i$ is defined according to \eref{eps}. 
The parameter $x_0$ (or equivalently $\kappa_0$) can thus be considered 
as a parameter adjusted to guarantee that $\tan \delta_1$ 
behaves like $k^3$ for $k \to 0$. 

The minimal number of transformations required to get 
a non-trivial potential is $3$, 
as for $N=2$ the two terms in condition \eref{cond1} 
exactly compensate each other, 
which implies that the two transformations also compensate each other. 
For $N=3$, the effective-range function \eref{ere} reads 
\begin{eqnarray}
 K_1(k^2) & = & \rmi k^3 \ \frac{\prod_{i=0}^2(\rmi\epsilon_i \kappa_i+k)%
  + \prod_{i=0}^2(\rmi\epsilon_i \kappa_i-k)}
{\prod_{i=0}^2(\rmi\epsilon_i\kappa_i+k)%
 - \prod_{i=0}^2(\rmi\epsilon_i\kappa_i-k)} \\[.5em]
& = & \epsilon_0 \epsilon_1 \epsilon_2 \kappa_0 \kappa_1 \kappa_2 -
(\epsilon_0 \kappa_0 + \epsilon_1 \kappa_1 + \epsilon_2 \kappa_2) k^2\,,
\end{eqnarray}
provided condition \eref{cond1} is satisfied. 
This is an exact two-term effective-range Taylor 
expansion with arbitrary scattering length and effective range. 
For a larger number of transformations, 
the orders of both the denominator and numerator increase: 
for $M$ transformations, the order of the Padé expansion in wave number 
of $K_1(k^2)$ is $[M/M-4]$ for even $M$ and $[M-1/M-3]$ for odd $M$. 

Potential \eref{Vr0} can also be obtained with a non-conservative
transformation of $V_0 = 0$ with a vanishing factorization energy $\cE=0$, 
i.e.\ a factorization solution finite at the origin, 
\begin{equation}
 u(r)=r+x_0\,,
\end{equation}
and hence the superpotential 
\begin{equation}
 w(r)=\frac{1}{r+x_0}\,.
\end{equation}
This procedure, proposed in \cite{samsonov:03}, 
is strictly equivalent to the method based on conservative transformations: 
one has 
\begin{equation}
 T_\mathrm{nc}(0,x_0) \equiv \bigg\{T_\mathrm{l}(0)\,,%
 T_\mathrm{r}(-\kappa_0^2)\bigg\}\,.
\end{equation}
For $V_0 = 0$, a non-conservative transformation is thus always 
equivalent to a pair of conservative transformations but this property should 
not be true for other initial potentials. 
The non-conservative transformation is the simplest way 
to establish potential \eref{Vr0}. 
In contrast, calculating the corresponding phase shift 
\eref{d1Vr0} is more complicated than with conservative transformations, 
as it requires to evaluate the Jost function. 
Using \eref{nonconsF1} with 
\begin{equation}
F_0(k)=1\,, \quad G_0(k)=-\rmi k\,, \quad w(0)=1/x_0\,, \quad w_\infty=0,
\end{equation}
one gets 
\begin{equation}
 F_1(k)=\frac{k+\rmi \kappa_0}{k}\,,
\end{equation}
in agreement with \eref{F1F1free}.

As an example, let us finally give alternative expressions for the potential obtained by
three conservative transformations of $V_1^\mathrm{free} (r)$ with $\epsilon_1, \epsilon_2 = +1$,
which implies that $\epsilon_0=-1$,
i.e. with two $T_\mathrm{l}$ transformations and one $T_\mathrm{r}$ transformation.
Condition \eref{cond1} then implies that $\kappa_1, \kappa_2 > \kappa_0$.
This chain of three transformations generates an effective potential that can also be obtained
from the vanishing potential, either with a zero-energy $T_\mathrm{l}$ transformation followed
by three conservative transformations, or directly with three non-conservative transformations.
One has thus
\begin{eqnarray}
\fl
V_4(r) & = & \frac{2}{r^2} -2 \frac{\rmd^2}{\rmd r^2}
\ln \mathrm{W}\left[\left(1+\frac{1}{\kappa_0 r}\right) e^{-\kappa_0 r},
\cosh(\kappa_1 r)-\frac{\sinh(\kappa_1 r)}{\kappa_1 r},
\cosh(\kappa_2 r)-\frac{\sinh(\kappa_2 r)}{\kappa_2 r}\right] \nonumber \\
\fl  & = & -2 \frac{\rmd^2}{\rmd r^2} \ln
\mathrm{W}[r, e^{-\kappa_0 r}, \sinh(\kappa_1 r), \sinh(\kappa_2 r)] \nonumber \\
\fl  & = & -2 \frac{\rmd^2}{\rmd r^2} \ln
\mathrm{W}\left[1+\kappa_0 r,
\cosh\left(\kappa_1 r +\mathrm{arctanh}\frac{\kappa_0}{\kappa_1}\right),
\cosh\left(\kappa_2 r +\mathrm{arctanh}\frac{\kappa_0}{\kappa_2}\right)\right],
\label{V4pwave}
\end{eqnarray}
where the elegance of expressions derived from the vanishing potential can again be seen
on the last two expressions,
while the first expression directly provides the central potential
in terms of modified spherical Bessel functions.
\subsubsection{$l=2$ potentials with conservative and non-conservative transformations}
Let us now establish the explicit correspondence between the conservative 
and non-conservative approaches for the $d$-wave, 
further revisiting reference \cite{samsonov:03}. 

We first construct solutions of $V^\mathrm{\,free}_2$ regular at infinity, 
\begin{eqnarray}
 u(\cE_i, r) & = & \left(-\frac{\rmd}{\rmd r}+\frac{2}{r}\right)%
 \left(-\frac{\rmd}{\rmd r}+\frac{1}{r}\right) \exp(-\kappa_i r) \\
 & \propto & \left(1+\frac{3}{\kappa_i r} + \frac{3}{\kappa_i^2 r^2}\right) e^{-\kappa_i r},
\quad i=0,1\,,
\end{eqnarray}
as obtained by two successive zero-energy transformations 
of the zero-potential solutions, 
with superpotentials $w_0(r)=1/r$ and $w_1(r)=2/r$. 
These solutions are related to the modified spherical Bessel functions of the second kind (order 2).
When used as factorization solutions for two conservative 
$T_\mathrm{r}$ transformations of $V^\mathrm{free}_2$, 
these provide the potential 
\begin{eqnarray}
 V_2(r) & = & \frac{6}{r^2}-2\frac{\rmd^2}{\rmd r^2} %
 \ln \mathrm{W}[u(\cE_0, r),u(\cE_1, r)] \nonumber \\[.5em]
& = & -2\frac{\rmd^2}{\rmd r^2}
\ln\left[\left(r+\frac{1}{\kappa_0}+\frac{1}{\kappa_1}\right)^3
-\frac{1}{\kappa_0^3}-\frac{1}{\kappa_1^3}\,\right]
\label{Vr0c}
\end{eqnarray}
which coincides with equation (34) of \cite{samsonov:03} 
with parameters $x_0\equiv \kappa_0^{-1}+\kappa_1^{-1}$ 
and $c\equiv -(\kappa_0^{-3}+\kappa_1^{-3})$. 

Since these are conservative transformations, 
the phase shift due to the central potential reads 
\begin{eqnarray}
 \delta_2(k) & \equiv & \delta_2^\mathrm{\,tot}(k)-\delta_2^\mathrm{\,free}(k)
=\arctan\frac{k}{\kappa_0}+\arctan\frac{k}{\kappa_1}\,,
%\nonumber \\
%& \equiv & \arctan \frac{3kr_0^2}{3 r_0 - k^2(r_0^3+c)}.
\label{d2Vr0c}
\end{eqnarray}
which can be shown to coincide with equation (35) of \cite{samsonov:03}. 
As for the $l=1$ case, the potential is long ranged 
and the phase shift does not satisfy 
the $l=2$ effective-range expansion. 
To restore it, more transformations have to be performed, 
with factorization wave numbers satisfying the conditions \eref{cond1} and 
\begin{equation}
\sum_{i=0}^{N-1} \frac{\epsilon_i}{\kappa_i^{\,3}} = 0\,.
\label{cond2}
\end{equation}
The parameters $x_0$ and $c$ (or $\kappa_0$ and $\kappa_1$) 
can thus be considered 
as parameters adjusted to satisfy the effective-range expansion.

The minimal number of transformations required to get 
a non-trivial potential is now 5. 
Indeed, with 5 scattering-matrix poles, the effective-range function 
\begin{equation}
 K_2(k^2) = \rmi k^{\,5} \ \frac{\prod_{i=0}^4(\rmi\epsilon_i\kappa_i+k)%
  + \prod_{i=0}^4(\rmi\epsilon_i\kappa_i-k)}
{\prod_{i=0}^4(\rmi\epsilon_i\kappa_i+k) -%
 \prod_{i=0}^4(\rmi\epsilon_i\kappa_i-k)}
\end{equation}
is a Padé approximant of order [4/0] in wave number, 
provided conditions \eref{cond1} and \eref{cond2} are satisfied. 
For a smaller number of poles, this effective-range function 
would vanish at zero energy, 
which would correspond to an infinite scattering length, 
according to \eref{ere}. 
With 5 poles, one has instead an exact effective-range 
Taylor expansion with three terms and 
\begin{equation}
 a_2 = \frac{1}{\prod_{i=0}^4 \epsilon_i \kappa_i}\,, \quad
 r_2 = 2\sum_{i,j=0}^4 \prod_{\scriptsize %
 \begin{array}{c} m=0 \\ m \neq i,j \end{array}}^4\!\!\!\!\! \epsilon_m\kappa_m\,,
  \quad
 r_2^{\,3} P_2 = \sum_{i=0}^4 \epsilon_i \kappa_i\,.
\end{equation}

Potential \eref{Vr0c} can also be obtained with a pair of 
non-conservative transformations of $V_0=0$, 
with vanishing factorization energies and factorization solutions 
finite at the origin 
\begin{equation}
 u(r)=r+x_0, \quad v(r) = u^2(r) + c \,u^{-1}(r)\,.
\end{equation}
With this approach, it is simpler to establish the potential 
expression \eref{Vr0c} but more complicated to calculate 
the corresponding phase shift \eref{d2Vr0c}.

This procedure can be generalized to an arbitrary partial wave: 
the chain of $2l$ conservative transformations allowing one
(i) to build the purely centrifugal potential from $V_0=0$ 
and (ii) to modify its phase shifts with $l$ arctangent terms, 
is equivalent to a chain of $l$ non-conservative transformations. 
These chains contain $l$ parameters which can be tuned to satisfy 
the effective-range expansion, 
$l$ terms of which have to vanish. 
The effective-range function expands as a Padé approximant. 
By generalizing the study made above for $l=1$ and 2, 
one shows that, for $M$ poles, the order of this expansion 
in wave number is $[M/M-2l-2]$ for even $M$ and 
$[M-1/M-2l-1]$ for odd $M$. 
This shows that the minimal number of transformations is $2l+1$.

For larger number of transformations,
the potential can probably still be expressed compactly in terms of solutions of the vanishing potential
[see \eref{V4pwave} for $l=1$].
However, this construction is less direct here since $l$ transformations have to be performed
at the same vanishing energy.
Hence, compact expressions would require to combine Wronskian expressions for distinct factorization
energies and for equal factorization energies,
which requires further investigation.
\subsection{Application to the neutron-proton $^3D_1$ wave}
The triplet $D$-wave neutron-proton potential of reference 
\cite{pupasov:11} was constructed along these lines: 
it has five $S$-matrix poles, i.e.\ the minimal number for a $D$ wave. 
To obtain the positions of these poles, 
we used as scattering data the phase shifts of the Reid93 potential 
\cite{stoks:94}. 
We considered 5-, 6- and 7-pole expressions for the phase shifts. 
Conditions \eref{cond1} and \eref{cond2} provide the correct low-energy 
expansions. 
The desired asymptotic behaviour of the potential suggests to choose 
the smallest $|\k|$ value close to 0.35 fm$^{-1}$. 
For each number of poles, we made a series of fits by the least-square method 
over energy intervals ranging from [0-50] MeV to [0-150] MeV.
We found that the 6- and 7-pole models give the same quality of phase-shift fits
as the 5-pole one
but the additional poles have no clear physical meaning
and complicate the control of the potential.
The most satisfactory result is thus the 5-pole model 
(as in the case of the $S$ wave). 
These poles $\rmi \epsilon_m \k_m$ lie on the imaginary wave number axis, 
with $\epsilon_m\k_m$ values $-0.36719, -0.54420, 0.34828, 0.71766, 3.3758$ fm$^{-1}$, $m=0-4$.
The corresponding potential $V_{D;5}$ reads 
\begin{equation}
 V_{D;5}(r) = \frac{6}{r^2}-2\frac{\rmd^2}{\rmd r^2} \ln \mathrm{W}[u_0,\ldots,u_4]\,,
\label{n-p-Dwave-diag-potential}
\end{equation}
with the set of solutions 
\begin{equation}
u_m(r) = \rme^{-\kappa_m r} \left[1+\frac{3}{\kappa_m r} + \frac{3}{\kappa_m^2 r^2} \right]\,,
\quad m=0,1\,,
\end{equation}
and 
\begin{equation}
u_m(r) = \left[3 \kappa_m r \cosh (\kappa_m r) - (3+\kappa_m^2 r^2) \sinh (\kappa_m r) \right]/r^2\,, 
\quad m=2,3,4\,.
\end{equation}

The effective-range parameters of our model read $a_2=5.931$ fm$^5$, 
$r_2=-3.552$ fm$^{-3}$ and $P_2=-0.07878$ fm$^8$. 
They are able to fit the elastic-scattering phase shifts over a wide energy range. 
\subsection{Discussion}
In the single-channel case, supersymmetric quantum mechanics
provides a complete inversion scheme,
including both scattering- and bound-state properties,
which are inverted independently of each other.

For scattering states, two efficient inversion schemes have been found:
the phase shifts can be directly fitted as a sum of arctangent terms,
which allows one to constrain scattering-matrix poles
to stay on the imaginary axis,
an important condition to get non-oscillating potentials.
Another option is to fit the effective-range function,
either as a Taylor expansion or as a Padé approximant.
The scattering-matrix poles can then be directly deduced from this expansion.
Once the $S$-matrix poles are found, by either method,
the corresponding potential is constructed by a sequence
of supersymmetric transformations,
with one transformation associated to each pole.

For bound states,
it is in general more convenient to construct a bound-state-less potential first
and subsequently add the bound states with the required properties
(binding energy and asymptotic normalization constant)
using a phase-equivalent transformation pair for each bound state
\cite{sparenberg:97a}.
For weakly bound states, both the binding energy and the ANC
can be related to the position and residue of one scattering-matrix pole.
The corresponding potential is then shorter ranged
than the phase-equivalent potentials
displaying different binding energy and ANC.
The simplest construction scheme for this
shortest-range potential is based on a single transformation
that adds a bound state and modifies the phase shift at the same time.

Conservative transformations are the only necessary blocks
to fulfil this inversion scheme.
Though non-conservative transformations allow one to derive
some potentials in a simpler way and provide elegant and compact expressions for them,
the complexity introduced by their non-conservativeness makes
them less convenient in an
inverse-scattering perspective.
Moreover, we have shown on several examples that non-conservative
potentials are an exact subset
of conservative potentials.
We conjecture that this is always the case when the potential
family rests on the zero potential
but this conjecture would deserve further study.
We will show below that the situation is quite different
in the coupled-channel case,
where non-conservative transformations seem to play an essential role,
at least in the case of different thresholds.
\section{Coupled-channel supersymmetric quantum mechanics}
\label{sec:cc}
In section \ref{sec:sc}, we reviewed the application of supersymmetric transformations to the single-channel radial Schrödinger equation.
Let us now generalize these results to coupled channels,
by following references \cite{amado:88a,amado:88b,amado:90,cannata:92,cannata:93,andrianov:95,andrianov:97}
and by reviewing and extending our own contributions.
Our main motivation is to introduce a non-trivial coupling (see \S \ref{sec:ms})
in the simplest possible way.

\subsection{General properties of coupled-channel transformations} 
\label{sec:cc1}

A solution
of the initial coupled-channel Schrödinger
equation with matrix potential $V_0$, 
\begin{equation}
H_0 \psi_0(k,r) = k^2 \psi_0(k,r)\,,
\label{cc.1}
\end{equation}
may be mapped into a solution 
\begin{equation}
\psi_1(k,r) = L\psi_0(k,r)
\label{cc.2}
\end{equation}
of the transformed equation 
\begin{equation}
H_1 \psi_1(k,r) = k^2 \psi_1(k,r)\,,
\label{cc.3}
\end{equation}
with the help of an operator $L$ satisfying the intertwining relation 
\begin{equation}
L H_0 = H_1 L.
\label{cc.3a}
\end{equation}
The differential matrix operator $L$ can be written as 
\begin{equation}
L = -I_N\frac{\rmd}{\rmd r} + w(r)
\label{cc.4}
\end{equation}
as a function of the superpotential matrix $w(r)$.
This superpotential is expressed as 
\begin{equation}
w(r) = u'(r) u^{-1}(r)\,,
\label{cc.6}
\end{equation}
where $u$ is a matrix solution of the initial Schrödinger equation 
\begin{equation}
H_0 u(r) = -\kappa^2 u(r)
\label{cc.7}
\end{equation}
and $\kappa$ is a diagonal matrix made of channel wave numbers $\kappa_i$
related to the factorization energy $\cE$, complex in general, by 
\begin{equation}
\kappa_i = \pm \sqrt{\Delta_i-2\mu_i\cE}\,.
\label{Thr_condit}
\end{equation}
The transformed Schrödinger equation contains a new potential 
\begin{equation}
V_1(r) = V_0(r) - 2 w'(r)\,.
\label{cc.5}
\end{equation}
In matrix form, $(u^{-1})' = -u^{-1}u'u^{-1}$
and the superpotential still verifies a Riccati equation
similar to \eref{wp}, 
\begin{equation}
w'(r) = V_0(r) + \kappa^2 - w(r)^2.
\label{cc.6a}
\end{equation}
From this relation, one deduces 
\begin{equation}
L^T L = H_0 + \kappa^2
\label{cc.6b}
\end{equation}
and 
\begin{equation}
L L^T = H_1 + \kappa^2,
\label{cc.6c}
\end{equation}
where the transposed operator $L^T = I_N \rmd/\rmd r + w(r)$
is the complex conjugate of the adjoint of $L$.

In order to have a Hermitian Hamiltonian, the potential
must be real and symmetric.
For a single transformation, reality imposes to choose
a real factorization energy, below all thresholds;
the factorization wave numbers $\kappa_i$ are then chosen strictly positive.
From now on, we assume that these conditions are satisfied,
except in \S \ref{sec:cfemc} and \S \ref{sec:pct}, where we shall meet
complex factorization energies in pairs of transformations.
The symmetry of the potential is realized for real
self-conjugate factorization solutions \cite{amado:88b},
i.e.\ real solutions with a vanishing self-Wronskian 
\begin{equation}
\mathrm{W}[u,u] = 0\,.
\label{cc.8}
\end{equation}
The most general transformation matrix may be expressed
in terms of the Jost solutions as 
\begin{equation}
u(r) = f_0(-\rmi \k ,r)\,C + f_0(\rmi \k ,r)\,D\,,
\label{cc.10}
\end{equation}
where the real constant matrices $C$ and $D$ are arbitrary,
except for the condition \eref{cc.8}
imposing that matrix $C^TD$ be symmetric \cite{samsonov:07}, 
\begin{equation}
D^T C = C^T D\,.
\label{cc.11}
\end{equation}
The first term of \eref{cc.10} is increasing at infinity while
the second term is decreasing.
Both terms are singular at the origin in general.

As shown in Lemma 1 of \cite{samsonov:07},
canonical matrices $C$ and $D$ with a maximal number of independent parameters
guaranteeing the Hermitian character of the superpotential \eref{cc.6}
have the following form, 
\begin{equation}
C = \left( \begin{array}{cc} I_{M} & 0 \\ Q & 0 \end{array} \right),
\qquad
D=\left( \begin{array}{cc} X_0 & -Q^T \\ 0 & I_{N-M} \end{array}\right),
\label{cc.12}
\end{equation}
where $M$ is the rank of $C$,
$X_0$ is a real symmetric nonsingular $M\times M$ matrix
and $Q$ is an $(N-M)\times M$ real matrix.
Matrix $X_0$ affects the potential shape but has no effect
on the asymptotic form of $u$.

The asymptotic behaviour of the superpotential 
\begin{equation}
w(r\to\infty) = w_\infty+{\rm o}(1)
\label{cc.13a}
\end{equation}
determines the transformed Jost solution and, hence,
the Jost and scattering matrices.
The asymptotic matrix $w_\infty$ is determined by the behaviour of
the transformation function \eref{cc.10} at large distances, 
\begin{equation}
u(r\to\infty) \to A
\left( \begin{array}{cc} \rme^{\kappa' r} &0 \\
0 & \rme^{-\kappa'' r} \end{array}\right),
\label{cc.13}
\end{equation}
where $\kappa'$ is a diagonal matrix of rank $M$ containing
the $M$ largest $\kappa_i$ wave numbers,
$\kappa''$ is a diagonal matrix of rank $N-M$ containing
the $N-M$ remaining wave numbers 
and $A$ is a constant matrix.

Here, an important distinction has to be made between
the equal-threshold and different-threshold cases.
For different thresholds, as shown in \cite{samsonov:07}, 
matrix $u$ becomes diagonal at large distances,
hence implying a diagonal form for the asymptotic superpotential 
\begin{equation}
 w_\infty = \left( \begin{array}{cc} \kappa' & 0 \\
0 & -\kappa'' \end{array}\right) \qquad \mbox{(different thresholds)}.
\label{cc.14bis}
\end{equation}
For the equal-threshold case, in contrast, one deduces from
\eref{cc.6} and \eref{cc.13} that 
\begin{equation}
w_\infty = \kappa\, A \left( \begin{array}{cc} I_M & 0 \\
0 & -I_{N-M} \end{array}\right) A^{-1} \qquad \mbox{(equal thresholds)},
\label{cc.14}
\end{equation}
where 
\begin{equation}
A = \left( \begin{array}{cc} I_M & -Q^T \\ Q & I_{N-M} \end{array}\right)
\label{cc.13b}
\end{equation}
with the property $AA^T = A^TA$. 
The asymptotic form of the superpotential is symmetric as expected 
and has a much richer structure than \eref{cc.14bis}.
Note that superpotential $w_\infty$ also has a richer structure than that
previously reported by Amado, Cannata and Dedonder \cite{amado:88b} for the equal-threshold case.
Their result corresponds to the choice $M=1$ when $w_\infty$
is expressed in terms of a single $(N-1)$-vector $Q=(q_1,\ldots,q_{N-1})^T$.
An interesting particular case occurs when $Q$ is such that $A$ is
proportional to an orthogonal matrix as we shall see in \S \ref{sec:single}.
In both cases \eref{cc.14bis} and \eref{cc.14},
$w_\infty$ and $\kappa$ commute and one has the property 
\begin{equation}
w_\infty^2 = \kappa^2
\label{cc.14c}
\end{equation}
in agreement with the asymptotic form of \eref{cc.6a}.

Once $w_\infty$ is determined, one can calculate
the Jost solution $f_1(k,r)$ and the Jost matrix $F_1(k)$
for the transformed potential $V_1$.
The Jost solution takes the form 
\begin{equation}
f_1(k,r) = L f_0(k,r) (w_\infty-\rmi k)^{-1}.
\label{cc.15}
\end{equation}
The right multiplication by the matrix $(w_\infty-\rmi k)^{-1}$
is introduced to guarantee the correct
asymptotic behaviour \eref{mc.13} of $f_1(k,r)$.

These general properties are valid for any type of factorization solution.
To continue the discussion, one needs now to be more specific about
some properties of the transformation at the origin.
\subsection{Conservative and non-conservative transformations}
In order to find the transformed Jost matrix, one needs to use
relation \eref{mc.15} between the regular solution and the Jost solutions.
One must thus derive the transformed regular solution $\varphi_1(k,r)$.
This is simple only when the transformation is conservative,
i.e.\ when it transforms a regular solution $\varphi_0$ into
a matrix equivalent to the regular solution $\varphi_1$, 
\begin{equation}
(L\varphi_0)(0)= 0\,.
\label{cc.16}
\end{equation}
To study this, we first consider the behaviour of the superpotential
in a vicinity of $r=0$.

Let us assume that $\cE$ does not correspond to a bound state
so that $F_0(\rmi \k )$ is invertible.
It is preferable to use an equivalent variant of expression \eref{cc.10}
for the factorization solution matrix, 
\begin{equation}
u(r) = 2\varphi_0(\rmi \kappa,r) F_0^{-1}(\rmi \kappa)\, \kappa\, C'%
 + f_0(\rmi \kappa,r) D',
\label{cc.26}
\end{equation}
where matrices $C'$ and $D'$ are real and constant.
Matrices $C'$ and $D'$ have the same canonical forms \eref{cc.12} as $C$ and $D$
with the same matrix $Q$ and a different, not necessarily invertible,
matrix $X'_0$.
This is easily shown by writing \eref{cc.26} in the form \eref{cc.10}
with \eref{mc.15} and by right multiplying the result by a matrix 
\begin{equation}
\left( \begin{array}{cc} I_M & 0 \\ Y & I_{N-M} \end{array}\right)
\label{cc.27}
\end{equation}
with block $Y$ properly chosen.

The superpotential requires that $u(r)$ be invertible.
Let us first assume that 
\begin{equation}
{\rm det\,} D' \neq 0
\label{cc.28}
\end{equation}
or $\det X'_0 \ne 0$.
This is equivalent to assuming that matrices $C$ and $D$ of \eref{cc.10} verify 
\begin{equation}
{\rm det} \left[F_0(-\rmi \k )F_0^{-1}(\rmi \k ) C + D \right] \neq 0\,.
\label{cc.19}
\end{equation}
Condition \eref{cc.19} is not satisfied when matrices $C$ and $D$ are
respectively proportional to $F_0(\rmi \kappa)$ and $-F_0(-\rmi \kappa)$.

From \eref{mc.17}, the Laurent series for the Jost solution is given by 
\begin{equation}
f(k,r \to 0) = r^{-\nu_0} \left[ (2\nu-I_N)!! F^T(k) + {\rm o}(1) \right].
\label{cc.20}
\end{equation}
With \eref{cc.6} and \eref{cc.20},
the leading term of the superpotential at $r\to 0$ reads
\begin{equation}
w(r\to 0) = -r^{-1} \nu_0 + {\rm o}(1).
\label{cc.21}
\end{equation}
The transformed regular solution has the behaviour at the origin 
\begin{equation}
L \varphi_0(k,r \to 0) = r^{\nu_0} (2\nu_0-I_N)!! + {\rm o}(r^{\nu_0})\,.
\label{cc.21a}
\end{equation}
This matrix is proportional to a regular solution, i.e.\
the transformation is conservative,
if all components $\nu_{0i}$ of $\nu_0$ are strictly positive.

For such a conservative transformation,
the singularities at the origin of the transformed potential, 
\begin{equation}
V_1(r\to 0) \to r^{-2} \nu_1 (\nu_1+I_N) = r^{-2} \nu_0(\nu_0-I_N)\,,
\label{cc.22}
\end{equation}
decrease by one unit, $\nu_0 \to \nu_1=\nu_0-I_N$.
Hence one can only apply this conservative transformation to potentials
for which matrix $\nu_0$ is positive definite, $\nu_0>0$,
i.e.\ to potentials singular at the origin in all channels.

Let us now consider the extreme opposite case $D' = 0$.
This implies $M = N$.
In this case, $u$ is equivalent to the regular solution
$\varphi_0(\rmi \kappa,r)$.
With \eref{mc.9}, the leading term of the superpotential at $r\to 0$ reads 
\begin{equation}
w(r\to 0) = r^{-1} (\nu_0 + I_N) + {\rm o}(1)\,.
\label{cc.21b}
\end{equation}
The transformed regular solution has the behaviour at the origin 
\begin{equation}
L \varphi_0(k,r \to 0) = {\rm o}(r^{\nu_0})
\label{cc.29}
\end{equation}
and the transformation is conservative.
The singularities at the origin of the transformed potential, 
\begin{equation}
V_1(r\to 0) \to r^{-2} (\nu_0+I_N)(\nu_0+2I_N)\,,
\label{cc.30}
\end{equation}
increase by one unit, $\nu_0 \to \nu_1=\nu_0+I_N$.
In this case, there is thus no particular condition on $\nu_0$.

Intermediate cases where $\det D' = 0$ but $D' \neq 0$
have a more complicated structure of subcases and must be discussed according to
the precise values of $\nu_0$, $C'$ and $D'$:
the transformation can be conservative for some channels
and non conservative for others.
A general study of the conservativeness of these
transformations would be useful.
In the following, we shall encounter examples of such conservative
transformations in \S \ref{sec:pct}
and of such non-conservative transformations in section \ref{sec:ipwt}.
For instance, in section \ref{sec:ipwt}, one has $\nu_0 = 0$
and the transformation is non conservative in all channels.
One has then the property 
\begin{equation}
L [\eta_0 + \varphi_0 w(0)](k,r \to 0) = r (k^2 + \kappa^2) + O(r^2)\,,
\label{cc.29a}
\end{equation}
where $\eta_0(k,r)$ is the irregular matrix solution at energy $\cE$
with initial conditions \eref{mc.25}.
\subsection{Transformed Jost matrix}
The transformed Jost matrix can be obtained from expression \eref{mc.15}
of the regular solution $\varphi_1(k,r)$ corresponding to $V_1$
or through \eref{mc.17} from the behaviour at the origin of the Jost solution \eref{cc.15} corresponding to $V_1$.

First, we consider conservative transformations with ${\rm det\,} D' \neq 0$.
The regular solution of the transformed potential $\varphi_1(k,r)$
is determined by \eref{mc.9} with the singularity parameter $\nu_1$.
To derive it, we act on both sides of expression \eref{mc.15}
of the regular solution $\varphi_0(k,r)$ for potential
$V_0$ with the transformation operator $L$.
From \eref{cc.4}, \eref{mc.9} and \eref{cc.21}, it follows that 
\begin{equation}
\varphi_1(k,r) = - L \varphi_0(k,r)\,,
\label{cc.23}
\end{equation}
as in the single-channel case \eref{sct.12}.
By taking \eref{cc.15} into account, \eref{cc.23} can be rewritten as 
\begin{equation}
\fl \varphi_1(k,r) =
-\frac{\rmi}{2k} \left[\, f_1(-k,r) (w_\infty+\rmi k) F_0(k)
- f_1(k,r) (w_\infty-\rmi k) F_0(-k) \right].
\label{cc.24}
\end{equation}
Comparing \eref{mc.15} and \eref{cc.24} leads to a relation
between the initial and transformed Jost matrices 
\begin{equation}
F_1(k)  = -(\rmi k + w_\infty) F_0(k)
\label{cc.25}
\end{equation}
with $w_\infty$ given by \eref{cc.14bis} or \eref{cc.14}.
The situation is thus very different in the equal-threshold
and different-threshold cases.
For equal thresholds, a single conservative
supersymmetric transformation is able
to introduce a non-trivial coupling to the potential,
as $w_\infty$ in \eref{cc.14} can be non diagonal;
this will be exploited in section \ref{sec:ipnot}.
In the presence of different thresholds, in contrast,
a single conservative transformation can only multiply
the Jost matrix by a diagonal matrix
according to \eref{cc.14bis};
hence, the scattering matrix \eref{mc.21} is modified in a simple way,
best seen in the Stapp parametrization \eref{Stapp1} of the two-channel problem:
only the phase shifts are modified by simple arctangent terms,
as in the single-channel case,
%{\bf (check and give explicit expression?)},
while the coupling matrix is left unchanged.

For conservative transformations in the case $D' = 0$,
the Jost solutions are still related by \eref{cc.15}.
With the behaviour \eref{cc.21b} of the superpotential,
one obtains at small distances 
\begin{equation}
f_1(k,r \to 0) =
r^{-\nu_0 + I_N} (2\nu_0 + I_N)!! F_0^T(k) (w_\infty - \rmi k)\,.
\label{cc.29b}
\end{equation}
Hence \eref{mc.17} leads to 
\begin{equation}
F_1(k)  = (-\rmi k + w_\infty) F_0(k)\,.
\label{cc.25a}
\end{equation}
For equal and different thresholds,
this conservative transformation can only multiply
the Jost matrix by a diagonal matrix.

When the transformation is non conservative ($\nu_0=0$), 
equation \eref{cc.29a} implies that
the regular solution of the transformed potential reads 
\begin{equation}
\varphi_1(k,r) = L\, [\,\eta_0(k,r) +  \varphi_0(k,r) w(0)\,]\,
 (k^2 + \kappa^2)^{-1},
\label{cc.23a}
\end{equation}
where 
\begin{equation}
k^2 + \kappa^2 = (E - \cE) I_N
\end{equation}
is a scalar matrix when all reduced masses are equal.
Similarly to the single-channel case \eref{nonconsF1},
the transformed Jost solution is obtained by comparing
its asymptotic expression
deduced from \eref{mc.9} and \eref{mc.26}
with the asymptotic expression \eref{mc.9} for $\varphi_1(k,r)$ as 
\begin{equation}
F_1(k)  = (-\rmi k + w_\infty)^{-1} [\,G_0(k) + F_0(k)\, w(0)\,],
\label{cc.25b}
\end{equation}
where \eref{cc.14c} has been used.
Here a coupling can be introduced even when the thresholds are not all equal.
Hence, non-conservative transformations will be needed
in the different-thresholds case
to obtain non-trivial couplings.
This possibility will be explored in section \ref{sec:ipwt}.
\subsection{Pairs of transformations}
\subsubsection{General properties}
Two-fold supersymmetric transformations lead to a number of interesting
quantum models with unusual properties \cite{andrianov:95}.
It is natural to consider a two-fold supersymmetric transformation
of the Schrödinger equation \eref{cc.1} as a chain of usual
(i.e.\ one-fold) supersymmetric transformations
\cite{samsonov:04,pupasov:10}.
However, a pair is less restrictive than a single transformation
since the intermediate
Hamiltonian may be chosen unphysical, and in particular complex.

The chain of two supersymmetric transformations
$H_0\rightarrow H_1\rightarrow H_2$
with factorization energies $\cE_0$ and $\cE_1$
results from the intertwining relations
\begin{eqnarray}
L_0 H_0 = H_1 L_0\,,
\label{et3.1} \\[.5em]
L_1 H_1 = H_2 L_1\,,
\label{et3.1a}
\end{eqnarray}
where the operators $L_j$ map solutions
of the Schrödinger equations to each other
as $\psi_1 = L_0 \psi_0$ and $\psi_2 = L_1 \psi_1$.
These operators can be combined into an operator
\begin{equation}
L = L_1 L_0
\label{et3.2a}
\end{equation}
defining the two-fold supersymmetric transformation
\begin{equation}
L H_0 = H_2 L\,.
\label{et3.2}
\end{equation}
It directly maps solutions of the initial Schrödinger equation
to solutions of the trans\-formed Schrödinger equation as
\begin{equation}
\psi_2 = L \psi_0\,.
\label{et3.2b}
\end{equation}

The first-order differential operators $L_j$ read
\begin{eqnarray}
L_0 = - I_N \frac{\rmd}{\rmd r} + w_0(r)\,,
\label{et3.3a} \\
L_1 = - I_N \frac{\rmd}{\rmd r} + \tilde{w}_1(r)\,.
\label{et3.3}
\end{eqnarray}
The superpotentials
\begin{eqnarray}
&& w_i(r) = u'(\cE_i, r)u^{-1}(\cE_i,r)\,, \quad (i=0,1),
\label{et3.4} \\[.5em]
&& \tilde{w}_1(r) = v'(\cE_1,r) v^{-1}(\cE_1,r)\,,
\label{et3.5}
\end{eqnarray}
are expressed in terms of the matrix factorization solutions $u(\cE_i,r)$
and of the transform
\begin{eqnarray}
v(\cE_1,r) = L_0 u(\cE_1,r)
\label{et3.5a}
\end{eqnarray}
of $u(\cE_1,r)$ through the first transformation.
The solutions $u(\cE_i,r)$ and $v(\cE_1,r)$ satisfy the Schrödinger equations
\begin{eqnarray}
H_0 u(\cE_i,r) = \cE_i u(\cE_i, r), \quad i=0,1\,,
\label{et3.6a} \\[.5em]
H_1 v(\cE_1,r) = \cE_1 v(\cE_1,r)\,.
\label{et3.6}
\end{eqnarray}

The twofold operator $L$ and its adjoint $L^\dagger$
obey the factorization properties
\begin{eqnarray}
L^\dagger L = (H_0-\cE_0 I_N) (H_0-\cE_1 I_N)\,,
\label{et3.6b} \\[.5em]
L L^\dagger = (H_2-\cE_0 I_N) (H_2-\cE_1 I_N)\,.
\label{et3.6c}
\end{eqnarray}
These relations can be obtained from the properties \eref{et3.1}, \eref{et3.1a},
\eref{cc.6b} and \eref{cc.6c} applied to $L_0$ and $L_1$.

The Hamiltonians in \eref{et3.1} and \eref{et3.1a} correspond to potentials related to each other
through superpotentials
\begin{eqnarray}
V_1(r) = V_0(r)-2w_0(r)\,,
\label{et3.11} \\[.5em]
V_2(r) = V_1(r)-2\tilde{w}_1(r)\,.
\label{et3.11a}
\end{eqnarray}
The sum of the two superpotentials $w_0$ and $\tilde{w}_1$ defines
the two-fold superpotential
\begin{equation}
W(r) = w_0(r)+\tilde{w}_1(r)
\label{et3.12}
\end{equation}
which directly connects $V_0$ to $V_2$,
\begin{equation}
V_2(r) = V_0(r) - 2W'(r)\,.
\label{et3.12a}
\end{equation}
\subsubsection{Equal factorization energies and phase-equivalent potentials}
This paragraph summarizes results from references \cite{sparenberg:97b} and \cite{leeb:00}, 
where phase-equivalent deformations of coupled potentials were introduced 
with the aid of supersymmetric transformations. 
A pair of supersymmetric transformations with equal factorization energies may modify bound-state properties 
without affecting the scattering matrix of the system in analogy with the single-channel case 
(see \S \ref{sect_equal_fact} and \S \ref{sec:chatpe}). 

Let us consider the possible choices of transformation functions in a chain of two successive 
supersymmetric transformations with coinciding factorization energies. 
A transformation function $u(\cE_0,r)$ determines operator $L_0$ and belongs to its one-dimensional kernel 
\begin{eqnarray} 
L_0 u(\cE_0,r)=0\,.
\label{cernel.ccSUSY} 
\end{eqnarray}
The transformation function $v(\cE_0,r)$ of the second supersymmetric transformation 
can be calculated as follows 
\begin{eqnarray}
v(\cE_0,r) = \left[u^\dagger(\cE_0,r)\right]^{-1} 
\left(A+\int\limits_{r_0}^r u^\dagger(\cE_0,t)u(\cE_0,t)B \rmd t\right)\,.
\label{second-tr-function.ccSUSY}
\end{eqnarray}
where $A$ and $B$ are $N\times N$ constant matrices. 
The choice of $u(\cE_0,r)$, $A$ and $B$ should provide the regularity (for $r>0$) 
and hermiticity of potential $V_2$. 
References \cite{sparenberg:97b} and \cite{leeb:00} give $u(\cE_0,r)$, $A$ and $B$ 
realizing the phase-equivalent bound-state removal and addition. 

Let us for instance detail the removal
of a non-degenerate bound state at energy $\cE_0$. 
Then the Jost matrix has rank $N-1$ at the bound-state energy, 
${\rm rank\,} F_0(i\kappa_0)=N-1$. 
The normalized zero eigenvector $\ve{v}_0$ of the Jost matrix 
$F_0(i\kappa_0)\ve{v}_0=0$ defined in \eref{mc.18} allows one 
to extract the bound-state wave function from the Jost solution 
$\psi(\cE_0,r)=f_0(i\kappa_0,r)F_0(-i\kappa_0)\ve{v}_0$. 
The vector solution $\psi(\cE_0,r)$ vanishes both at the origin and at infinity. 
To remove this non-degenerate bound state and obtain a phase-equivalent Hamiltonian, 
one has to choose $u(\cE_0,r)=f_0(i\kappa_0,r)$, $A=I_N$, 
$B=F_0(-i\kappa_0)\ve{v}_0 \ve{v}_0^\dagger F_0^\dagger(-i\kappa_0)^{-1}$. 
A phase-equivalent removal of a bound state increases the singularity 
of the potential at the origin.
Inverting this chain of transformations, 
we obtain a recipe for a phase-equivalent bound-state addition. 
The choice of vector $\ve{v}_0$ then determines the ANCs of the bound state
without affecting the scattering matrix. 

The phase-equivalent bound-state addition requires a potential $V_0$ 
singular at the origin [see \eref{mc.8}], 
whereas the bound-state removal may be applied to any regular potential with bound states. 
In a simple case where $\nu_0={\rm diag}(\nu_{0,1},\nu_{0,2},\ldots,\nu_{0,N})$, 
$\nu_{0,1}<\nu_{0,j}, j=2,N$, a bound-state removal leads to 
$\nu_2={\rm diag}(\nu_{0,1}+2,\nu_{0,2},\ldots,\nu_{0,N})$ \cite{sparenberg:97b}. 
Other examples and numerical studies suggest a topological invariant 
which counts the number of bound states $n_b$ 
and the order of singularity at the origin 
$\Delta=\pi(n_b+\frac{1}{2}{\rm Tr\,}\nu)$.
This topological invariant could be related 
with the coupled-channel generalization of the Levinson theorem. 
\subsubsection{Different factorization energies}
For compactness, we adopt the notations $u_0\equiv u(\cE_0,r)$ and $u_1\equiv u(\cE_1,r)$
for the factorization solutions.
We only consider self-conjugate factorization solutions,
i.e.\ solutions with a vanishing self-Wronskian $\mathrm{W}[u_i,u_i] = 0$, $i=0, 1$.
This implies that the superpotentials $w_i$ are symmetric,
which in turn implies that the Wronskian
of the factorization solutions can be written as
\begin{equation}
\mathrm{W}[u_0,u_1] =
 u_0^T \left[\,w_1(r)-w_0(r)\,\right] u_1
\label{et3.8}
\end{equation}
in the case of different factorization energies.
From \eref{et3.5a}, solution $v(\cE_1,r)$ then reads
\begin{equation}
v(\cE_1,r) = [\,w_0(r)-w_1(r)\,] u_1
= -\left[u_0^T\right]^{-1} \mathrm{W} [u_0, u_1]\,.
\label{et3.9}
\end{equation}

Using \eref{et3.5} and \eref{cc.6a}, one can then rewrite $W$ in the compact form
\begin{equation}
W(r) = (\cE_0-\cE_1) [\,w_1(r) - w_0(r)\,]^{-1},
\label{et3.13a}
\end{equation}
which generalizes \eref{W2}.
With \eref{et3.8}, it can be rewritten as
\begin{equation}
W(r) =
 (\cE_0-\cE_1)\, u_1 \mathrm{W}[u_0, u_1]^{-1} u_0^T
 \,.
\label{et3.13}
\end{equation}
This expression may be used in cases where the individual
superpotentials $w_i$ are singular.
Using the Wronskian derivative \eref{Wder},
one gets the expression for the final potential 
\begin{equation}
V_2(r)=V_0(r)-2\,\frac{\rmd}{\rmd r}
 \left\{\left[u_0^T\right]^{-1} \frac{\rmd \mathrm{W}[u_0, u_1]}{\rmd r} \mathrm{W}[u_0, u_1]^{-1} u_0^T \right\},
\end{equation}
where the superpotential expression generalizes the logarithmic derivative \eref{W} appearing in the singe-channel case.

Similarly, the action of the second order transformation
operator $L$ on $\psi_0(k,r)$,
\begin{equation}
\psi_2(k,r) =
\left(\tilde{w}_1-I_N\frac{\rmd}{\rmd r}\right)%
 \left(w_0-I_N\frac{\rmd}{\rmd r}\right) \psi_0(k,r)\,,
\label{et3.14}
\end{equation}
can be rewritten by expressing the second derivative
of the matrix solution $\psi_0(k,r)$ as 
\begin{equation}
\psi_2(k,r) = \left[ -k^2-\frac{\kappa_0^2+\kappa_1^2}{2}
+ W(r) \left(\frac{w_0+w_1}{2} -
I_N\frac{\rmd}{\rmd r}\right)\right]\psi_0(k,r)\,.
\label{et3.15}
\end{equation}
\subsubsection{Mutually conjugate factorization energies}
\label{sec:cfemc}
For two successive transformations with respective
complex factorization energies
$\cE \equiv \cE_R+\rmi \cE_I$ and $\cE^*\equiv \cE_R-\rmi \cE_I$,
the results of the previous paragraph apply and simplify with $w_0 = w\equiv w_R + \rmi w_I$ and $w_1 = w^*$.
In particular, the two-fold superpotential reads
\begin{equation}
W(r) = -\cE_I [\Imag w(r)]^{-1},
\label{W2conj}
\end{equation}
which generalizes \eref{sct.35}.
It can also be written as
\begin{equation}
W(r) = 2 \cE_I u^* {\mathrm{W}}[u,u^*]^{-1} u^T,
\label{et3.13conj}
\end{equation}
leading to the potential
\begin{equation}
  V_2(r)=V_0(r)-2\,\frac{\rmd}{\rmd r}
 \left([u^T]^{-1} \frac{\rmd \mathrm{W}[u,u^*]}{\rmd r} \mathrm{W}[u,u^*]^{-1} u^T \right).
\label{V2conj}
\end{equation}
For the matrix solutions, one gets
\begin{equation}
\psi_2(k,r) = \left[ -k^2 -\cE_R
+ W(r) \left(w_R - I_N\frac{\rmd}{\rmd r}\right)\right]\psi_0(k,r).
\label{et3.15conj}
\end{equation}
\subsection{Chains of transformations \label{sec:chains}}
For the scalar Schrödinger equation,
supersymmetric (Darboux) transformation chains can be formulated in two alternative ways
(see \S \ref{sec:chat}):
either by constructing iterative chains of first-order transformations
or by using Crum-Krein formulas which describe high-order transformations in a closed form.
The same options are available for the matrix Darboux transformations,
though a systematic study of all possible chains is not done yet for the coupled-channel case.

For the case of equal thresholds and different factorization energies, 
an extension of the Crum-Krein formulas to the matrix case was developed in \cite{samsonov:04}.
For the purpose of inversion problems,
it is often necessary to combine one-channel transformations
which are able to invert a single phase shift
with matrix transformations which are able to invert a full scattering matrix.
Such single-channel transformations cannot be directly incorporated
into the usual chain of matrix supersymmetric transformations considered in \cite{samsonov:04}.
Nevertheless, theorems proved in \cite{samsonov:04} have a more general
validity than supersymmetric transformations of the matrix Schrödinger equation.
Actually, they represent a closure of a special recursion procedure
which can easily be generalized to incorporate both single-channel and matrix transformations.
This generalization of the coupled-channel Crum-Krein formulas established in \cite{samsonov:04}
requires to formulate single-channel transformations as singular matrix transformations \cite{pecheritsin:11}.

In the following, we make the results of \cite{pecheritsin:11} explicit for the two-channel case:
we define singular transformations and provide the explicit expression of a potential
constructed with a chain of singular and non-singular transformations.
These results will be illustrated in section \ref{sec:ipnot} for the neutron-proton triplet case.
We show there that Crum-Krein formulas are particularly useful
when the initial potential vanishes or is purely centrifugal,
as they allow one to express arbitrary matrix potentials with a rational scattering matrix in terms of spherical Bessel functions
(solutions of the free radial Schrödinger equation).

Consider a two-channel Schrödinger equation
with a diagonal potential matrix $V_0(r)$,
\begin{equation}
V_0(r) =\left(
\begin{array}{cc}
V_0^{(I)}(r) & 0 \\ 0 & V_0^{(II)}(r)
\end{array}
   \right).
%\label{V-block}
\end{equation}
Here $V_0^{(I)}$ and $V_0^{(II)}$ are scalar potentials in the first and second channels.
In this case, the matrix Schrödinger equation \eref{sc.2} with potential matrix $V_0$
splits into two independent scalar equations
\begin{equation}
\left\{ \begin{array}{ll}
         H_0^{(I)}\Psi_E^{(I)}(r) = E \Psi_E^{(I)}(r), &  H_0^{(I)}=- \frac{\rmd^2}{\rmd r^2} + V_0^{(I)}(r) \\
	 H_0^{(II)}\Psi_E^{(II)}(r) = E \Psi_E^{(II)}(r), & H_0^{(II)}=- \frac{\rmd^2}{\rmd r^2} + V_0^{(II)}(r).
        \end{array}
        \right.
\label{shred-I}
\end{equation}

Let us first consider a transformation that only affects channel $I$.
This transformation can be written in matrix form as
\begin{eqnarray}\label{darbu-m}
L & \equiv & \left(
\begin{array}{cc}
L^{(I)} & 0 \\ 0 & 1
\end{array}
\right) \\
& = & - I_+ \frac{\rmd}{\rmd r} +
\left(
\begin{array}{cc} w^{(I)}(r) & 0 \\
0 & 1 \end{array}\right),
\label{darboux-sing}
\end{eqnarray}
where matrix $I_+$ reads
\begin{equation}
I_+=
\left(
\begin{array}{cc}
1 & 0 \\ 0 & 0\end{array}
\right). 
\end{equation}
This rank-1 matrix is singular;
we thus call such transformations {\em singular supersymmetric transformations} \cite{andrianov:97}.
The function $w^{(I)}(r)$ is a usual superpotential \eref{w0} for the first channel,
defined by the scalar transformation solution
\begin{equation}
w^{(I)}(r)=\frac{\rmd}{\rmd r}\ln u^{(I)}(r)\,,\qquad
H_0^{(I)} u^{(I)} =  {\cal E}^{(I)} u^{(I)}\,.
%\label{tildeU}
\end{equation}

Similarly, a transformation only affecting channel $II$ can be defined,
introducing operator $L^{(II)}$.
Operators $L^{(I)}$ and $L^{(II)}$ can be combined in a single matrix operator
\begin{eqnarray}\label{darbu-m-2}
L \equiv \left(
\begin{array}{cc}
L^{(I)} & 0 \\ 0 & L^{(II)}
\end{array}
\right) 
\\
= - I_2\frac{\rmd}{\rmd r} +
\left(
\begin{array}{cc} w^{(I)}(r) & 0 \\
0 & w^{(II)}(r) \end{array}\right),
\end{eqnarray}
where $I_2$ is the $2\times2$ identity matrix.
In contrast with operator \eref{darboux-sing},
this operator is thus regular.
It has the same structure as operator \eref{cc.4}
but with a diagonal superpotential and
with a factorization constant defined
by the diagonal matrix ${\rm diag}({\cal E}^{(I)}, {\cal E}^{(II)})$.
This diagonal matrix generalizes matrix $-\kappa^2$ appearing in \eref{cc.7}
as the factorization energies ${\cal E}^{(I)}$ and ${\cal E}^{(II)}$ are now independent of each other
and do not have to satisfy the threshold condition \eref{Thr_condit}.

Iterating single-channel transformations leads to operators $L^{(I)}$ and $L^{(II)}$ of the form \eref{Ln},
where the numbers $N^{(I)}$ and $N^{(II)}$ of transformations in channels $I$ and $II$ are independent of each other.
Similarly, $M$ regular matrix transformations, whether diagonal or not,
can be iterated,
which defines the $M$th-order matrix operator $L^{(M)}$
\begin{eqnarray}\label{secSDT-matrix-operator-ns}
L^{(M)}=(-1)^{M}I_2\frac{\rmd^M}{\rmd r^M}+A_{M-1}(r)\frac{\rmd^{M-1}}{\rmd r^{M-1}}+\ldots+A_0(r)\,,
\end{eqnarray}
where $A_j(r)$ are some matrix coefficients.
It is natural to write the composition of $L^{(I)}$,  $L^{(II)}$ and  $L^{(M)}$ in matrix form as follows
\begin{equation}\label{secSDT-single-ch-matrix-composition}
L=L^{(M)}
\left(
\begin{array}{cc}
L^{(I)} & 0 \\ 0 & 1
\end{array}
\right)
\left(
\begin{array}{cc}
1 & 0 \\ 0 & L^{(II)}
\end{array}
\right)=L^{(M)}
\left(
\begin{array}{cc}
L^{(I)} & 0 \\ 0 & L^{(II)}
\end{array}
\right).
\end{equation}
When the numbers of single-channel transformations are different in channels $I$ and $II$,
operator $L$ cannot be written in the form \eref{secSDT-matrix-operator-ns}
since its first matrix coefficients are singular matrices.
To fix ideas, let us assume that more single-channel transformation are applied to channel $I$ than to channel $II$
and let us denote the difference between these numbers by $m=N^{(I)}-N^{(II)}> 0$.
Let us denote by $N=M+N^{(I)}$ the total number of transformations,
$m$ of which not reducing to regular transformations.
Operator $L$ then takes the form
\begin{eqnarray}\label{secSDT-singular-matrix-composition}
L =  (-1)^{N} I_+ \frac{\rmd^{N}}{\rmd r^{N}} + A_{N-1}(r) \frac{\rmd^{N-1}}{\rmd r^{N-1}} + \ldots + A_0(r),
\end{eqnarray}
where matrices $A_j(r)$ vanish except for their upper left coefficient when $j\ge N-m+1$
and are regular for $j< N-m+1$.

To incorporate singular transformations of the type \eref{darboux-sing}
into the chain of usual matrix supersymmetric transformations considered in \cite{samsonov:04},
we need auxiliary transformation matrices $\cU$
which are built from single-channel transformation functions
\begin{equation}\label{def-singular-transromation-solution}
\cU = \left(
\begin{array}{cc} u^{(I)} & 0 \\ 0 & 1
\end{array}\right).
\end{equation}
These matrices do not satisfy the matrix Schrödinger equation itself.
Nevertheless, they properly define the superpotential
\begin{equation}
\cU'\cU^{-1}  =
\left(
\begin{array}{cc} (u^{(I)})' (u^{(I)})^{-1} & 0 \\ 0 & 0
\end{array}
\right),
\end{equation}
and the matrix valued operator \eref{darboux-sing}
\begin{equation} \label{darbu-pm}
L =- D^{(1,0)} + \cU'\cU^{-1},
\end{equation}
where
\begin{equation}\label{def-singular-D-operator}
D^{(\alpha, \beta)}=
\left( \begin{array}{cc} \frac{\rmd^\alpha}{\rmd r^\alpha} & 0 \\ 0  & \frac{\rmd^\beta}{\rmd r^\beta} \end{array} \right).
\end{equation}

Consider the transformation defined by operator \eref{secSDT-singular-matrix-composition}.
Let $\cU_k$, $k = 1, \ldots, m$, be $2\times 2$ matrices 
having the structure fixed by \eref{def-singular-transromation-solution}, 
i.e.\ only the first row of these matrices contains solutions of the 
first-channel Schrödinger equation, 
\begin{equation}%\label{Uk-def}
\cU_k= \left(
\begin{array}{cc} u_k^{(I)} & 0 \\ 0 & 1
\end{array}\right).
\end{equation}
Let $\cU_l$, $l = m+1, \ldots, N$, be $2\times 2$ matrices 
where each row is composed of solutions of the Schrödinger equation, 
\begin{equation}%\label{Uk-def}
\cU_l= \left(
\begin{array}{cc} u_{l;11} & u_{l;12} \\ u_{l;21} & u_{l;22}
\end{array}\right) .
\end{equation}
This set of $N$ matrices thus involves singular and usual matrix Darboux transformations. 

Let us introduce matrix ${\cal W}$ with $N$ rows and $N$ columns of $2 \times 2$ block elements
\begin{equation}\label{W-def}
{\cal B}_{ij} = D^{(i-1, p)} \cU_j, \qquad i, j = 1, \dots, N
\end{equation}
where
\begin{equation}
p = \left\{ 
\begin{array}{ccc} 
i-1 & \textrm{for} & j<i,\ j\le m, \\
 0  & \textrm{for} & j\ge i,\ i\le m, \\
 m  & \textrm{for} & i>m,\ j>m.
\end{array} \right.
\end{equation}
The $2m \times 2m$ left upper block in this $2N \times 2N$ matrix (with $N>m$) 
corresponds to the $m$ first singular Darboux transformations.  

We also need a set of four $2N \times 2N$ matrices ${\cal W}^{k,l}$, $k,l = 1,2$ 
obtained from matrix \eref{W-def} by replacing its last row of blocks ${\cal B}_{Nj}$ 
by new blocks ${\cal B}^{k,l}_{Nj}$ defined by
\begin{eqnarray}
{\cal B}_{Nj}^{1,1} = D^{(1,0)} {\cal B}_{Nj}, 
\qquad {\cal B}_{Nj}^{2,2} = D^{(0,1)} {\cal B}_{Nj},
\nonumber \\
{\cal B}_{Nj}^{1,2} = 
\left( \begin{array}{cc} 0 & \frac{\rmd}{\rmd r} \\ 0  & 1 \end{array} \right) {\cal B}_{Nj},
\qquad {\cal B}_{Nj}^{2,1} = 
\left( \begin{array}{cc} 1 & 0 \\ \frac{\rmd}{\rmd r} & 0 \end{array} \right) {\cal B}_{Nj}.
\end{eqnarray}
Explicit expressions can be found in \cite{pecheritsin:11} 
and an example is given in \S \ref{sec:eptnp}.

The main result of \cite{pecheritsin:11} may then be formulated as follows.
A chain of $m$ Darboux transformations in the first channel and $N-m$ matrix Darboux transformations, 
applied to the initial diagonal potential $V_0$,
leads to the new matrix potential
\begin{equation}
V_N = V_0 - 2 \frac{\rmd}{\rmd r} W_N\,,
\label{VN}
\end{equation}
where the matrix elements of the $N$-fold ``superpotential'' $W_N$
are defined as ratios of determinants
\begin{equation}
\label{secSDT-WNij-def}
W_{N; ij} = \frac{\det {\cal W}^{i, j}}{\det {\cal W}}\,.
\end{equation}
Using similar formulas one can also find the components of the vector solutions for potential $V_N$ \cite{pecheritsin:11}.

%%%%%%%%%%%%%%%%%%%%%%%%%%%%%%%%%%%%%%%%%%%%%%%%%%%%%%%%%%%%%%%%%%%%%%%%%%%%%%%%%%
\section{Two-channel inverse problems with equal thresholds}
\label{sec:ipnot}
We now particularize the contents of section \ref{sec:cc} to a two-channel case.
As discussed in \S \ref{sec:ms}, the equal-threshold case
corresponds to a collision of particles with spin.
The two-channel case is for example useful for elastic collisions between
particles with a spin 1/2.
The identity matrix $I_2$ is here written as $I$.
\subsection{Potentials with a generalized asymptotic behaviour}
\label{sec:gc}
One-fold transformations for coupled-channel radial
Schrödinger equations were first introduced in \cite{amado:88a,amado:88b,amado:90,cannata:92,cannata:93,andrianov:95}.
These transformations were generalized in \cite{pupasov:09}
to allow for the introduction of coupling between channels.
The additional requirement that the transformed potential be
physical was shown to result in a strong constraint
on the transformation parameters.
In this section, we generalize the results of \cite{pupasov:09,pupasov:10}
in such a way that this constraint can be avoided.
Moreover we show that the technique remains valid when supersymmetric
transformations are iterated.
Iterations are necessary in realistic inversions.
This generalization for pairs of complex conjugate
transformations was employed in \cite{pupasov:11}.

In the two-channel case, according to \eref{mc.5},
a physical potential has the diagonal long-range behaviour \eref{mc.7}, 
\begin{equation}
\tilde{V}(r\to\infty) \to r^{-2} l(l+I)
= r^{-2} \left( \begin{array}{cc} l_1(l_1+1) & 0 \\
0 & l_2(l_2+1) \end{array} \right).
\label{et1.1}
\end{equation}
This behaviour is encountered in various physical problems.
Because of parity conservation,
the physical cases correspond to $l$ values with the same parity.
The most interesting cases are $l_1 = l_2$ and $l_1 = l_2 \pm 2$.

In supersymmetric transformations, however, the asymptotic behaviour \eref{et1.1}
is not necessarily conserved as we shall see below.
This raises two problems.
(i) Iterations must be able to start from a more general case.
(ii) A correction to the final potential must
be introduced to comply with \eref{et1.1}.
The principle of the resolution of problem (ii) was described
in \cite{pupasov:11}.

We thus consider a more general asymptotic behaviour of the potential 
\begin{equation}
V(r\to\infty) \to r^{-2} \Lambda,
\label{et1.2}
\end{equation}
where $\Lambda$ is a real symmetric matrix 
\begin{equation}
\Lambda = \left( \begin{array}{cc} \Lambda_{11} & \Lambda_{12} \\
\Lambda_{12} & \Lambda_{22} \end{array} \right)
\label{et1.3}
\end{equation}
with eigenvalues $\Lambda_i = l_i(l_i+1)$ with $l_i$ integer.
This matrix can be diagonalized with a rotation-like transformation 
\begin{equation}
R^T(\theta) \Lambda R(\theta) = \mathrm{diag} (\Lambda_1, \Lambda_2)
\label{et1.4}
\end{equation}
where $R(\theta)$ is an orthogonal matrix 
\begin{equation}
R(\theta) = \left( \begin{array}{cc}  \cos\theta & \sin\theta \\
-\sin\theta & \cos\theta \end{array} \right)
\label{et1.6}
\end{equation}
depending on a single constant $\theta$.

Matrix $\Lambda$ in \eref{et1.3} is a scalar matrix in the case $l_1 = l_2$
and the following considerations are then unnecessary.
For $l_1 \neq l_2$, let us consider the auxiliary potential 
\begin{equation}
\tilde{V} = R^T(\theta) V R(\theta).
\label{et1.5}
\end{equation}
The asymptotic form of this auxiliary potential is diagonal
like in \eref{et1.1}.
Since the thresholds are equal, the Hamiltonian 
\begin{equation}
\tilde{H} = R^T(\theta) H R(\theta)
\label{et1.7}
\end{equation}
is unitarily equivalent to $H$
and all matrix solutions of $H$ and $\tilde{H}$ are related by 
\begin{equation}
\psi = R(\theta) \tilde{\psi} R^T(\theta).
\label{et1.8}
\end{equation}
The final factor is needed to keep the correct asymptotic
behaviour \eref{mc.13}
of Jost solutions.
It can also be introduced for convenience in other cases.

The Jost solutions $\tilde{f}(k,r)$ of $\tilde{H}$ have the asymptotic form, 
\begin{equation}
\tilde{f}(k,r\to\infty) = \diag \left[h_{l_1}(kr),h_{l_2}(kr)\right]
\label{et1.9}
\end{equation}
in terms of the spherical Hankel functions \eref{sc.12}.
Their asymptotic behaviour is given by \eref{sc.13}.
Hence the Jost solutions $f(k,r)$ of $H$ are given by 
\begin{equation}
f(k,r) = R(\theta) \tilde{f}(k,r) R^T(\theta).
\label{et1.10}
\end{equation}
Since both regular and Jost solutions transform according to
\eref{et1.8}, one deduces from relation \eref{mc.15}
the transformation of the Jost matrix 
\begin{equation}
\tilde{F}(k) = R^T(\theta) F(k) R(\theta).
\label{et1.11}
\end{equation}
This equation can be used to transform a Jost matrix corresponding
to an unphysical asymptotic behaviour \eref{et1.2} of the potential
into a physical Jost matrix.
The scattering matrix \eref{mc.16} then transforms according to 
\begin{equation}
\tilde{S}(k) = R^T(\theta) S(k) R(\theta).
\label{et1.12}
\end{equation}
An orthogonal transformation can transform a potential
with an unphysical asymptotic behaviour \eref{et1.2} into a physical potential.
In the following paragraphs, we shall use this procedure to eliminate
unphysical asymptotic behaviours.

Notice that the singularity at the origin becomes more
complicated than in \eref{mc.9}.
The regular solution behaves as 
\begin{eqnarray}
\tilde{\varphi}(k,r\to 0) \to R^T(\theta) r^{\nu+I}[(2\nu+I)!!]^{-1}
R(\theta)\,.
\label{et1.13}
\end{eqnarray}
The singularity may also occur in the non-diagonal terms.
This modification does not generate any practical difficulty.
\subsection{Single transformation}
\label{sec:single}
For the transformation,
we choose transformation function \eref{cc.10} with matrices \eref{cc.12}, 
\begin{equation}
C = \left(
\begin{array}{cc}1&0\\ q &0\end{array}\right),
\qquad D=\left(
\begin{array}{cc}x&-q\\ 0 &1\end{array}\right),
\label{et2.1}
\end{equation}
which contain only two independent real parameters $x \ne 0$ and 
\begin{equation}
q \equiv \tan \tfrac{\alpha}{2}\,.
\label{et2.3}
\end{equation}
These parameters are restricted by condition \eref{cc.19}.
Matrix $A$ of \eref{cc.13b} then becomes 
\begin{equation}
A =
\left(\cos{\textstyle \frac{\alpha}{2}}\right)^{-1}%
 R\left(-{\textstyle \frac{\alpha}{2}}\right),
\label{et2.3a}
\end{equation}
i.e., it is proportional to an orthogonal matrix.

With the help of the Pauli matrices $\sigma_x$, $\sigma_y$ and $\sigma_z$,
one can write 
\begin{equation}
R(\theta) =  I \cos \theta+ \rmi\,\sigma_y \sin \theta \,.
\label{et2.3b}
\end{equation}
From the general expression \eref{cc.14}, one obtains 
\begin{equation}
w_\infty =
 \k R\left(-{\textstyle \frac{\alpha}{2}}\right)%
  \sigma_z R\left({\textstyle \frac{\alpha}{2}}\right)
\label{et2.4a}
\end{equation}
or \cite{pupasov:09} 
\begin{equation}
w_\infty = \k R(-\alpha) \sigma_z
= \k\left( \begin{array}{cc}\cos \alpha& \sin \alpha\\
\sin\alpha & -\cos\alpha\end{array}\right).
\label{et2.4}
\end{equation}
However, one needs to go to the order $\Or(r^{-1})$ to find the modification
of the asymptotic form of the transformed potential.

The asymptotic behaviour of \eref{cc.10} reads with
\eref{et2.1} and \eref{sc.13}, 
\begin{equation}
u(r\to\infty) \to \left(
\begin{array}{cc} \rme^{\k r}\left(1-\frac{\Lambda_{11}+
q\Lambda_{12}}{2\k r}\right) &
-q \rme^{-\k r}\left(1+\frac{\Lambda_{11}-q^{-1}\Lambda_{12}}{2\k r}\right) \\
q \rme^{\k r}\left(1-\frac{\Lambda_{22}+q^{-1}\Lambda_{12}}{2\k r}\right) &
\rme^{-\k r}\left(1+\frac{\Lambda_{22}-q\Lambda_{12}}{2\k r}\right) \end{array}
\right).
\label{et2.4b}
\end{equation}
This asymptotic behaviour can be written more compactly as 
\begin{equation}
u(r\to \infty)\to \left(Q-\frac{1}{2\k r}\Lambda Q\sigma_z \right)%
\rme^{\k r\sigma_z},
\label{et2.4c}
\end{equation}
where 
\begin{equation}
Q = I - \rmi\, q\, \sigma_y
\label{et2.4d}
\end{equation}
generalizes operators $Q_{\pm}$ of \cite{pupasov:10}.
With 
\begin{equation}
\Lambda = \Lambda_+\, I + \Lambda_-\, \sigma_z + \Lambda_{12}\, \sigma_x
\label{et2.4e}
\end{equation}
where $\Lambda_\pm = \dem (\Lambda_1 \pm \Lambda_2)$ and \eref{et2.4c}
introduced in \eref{cc.6},
the superpotential reads 
\begin{equation}
w(r\to\infty) \to w_\infty - \frac{1}{r}
(\sin\alpha \,\Lambda_- - \cos \alpha\, \Lambda_{12})
(\sin\alpha\, \sigma_z - \cos\alpha \,\sigma_x)\,.
\label{et2.5}
\end{equation}
Equation \eref{cc.5} leads to 
\begin{equation}
\fl V_1(r\to\infty) \to r^{-2} \left[\, \Lambda_+ I
+ \Lambda_- (\cos 2\alpha \sigma_z + \sin 2\alpha \sigma_x)
- \Lambda_{12} (\cos 2\alpha\, \sigma_x - \sin 2\alpha \,\sigma_z)\,
\right]\,.
\label{et2.6}
\end{equation}
This result is a generalization of equation (43) of \cite{pupasov:09}.
A similar asymptotic behaviour of the matrix potential
had been obtained from the Gel'fand-Levitan equation in \cite{fulton:56}.
Expression \eref{et2.6} can be written more compactly as 
\begin{equation}
V_1(r\to\infty) \to r^{-2}  R^T(\alpha)\, \sigma_z\, \Lambda\, \sigma_z R(\alpha)
\,.
\label{et2.8}
\end{equation}
The asymptotic behaviour of the potential is thus modified by the transformation.
It does not keep the general form \eref{et1.2} because of the $\sigma_z$ factors.
We shall see below that this problem can easily be solved.

Introducing $w_\infty$ as given in \eref{et2.4a} into \eref{cc.25}
for the two-channel case provides an explicit relation
between the transformed Jost matrix $F_1(k)$ and the initial Jost matrix $F_0(k)$, 
\begin{eqnarray}
F_1(k) & = & -R^T\left({\textstyle \frac{\alpha}{2}}\right)%
 \left( \begin{array}{cc} \rmi k + \k & 0 \\
0 & \rmi k - \k \end{array}\right)%
 R\left({\textstyle \frac{\alpha}{2}}\right) F_0(k)
\label{et2.9} \\
& = & -\left( \begin{array}{cc} \rmi k + \k\cos\alpha& \k\sin\alpha \\
\k\sin\alpha & \rmi k - \k\cos\alpha\end{array}\right) F_0(k)
\label{et2.10}
\end{eqnarray}
Expression \eref{et2.9} follows immediately from the factorization \eref{cc.14}
where $A$ is proportional to the orthogonal
matrix $R^T({\textstyle \frac{\alpha}{2}})$.
It shows that the determinants of the Jost matrices are related by 
\begin{equation}
{\rm det\,}F_1 = (k^2+\k^2)\, {\rm det\,}F_0.
\label{et2.11}
\end{equation}
The transformation adds one bound state and one virtual state \cite{pupasov:09}.
The scattering matrices are related by 
\begin{equation}
\fl \tilde{S}_1(k)
= \left( \begin{array}{cc} \rmi k \cos\alpha + \k & \rmi k\sin\alpha \\
\rmi k\sin\alpha & \rmi k \cos\alpha - \k \end{array} \right)
%\nonumber \\ & \times &
S_0(k) \left( \begin{array}{cc} \rmi k \cos\alpha + \k & \rmi k\sin\alpha \\
\rmi k\sin\alpha & \rmi k \cos\alpha - \k \end{array} \right)^{-1}.
\label{et2.13}
\end{equation}
The transformation thus contains an additional parameter with
respect to \cite{pupasov:09}.

Let us first discuss the case $\Lambda_{12} = 0$ considered in \cite{pupasov:09}
where one starts from a potential with
a physical asymptotic behaviour \eref{et1.1}.
From \eref{et2.8}, for $l_1 \neq l_2$, the transformed potential
has a non-zero long range coupling 
\begin{equation}
V_1(r\to\infty) = r^{-2}  R^T(\alpha)\, l(l+I) R(\alpha).
\label{et2.12}
\end{equation}
Moreover, the diagonal entries of $V_1$ are not usual centrifugal terms.
A choice $\alpha = 0$ is possible but does not introduce a coupling
in the Jost function when one starts with uncoupled potentials.
In \cite{pupasov:09}, the choice $\alpha=\pm\pi/2$ was made
to obtain both a coupling in the Jost function \eref{et2.10}
and a physically reasonable long-range behaviour \eref{et1.1}
with the locations of $l_1$ and $l_2$ exchanged.
The permutation of $l_1$ and $l_2$ obtained in \cite{pupasov:09}
can thus be reinterpreted as a $\pm \pi/2$ pseudorotation.

\subsection{Iterations}
Let us now consider the more general case $\Lambda_{12} \ne 0$,
for which potential $V_1$ does not have a physical asymptotic behaviour.
We shall show that a physical behaviour can be restored
by iterating the transformations studied above.

Indeed, if two successive transformations are performed
with parameters $(\k_1,\alpha_1)$
and $(\k_2,\alpha_2)$, the asymptotic behaviour of the potential 
\begin{equation}
V_2(r\to\infty) = r^{-2}  R^T(\alpha_2-\alpha_1)\,
 \Lambda\, R(\alpha_2-\alpha_1).
\label{et2.14}
\end{equation}
keeps the form \eref{et1.2}.
A transformation \eref{et1.5} with $\theta = \alpha_1-\alpha_2$
allows one to return to a physical situation.
It is thus possible to iterate the process and perform $2p$
successive transformations.
At the end of the iteration, an elimination of the unphysical behaviour
is obtained as explained in \S \ref{sec:gc} with 
\begin{equation}
\theta = \sum_{j=1}^{2p} (-1)^{j+1} \alpha_j.
\label{et2.15}
\end{equation}
Notice that the behaviour of $V_2$ at the origin is not
any more given by \eref{mc.8}.
A singularity also appears in the non-diagonal term.

An advantage of the present generalized expressions
is that they can be iterated.
With such iterations, one can expect to be able to approximate
realistic scattering matrices to a good approximation and hence
to construct a coupled-channel potential by inversion.
However, with this procedure it is not easy to find the different pairs of parameters $(\k_i,\alpha_i)$ 
that satisfactorily reproduce the data.
There are indeed three functions to adjust
(the different entries of the symmetric collision matrix 
or the two eigenphases $\delta_1$, $\delta_2$ and the mixing parameter $\epsilon$)
and they all depend on all the fitting parameters simultaneously.
It would thus be more convenient to split the coupled-channel inverse problem into simpler successive subproblems
by first fitting the eigenphases separately with some parameters,
then fitting the mixing parameter with other parameters.
In the following, we present different approaches following these lines.
First, in the next paragraph we give different examples of coupled potentials 
that can be obtained with a single coupling supersymmetric transformation 
starting from a diagonal matrix potential.
This leads to a separable inversion method, but for the case of channels with different parities only.
Then, in \S \ref{sec:pct},
we present a systematic method based on transformation pairs with conjugate solutions \cite{pupasov:10,pupasov:11}
also applicable to the case of equal-parity channels and hence physically more interesting.

\subsection{Different types of coupling}
\label{sect_diff_coupl}
It is remarkable that different levels of coupling exist. 
A coupling may exist in a potential but its Jost and 
scattering matrices are uncoupled. 
A coupling may exist in a potential and its Jost matrix 
but the scattering matrix is uncoupled. 
Finally all three quantities may be coupled. 
Here we illustrate these various possibilities. 
We consider the case of conservative transformations for equal thresholds. 
\subsubsection{Trivially-coupled potentials and $S$ matrices}
\label{subsec:trivial_coupling}
The coupled-channel supersymmetric inversion starts from simple diagonal potentials. 
We show that degenerate initial diagonal potentials with a high symmetry 
cannot be transformed into coupled potentials. 

The simplest example is the scalar Hamiltonian 
\begin{equation}
H_0 = h_0 I_N.
\end{equation}
In this case, the regular solution $\varphi_0(k,r)I_N$ and 
the Jost solution $f_0(k,r)I_N$ are scalar functions 
multiplied by the unit matrix. 
However, the transformation matrix \eref{cc.10} needs not be diagonal 
because matrices $C$ and $D$ can be non diagonal. 
Hence the superpotential \eref{cc.6} and the transformed potential 
are not necessarily diagonal. 
However, in this case, a constant orthogonal matrix $R$ can be found 
which diagonalizes the new potential \eref{cc.5} for any 
transformation matrix \eref{cc.10}. 

Indeed, a right multiplication of the transformation matrix \eref{cc.10} 
by $D^{-1}$ does not modify the superpotential. 
The transformation matrix can be rewritten as 
\begin{equation}
u(r)D^{-1} = f_0(-i\kappa,r) CD^{-1} + f_0(i\kappa,r) I_N
\label{tcp.1}
\end{equation}
where $CD^{-1}$ is symmetric because of \eref{cc.11}. 
Let $R$ be the orthogonal matrix which diagonalizes $CD^{-1}$. 
One readily sees that it diagonalizes $u(r)D^{-1}$. 
The superpotential is also diagonalized. 
Hence, the apparently coupled transformed potential is equivalent 
to a diagonal, uncoupled, potential. 
\subsubsection{Coupled potentials with uncoupled $S$ matrices}
\label{subsec:coupled_potentials_uncoupled_S-matrices}
In this paragraph, we start from a Hamiltonian  
\begin{equation}
H_0 = \diag (h_1, h_2, \dots, h_N).
\label{cpus.1}
\end{equation}
Its entries are chosen to have the same Jost function $f_0(k)$. 
The Hamiltonians $h_i$ are different members of a phase-equivalent family. 
The Jost function of $H_0$ is a scalar matrix,
\begin{equation}
F_0(k) = f_0(k) I_N
\label{cpus.2}
\end{equation}
and so is the $S$ matrix.
This is an example of quantum system for which the $S$ matrix obeys 
a richer symmetry than the Hamiltonian. 
We now show that it is possible to transform such a Hamiltonian into a coupled one
while keeping the $S$ matrix uncoupled.

As an example \cite{pupasov:09}, consider two members of the family of potentials 
\begin{eqnarray}
v_0(r;\alpha) = -2\frac{\rmd^2}{\rmd r^2}
\ln \mathrm{W} [\sinh(\k_0r),\exp(\k_1r)+\alpha\exp(-\k_1r),\sinh(\k_2r)],
\nonumber \\ 
\k_0 < \k_1 < \k_2\,, \qquad \alpha <-1.
\label{Vb}
\end{eqnarray}
obtained from the Bargmann potential \eref{BargV2} by applying 
an additional conservative supersymmetric transformation with parameter $\kappa_2$. 
These potentials depend on a parameter $\alpha$. 
They have the same Jost and scattering matrices 
and one bound state at energy $E=-\k_1^2$. 
The additional transformation raises the singularity index to $\nu=1$. 
The corresponding two-channel diagonal potential 
\begin{equation}
V_{0}(r) = \diag\left[v_0(r;\alpha_1),v_0(r;\alpha_2)\right]
\end{equation}
has a two-fold degenerate bound state at energy $E=-\k_1^2$.
Both its Jost and scattering matrices are scalar matrices, 
\begin{equation}
F_0(k) = -(\k_1+\rmi k)[(\k_0-\rmi k)(\k_2-\rmi k)]^{-1}I_2\,.
\label{JostM0-ex2}
\end{equation}

For the coupling transformation we choose the transformation matrix 
\eref{cc.10} with matrices \eref{et2.1}, where the Jost solutions $f_0(i\k,r)$ 
of the Schrödinger equation with potential $V_0$ are used. 
To avoid a singularity at finite distance in the transformed potential, 
we impose the restriction $\kappa > \k_2$. 
The Jost matrix $F_1(k)$ given by \eref{cc.25a}
can be diagonalized by the orthogonal matrix 
$R({\textstyle \frac{\alpha}{2}})$ defined in \eref{et1.6}, 
\begin{eqnarray}
F_1(k) & = & -\left( \begin{array}{cc} 
\rmi k+ \k\cos \alpha & \k\sin \alpha \\
\k \sin\alpha & \rmi k-\k \cos\alpha
\end{array} \right)
F_0(k)
\label{tc.example-2-JostM}
\\
& = & R^T({\textstyle \frac{\alpha}{2}}) \tilde{F}_1(k) 
R({\textstyle \frac{\alpha}{2}}),
\end{eqnarray}
where 
\begin{equation}
\tilde{F}_1(k) = F_0(k) \diag(\rmi k+ \k,\rmi k-\k). 
\label{JostM1-ex2}
\end{equation}
This just corresponds to a trivial coupling in both Jost and scattering matrices. 

\begin{figure}
\begin{center}
%\begin{minipage}{15cm}
\scalebox{0.7}{{\includegraphics{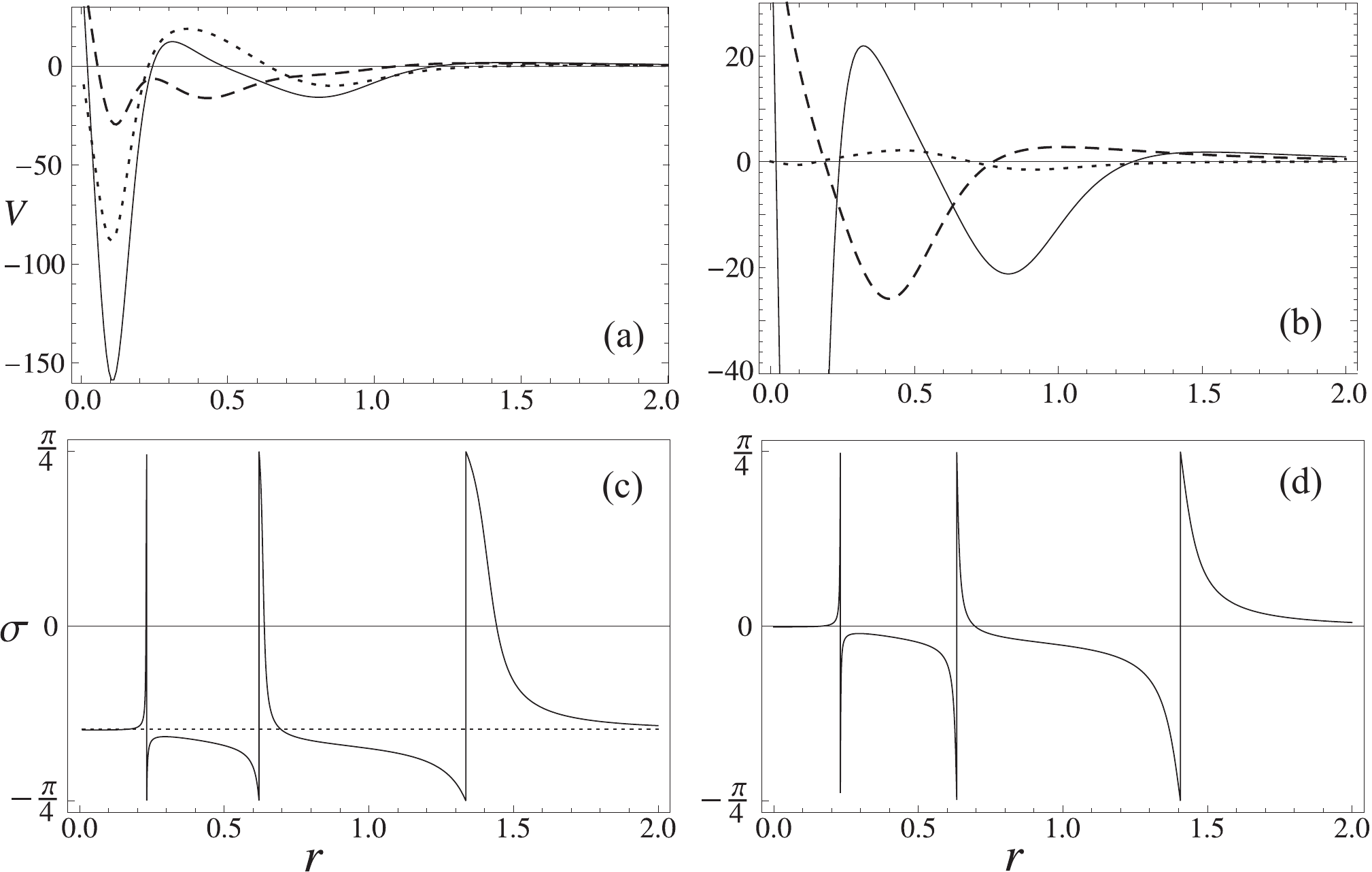}}} \\
\caption{
Example of a two-channel non-trivially coupled potential 
with diagonal Jost and scattering matrices \cite{pupasov:09}. 
(a): potential $V_1$ ($V_{1;11}$ - solid line, $V_{1;12}$ - dotted line, $V_{1;22}$ - dashed line),
(b): potential $\tilde{V}_1$, 
(c) and (d): corresponding mixing functions $\sigma(r)$ 
and  $\tilde{\sigma}(r)$. 
The potential parameters are chosen as 
$\k_0=1$, $\k_1=2.5$, $\k_2=3.5$, $\alpha_1=-2$, $\alpha_2=-1.5$,
$\k=6$, $q=0.5$, $x=25$. 
\label{fig2ntrivial-couplv-diag-fs}}
%\end{minipage}
\end{center}
\end{figure}

Figure \ref{fig2ntrivial-couplv-diag-fs}(a) shows the transformed potential $V_1$. 
In figure \ref{fig2ntrivial-couplv-diag-fs}(a), we also plot the physically 
equivalent potential $\tilde{V}_1 = R V_1R^T$ whose Jost and scattering 
matrices are diagonal. 
The simplest way to demonstrate that the potential is non-trivially coupled 
is to calculate the mixing function 
\begin{equation}
\sigma(r) = \tfrac{1}{2}\arctan [2V_{1;12}(r)/(V_{1;22}(r)-V_{1;11}(r))]\,.
\end{equation}
If $\sigma(r)$ is a constant, the potential matrix is globally diagonalizable. 
As we see from figures \ref{fig2ntrivial-couplv-diag-fs} (c) and (d), 
a non constant $\sigma(r)$ means that $V_1(r)$ and $\tilde{V}_1(r)$ 
have a non-trivial coupling. 

The scattering matrix $S_1$ has a constant mixing angle 
\begin{equation}
\epsilon(k)= -\frac{\alpha}{2}
\label{tma}
\end{equation}
and the matrix $\tilde{S}_1 = R S_1 R^T$ is diagonal.
The eigenphase shifts read 
\begin{eqnarray}
\delta_{1;1}(k) & = & 2\pi-\sum_{j=0}^2 \arctan\frac{k}{\k_j}+\arctan\frac{k}{\k}\,,
\\
\delta_{1;2}(k) & = & \pi-\sum_{j=0}^2 \arctan\frac{k}{\k_j}-\arctan\frac{k}{\k}\,.
\label{tcps}
\end{eqnarray}

The Jost and scattering matrices depend only on the factorization constants $\k_j$, $\k$, 
whereas the potential also depends on four additional parameters 
$\alpha_1$, $\alpha_2$, $q$, and $x$. 
These parameters provide a phase-equivalent deformation of the potential, 
preserving the diagonal character of $\tilde{F}_1$ and $\tilde{S}_1$. 
In the inverse-scattering theory, such deformations are related with the ANCs 
of the bound-state wave functions.

We may further relax the symmetry in \eref{cpus.2} by keeping 
the scattering matrix as the only scalar matrix.
In this case, only $S_1$ will be globally diagonalizable \cite{pupasov:09}. 
\subsubsection{Non-trivially coupled potentials, Jost and scattering matrices}
\label{subsec:vf_coupling_pjs_nontriv}
In the previous paragraph, we show examples of trivially 
coupled systems for equal partial waves. 
If the phase shifts of the initial scattering matrix differ from each other, 
the transformed system is non-trivially coupled at the level of the potential, 
Jost and scattering matrices, simultaneously. 
Reference \cite{pupasov:09} contains a number of examples in the case of the 
$S$-$S$ and $S$-$D$ partial waves
but they do not present a strong practical interest
since their scattering matrix has a complicated structure.
In contrast, an interesting property appears in the unphysical $S$-$P$ coupling \cite{pupasov:09},
for which a particular case could be found where a first-order coupling supersymmetric transformation 
modifies the mixing parameter while preserving the phase shifts.

\begin{figure}%[ht]
\begin{center}
\begin{minipage}{10cm}
\scalebox{0.9}{{\includegraphics{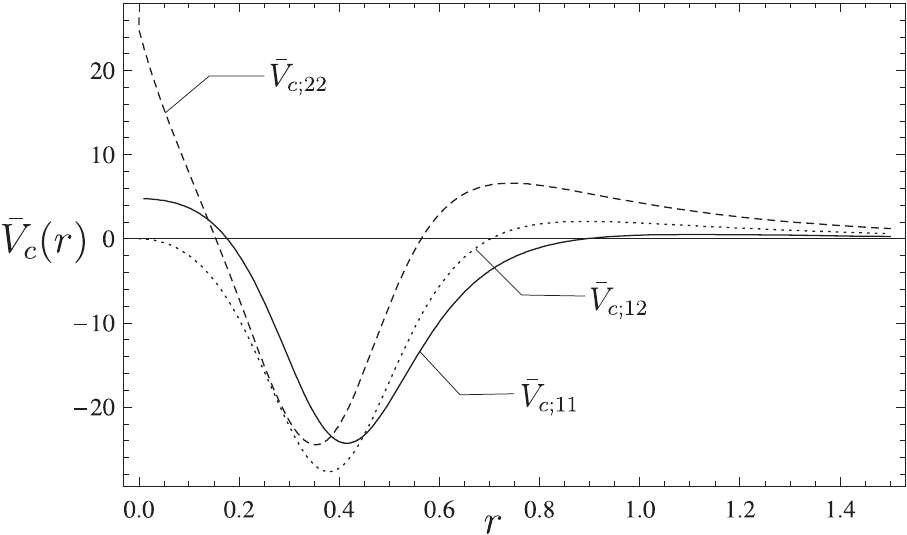}}} \\
\caption{\small Exactly solvable $S$-$P$ potential 
$\bar{V}_c=V_c-\tilde{l}(\tilde{l}+1)r^{-2}$ \cite{pupasov:09}. 
\label{figVps}}
\end{minipage}
\end{center}
\end{figure}

Figure \ref{figVps} shows an example of a coupled $S$-$P$ potential $V_1$ 
obtained from the initial diagonal potential 
\begin{equation}
V_0(r)= \diag \left( -2\frac{\rmd^2}{\rmd r^2}
\ln \mathrm{W}[\sinh \k_0 r,\sinh \k_1 r], 2r^{-2} \right).
\label{ipsp}
\end{equation}
The potential $V_0$ coincides in the $S$ channel with the Bargmann potential 
\eref{BargV2} with $\alpha=-1$. 
The potential in the $P$ channel is just the centrifugal term. 
The Jost solution corresponding to $V_0$ reads 
\begin{equation}
\fl f_0(k,r) = \diag \left[
\frac{\mathrm{W}[\sinh \k_0 r, \sinh \k_1 r, \rme^{\rmi kr}]}
{(k+\rmi\k_0)(k+\rmi\k_1) \mathrm{W}[\sinh \k_0 r, \sinh \k_1 r]}, 
\frac{(\rmi+kr)\rme^{\rmi kr}}{kr} \right]. 
\label{jssp}
\end{equation}
To define the matrix transformation solution \eref{cc.10}, 
we choose the coupling parameter $q=1$ (see \eref{et2.12} and the discussion below). 

In this case, the mixing parameter depends on the factorization constant 
of the coupling transformation $\kappa$ only and reads \cite{pupasov:09} 
\begin{equation}
\epsilon(k)=\arctan\frac{k}{\kappa}\,.
\label{mpsp}
\end{equation}
This expression is independent of the exact form 
of the initial diagonal $S$-$P$ potential. 
One can modify the diagonal potential and the corresponding phase shifts by 
methods described in section \ref{sec:sc}. 
\begin{figure}%[ht]
\begin{center}
%\begin{minipage}{10cm}
\scalebox{0.9}{{\includegraphics{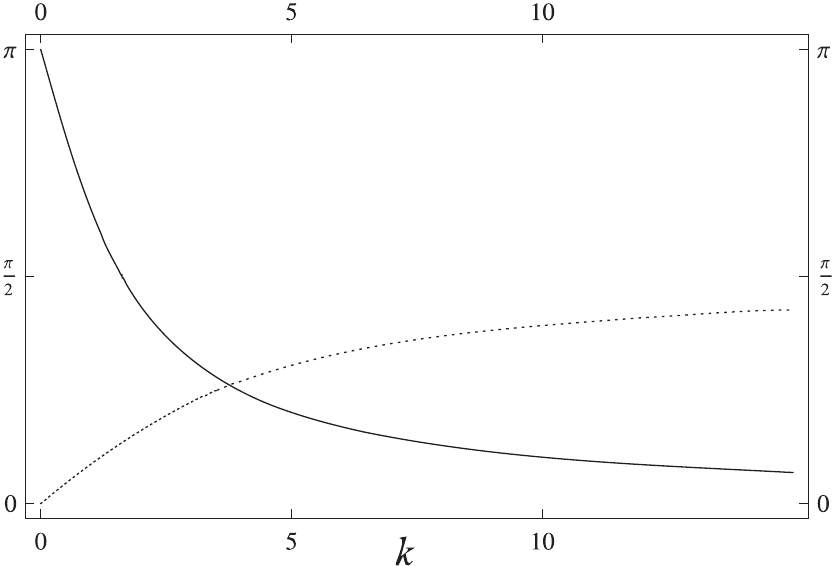}}} \\
\caption{\small 
The scattering matrix for the coupled $S$-$P$ potential \cite{pupasov:09}. 
The $S$-wave phase shift is shown as a solid line. 
The $P$-wave phase shift vanishes. 
The mixing angle $\epsilon$ is shown as a dashed line. 
\label{figSPps}}
%\end{minipage}
\end{center}
\end{figure}

Figure \ref{figSPps} shows the corresponding phase shifts and mixing angle. 
In our example, the phase shifts read before and after the transformation, 
\begin{equation}
\delta_{P}(k)=0\,,
\qquad \delta_{S}(k)=\pi-\arctan\frac{k}{\k_0}-\arctan\frac{k}{\k_1}\,,
\end{equation}
according to \eref{Bargd2}.
Parameters $\k_0$ and $\k_1$ allow one to fit the $S$-channel phase shift. 

Since the transformed eigenphase shifts coincide with the initial phase shifts, 
one can fit the phase shifts for the $S$ and $P$ waves separately 
before the coupling transformation. 
The factorization constant $\kappa$ of the coupling supersymmetric transformation 
can be used to fit the slope of the mixing angle \eref{mpsp} at zero energy. 
From the point of view of inversion,
the present example, based on a single transformation, is thus interesting.
However, it cannot be generalized to the physically interesting case of equal parities, 
where a single coupling supersymmetric transformation also modifies 
the phase shifts in a complicated way.
Hence, we now focus on particular pairs of transformations,
for which a separation of the inverse problem is also possible
but which in addition are still valid for equal-parity channels.
\subsection{Inverse problem with pairs of conjugate transformations}
\label{sec:pct}
A new type of transformation pair was introduced in \cite{pupasov:10},
that allows one to modify the coupling while keeping the eigenphase shifts unchanged
in the two-channel case.
Such pairs have a strong practical interest
and were applied to the neutron-proton system in \cite{pupasov:11}.
They were recently generalized to an arbitrary number of channels \cite{pupasov:13},
where they were shown to be applicable to an even number of channels only.
Here, we present them in more detail for the two-channel case,
stressing the particular choice of parameters for which they occur.
 
\subsubsection{Transformation of potential}
Let us define the complex wave number $\kappa = \k_R + \rmi \k_I$
from the complex energy $\cE = \cE_R + \rmi \cE_I$ through 
\begin{equation}
\cE = \kappa^2, \qquad \kappa_I > 0.
\label{et4.1}
\end{equation}
Notice that we do not introduce here a minus sign.
The transformation function is chosen as \eref{cc.26}
with $\k$ replaced by $-\rmi\k$, 
\begin{equation}
u(r)=\frac{2\k}{\rmi}\,
\varphi_0(\k,r)F_0^{-1}(\k)\left(\begin{array}{cc} 1 & 0 \\
q & 0 \end{array}\right)
+f_0(\k,r)\left(\begin{array}{cc} 0 & -q \\ 0 & 1 \end{array}\right).
\label{et4.2}
\end{equation}
It behaves near the origin as
\begin{equation}
u(r\to 0)=\left(
\begin{array}{cc} a_1r^{\nu_1+1} & b_1r^{-\nu_1} \\
a_2r^{\nu_2+1} & b_2r^{-\nu_2} \end{array}\right)[1+{\rm o}(r)]
\label{et4.5}
\end{equation}
and at infinity as
\begin{equation}\fl
u(r\to\infty) \to \left(
\begin{array}{cc} \rme^{-\rmi \k r}\left(1-\rmi \dfrac{\Lambda_{11}%
+q\Lambda_{12}}{2\k r}\right) &
-q \rme^{\rmi \k r}\left(1+\rmi \dfrac{\Lambda_{11}-%
q^{-1}\Lambda_{12}}{2\k r}\right) \\
q \rme^{-\rmi \k r}\left(1-\rmi \dfrac{\Lambda_{22}+%
q^{-1}\Lambda_{12}}{2\k r}\right) &
\rme^{\rmi \k r}\left(1+\rmi \dfrac{\Lambda_{22}-%
q\Lambda_{12}}{2\k r}\right) \end{array}\right).
\label{et4.3}
\end{equation}
The asymptotic behaviour of $u$ can be written more compactly as
\begin{equation}
u(r\to \infty)\to \left(Q-\frac{\rmi}{2\k r}\Lambda Q\sigma_z \right)%
\rme^{-\rmi \k r\sigma_z},
\label{et4.6}
\end{equation}
where $Q = I - \rmi\, q\, \sigma_y$ as in \eref{et2.4d} but $q$ is now complex.

The transformed potential is then given by \eref{W2conj} and \eref{V2conj}.
Property \eref{Wder} leads to the Wronskian asymptotics at large distances,
\begin{equation}\fl
\mathrm{W}[u,u^*](r\to\infty) \to 4\,\rmi\, \rme^{-\rmi \k r\sigma_z}
\left( p\, \k_R\,\sigma_z - q_I\k_I\sigma_x -
 \frac{\k_R \k_I}{2|\kappa|^2 r} \,\sigma_z Q^T \Lambda Q^* \sigma_z  \right)
\rme^{\rmi \k^*r\sigma_z},
\label{et4.7}
\end{equation}
with $p = \dem(1+|q|^2)$.
It can be inverted (up to $r^{-1}$) to give
\begin{eqnarray}
\mathrm{W}[u,u^*]^{-1} (r\to\infty) \to \frac{1}{4\,\rmi\, \gamma}\,
 \rme^{-\rmi\, \k^*r\,\sigma_z}
\bigg[ p\, \k_R\,\sigma_z - q_I\k_I\,\sigma_x
\nonumber \\
+ \frac{\k_R\, \k_I}{2\,\gamma\,|\kappa|^2 r}\,
(\,p\, \k_R +\rmi\, q_I\,\k_I\,\sigma_y)\, Q^T%
 \Lambda Q^* (p\, \k_R -\rmi\, q_I\,\k_I\,\sigma_y) \bigg]\,
\rme^{\rmi \k r\sigma_z}
\label{et4.8}
\end{eqnarray}
with $\gamma = (p \k_R)^2 + (q_I\k_I)^2$.
Using \eref{et2.4e}, the two-fold superpotential
\eref{et3.13conj} is given up to $r^{-1}$
after a long but straightforward calculation by
\begin{equation}\fl
W(r\to\infty) \to W_\infty
- \frac{\sin\phi}{r}\,
[\,\Lambda_- (\sin\phi \,\sigma_z + \cos\phi\, \sigma_x)
+ \Lambda_{12} (\sin\phi\, \sigma_x - \cos\phi\, \sigma_z)\,]
\label{et4.9}
\end{equation}
with
\begin{equation}
W_\infty = C_z \sigma_z - C_x \sigma_x
\label{et4.9d}
\end{equation}
and
\begin{eqnarray}
C_z = \frac{2\k_R \k_I}{\gamma} [p (1-p) \k_R + q_R q_I \k_I],
\label{et4.9a} \\
C_x = \frac{2\k_R \k_I}{\gamma} [q_I (1-p)\k_I - q_R p \k_R],
\label{et4.9b} \\
\phi = 2 \arctan \frac{2q_I \k_I}{(1+|q|^2) \k_R}.
\label{et4.9c}
\end{eqnarray}
Hence, the asymptotic form of the transformed potential \eref{V2conj} reads
\begin{equation}
V_2(r\to\infty) \to r^{-2} R^T(-\phi)\, \Lambda\, R(-\phi).
\label{et4.10}
\end{equation}
The form of this expression is identical to \eref{et2.14}
despite the fact that the angle $\phi$ has a quite different definition.
This unphysical asymptotic behaviour can thus be eliminated in the same way
with a rotation matrix which gives the potential
\begin{equation}
\tilde{V}_2 = R^T(\phi) V_2 R(\phi).
\label{et4.11}
\end{equation}

\subsubsection{Transformation of Jost matrix}

The Jost matrix and the collision matrix are obtained from \eref{mc.15}
and the asymptotic form of the Jost solution.
Let us apply $L$ to the initial Jost solution with \eref{et3.15conj},
\begin{eqnarray}
L f_0(k,r) & = & \left\{ (-k^2+\cE)\, I +
 W(r) \left[w(r)-I\frac{\rmd}{\rmd r}\right] \right\} f_0(k,r)
\label{et4.12} \\[.5em]
& \equiv & U(k,r)f_0(k,r).
\label{et4.12a}
\end{eqnarray}
The matrix $U_{\infty}(k) = \lim_{r\to\infty} U(k,r)$
determines the transformed Jost and scattering matrices.
With \eref{et4.9}, the limit of the product $W w$ is given by
\begin{equation}
W w \mathop{\rightarrow}_{r\rightarrow \infty} -\rmi \frac{\k_R \k_I \k}{\gamma}\, [(p^2 \k_R - \rmi q_I^2\k_I) I
- \rmi\, p\, q_I \k^* \sigma_y\,]
\label{et4.13}
\end{equation}
while the limit of the product $W f'_0$ is given by
\begin{equation}
W f'_0 \mathop{\rightarrow}_{r\rightarrow \infty} \rmi\, k\, W_\infty f_0.
\label{et4.13a}
\end{equation}
With \eref{et4.9c}, one obtains the expression
\begin{eqnarray}
U_{\infty}(k) = (|\cE| \cos\phi - k^2) I + \rmi\, |\cE| \sin\phi\, \sigma_y
+ \rmi\, k\, (C_z \sigma_z - C_x \sigma_x).
\label{et4.14}
\end{eqnarray}
From \eref{mc.13}, it follows that the function
\begin{equation}
f_2(k,r) = Lf_0(k,r) U_{\infty}^{-1}(k)
\label{et4.15}
\end{equation}
is the transformed Jost solution.

The regular solution of the transformed Schrödinger
equation is also transformed
according to \eref{et3.15}.
Since the first-order transformations involving $L_1$ and $L_2$ 
are conservative as shown by \eref{et4.5},
the result of the two-fold supersymmetric transformation
applied to $\varphi_0(k,r)$
can be written as \cite{pupasov:10}
\begin{equation}
L\varphi_0(k,r)=\varphi_2(k,r)\,U_0(k),
\label{et4.16}
\end{equation}
where $U_0$ is a invertible matrix that is constant with respect to $r$.
The precise value of $U_0$ is not important for the following.
The important property is that $U_0$ is an even matrix
function of wave number $k$,
\begin{equation}
U_0(k)=U_0(-k)
\label{et4.17}
\end{equation}
since $\varphi(k,r) = \varphi(-k,r)$.

Applying operator $L$ to the relation \eref{mc.15}
between the Jost solutions and the regular solution,
one obtains with \eref{et4.15} and \eref{et4.16}
\begin{equation}\fl
\varphi_2(k,r)
= \frac{\rmi}{2k}\left[\,f_2(-k,r)\,%
U_{\infty}(-k)\,F_0(k)-f_2(k,r)U_{\infty}(k)\,F_0(-k)\,\right] U_0^{-1}(k).
\label{et4.18}
\end{equation}
With \eref{et4.17}, the transformed Jost matrix thus reads
\begin{equation}
F_2(k)=U_{\infty}(-k)\,F_0(k)\,U_0^{-1}(k).
\label{et4.19}
\end{equation}
With the rotation leading to \eref{et4.11}, one obtains
\begin{equation}
\tilde{F}_2(k) = R(\phi)\, U_{\infty}(-k)\, F_0(k)\, \tilde{U}_0^{-1}(k).
\label{et4.20}
\end{equation}
where $\tilde{U}_0 = R(\phi) U_0$.

The transformation of the scattering matrix then follows
 from its definition \eref{mc.21},
\begin{equation}
\tilde{S}_2(k) = R(\phi)\, U_{\infty}(-k) S_0(k)\, U_{\infty}(k)^{-1} R^T(\phi).
\label{et4.21}
\end{equation}
Note that the transformed $S$ matrix does not depend on $U_0$
because of \eref{et4.17}.
This energy-dependent expression contains two complex parameters $\cE$ and $q$
or equivalently four real parameters $|\cE|$, $\phi$, $C_x$ and $C_z$.
In principle, they can be used to fit a given two-channel collision matrix
over some energy domain.
However, it is unlikely that this fit will be valid over a broad energy domain.
Iterating the procedure will be necessary.
Although such an iteration is technically feasible,
the fitting of parameters can be very tedious, especially
when several iterations are necessary.
A more practical method is thus welcome, where the fit is simpler
and the iteration is easily performed.
\subsubsection{Eigenphase-preserving transformations}
\label{sec:ept}
Equation \eref{et4.14} is general but not easy to use.
An important particular case does not share that drawback
\cite{pupasov:10,pupasov:11}
when $U_\infty$ is proportional to an orthogonal matrix.
This is realized when the diagonal elements of the
 $2 \times 2$ matrix are real and
equal and the off-diagonal elements are real and opposite,
hence if and only if the coefficients of $\sigma_x$ and $\sigma_z$ vanish, i.e.\
$C_x = C_z = 0$ or equivalently $q = \pm i$.
As shown below, the fact that matrix $U_\infty$ becomes proportional
 to an orthogonal matrix
has the important consequence that the eigenphases
are not modified by the transformation.

When $q = \pm i$, the two-fold superpotential
\eref{et4.9} is then simplified up to $r^{-1}$ as
\begin{equation}\fl
W(r\to\infty) \to - \frac{\sin\phi}{r}\,
 [\,\Lambda_- (\sin\phi\ \sigma_z +
 \cos\phi \,\sigma_x)
+ \Lambda_{12} (\sin\phi\, \sigma_x - \cos\phi \,\sigma_z)\,]
\label{et4a.9}
\end{equation}
where $\phi$ defined in \eref{et4.9c} is now related to the phase of
$\cE = |\cE| \rme^{\pm \rmi \phi}$.
The asymptotic form of the potential is still given by \eref{et4.10}
and is transformed with a rotation matrix as in \eref{et4.11}.

Using \eref{et2.3b} and the fact that coefficients $C_x$ and $C_z$ vanish,
one obtains for $U_{\infty}(k)$ the simple expressions
\begin{eqnarray}
U_{\infty}(k) & = & -k^2 I + |\cE| R(\phi)
\label{et4a.14a} \\[.5em]
& = & [(k^2-\cE)(k^2-\cE^*)]^{1/2} R[\pm \epsilon(k)],
\label{et4a.14}
\end{eqnarray}
where
\begin{equation}
\epsilon(k) = \arctan \frac{\cE_I}{\cE_R-k^2}.
\label{et4a.22}
\end{equation}
The fact that $U_{\infty}$ is proportional
to an orthogonal matrix is crucial to conserve the eigenphase shifts.

The transformation of the scattering matrix then follows from
its definition \eref{mc.21}
\cite{pupasov:11},
\begin{equation}
\tilde{S}_2(k) = R[\phi \mp \epsilon(k)] S_0(k) R^T[\phi \mp \epsilon(k)],
\label{et4a.21}
\end{equation}
where $\epsilon(k)$ is given by \eref{et4a.22}.
The proportionality factor $[(k^2-\cE)(k^2-\cE^*)]^{1/2}$ in
\eref{et4a.14} disappears.
The two collision matrices only differ by an orthogonal transformation.
Diagonalizing $\tilde{S}_2$ in the same way as $S_0$ in \eref{mc.22},
one sees that $S_0$ and $\tilde{S}_2$ have the same eigenvalues.
%This is the major difference with the general case or with
%the approach with a single transformation
%where the transformation matrices in \eref{et3.13} are complex.
The mixing angle of $S_2$ is given by the sum of
$\epsilon_0$, $\phi$ and $\epsilon(k)$ defined by \eref{et4a.22}.

\subsubsection{Iteration of eigenphase-preserving transformations}

The transformed potential $V_2$
can be used as a starting point for a next eigenphase-preserving transformation.
It conserves the form \eref{mc.8} at the origin
with possibly modified parameters $\nu_2$
(see \cite{pupasov:10} for a discussion).
It also conserves the form \eref{et1.2} at infinity.
This means that the two-fold supersymmetric
transformation considered above can be iterated
as long as desirable.
Iterating $P$ times such supersymmetric transformations, with
factorization energies $\cE_j = \cE_{jR} + \rmi\, \cE_{jI}$, $j=1,\ldots,P$,
one builds a chain of Hamiltonians, $H_0 \to H_2 \to \ldots \to H_{2P}$.
The final potential $V_{2P}$ corresponds to the
eigenphase-equivalent $S$ matrix
\begin{equation}
\tilde{S}_{2P}=R\,[\,\epsilon_{2P}(k)-
\epsilon_{2P}(0)\,]\, S_0\, R^T\,[\,\epsilon_{2P}(k)-\epsilon_{2P}(0)\,],
\label{et5.1}
\end{equation}
where the mixing parameter reads
\begin{equation}
\epsilon_{2P}(k) =
\epsilon_0(k)+\sum\limits_{j=1}^P\arctan \frac{\cE_{jI}}{\cE_{jR}-k^2}.
\label{et5.2}
\end{equation}
As above, to get a vanishing zero-energy mixing parameter,
we have performed a compensation rotation of angle
\begin{equation}
\theta = -\epsilon_{2P}(0) = -\sum_{j=1}^P \arctan \frac{\cE_{jI}}{\cE_{jR}}.
\label{et5.3}
\end{equation}
\subsection{Application to the neutron-proton $^3S_1$-$^3D_1$ waves}
\label{sec:eptnp}
In this section, we first apply the Crum-Krein formulas discussed in \S \ref{sec:chains}
to construct the simplest local $2\times2$ potential matrix describing the $^3S_1$-$^3D_1$
neutron-proton interaction at low energy.
Next, by using the iterative approach,
we construct a potential that fits the scattering data on the whole elastic-scattering energy range
\cite{pupasov:11}.

The scattering matrix corresponding to the simplest potential coincides, by construction,
with the scattering matrix proposed by Newton and Fulton \cite{newton:57}
and used by Kohlhoff and von Geramb \cite{kohlhoff:93}
as input of the Marchenko \cite{marchenko:55} inversion method.
We start with a diagonal matrix potential
\begin{equation}%\label{v-sd0}
V_0=\left(
   \begin{array}{cc}
   0 & 0 \\
   0 & 6/r^2
  \end{array}
   \right), \qquad r\in (0,\infty)
\end{equation}
describing two non-interacting particles in the $S$ and $D$ relative partial waves.
The Jost
\begin{equation}%\label{jost-sd}
\fl f_S(kr)={\rm e}^{\rmi kr}\,,\qquad f_D(kr)=
{\rm e}^{\rmi kr}\left(1+\frac{3\rmi}{kr}-\frac{3}{(kr)^2}\right),
\end{equation}
and regular
\begin{equation}%\label{reg-sd}
\fl
\varphi_S(kr)= \sin(kr)\,,\qquad
\varphi_D(kr)=\frac{(3-k^2 r^2) \sin(kr) - 3 k r \cos(kr)}{ k^2 r^2}\,.
\end{equation}
one-channel $S$- and $D$-wave solutions are well known for this potential.

We now realize a four-fold singular supersymmetric transformation with $m=2$, $N=4$.
The first two transformation matrices $\cU_1$ and $\cU_2$
with the structure
\begin{equation}
\cU_1=\left(
   \begin{array}{cc}
\rmi \varphi_S(\rmi\kappa_0r) & 0 \\
   0           & 1
  \end{array}
   \right),\qquad
   \cU_2=\left(
   \begin{array}{cc}
   f_S(-\rmi\kappa_1r) & 0 \\
   0           & 1
  \end{array}
   \right),
\end{equation}
provide the $S$-wave low-energy neutron-proton phase shifts
and create the bound state (see \S \ref{sec:npS}). 
The solutions entering matrices $\cU_1$ and $\cU_2$
are proportional to \eref{Bargsol0} and \eref{Bargtau0} with $\alpha=0$, respectively.
Their role is to construct the Bargmann potential \eref{Barg0}
with the two-pole scattering matrix \eref{BargS2} in the first channel ($S$-wave).

The two other transformation matrices
\begin{equation}\label{cu34}
\cU{}_3=\left(
   \begin{array}{cc}
   -\rmi P(-\kappa) \varphi_S(\kappa r)  & \rmi P(\kappa) f_S(\kappa r) \\
   \varphi_D(\kappa r)             & f_D(\kappa r)
  \end{array}
   \right),\qquad
   \cU{}_4=\cU{}_3^*\,,
\end{equation}
correspond to mutually-conjugated imaginary factorization energies ${\cal E}=\kappa^2$, with $\kappa=\chi(1+i)$,
and ${\cal E}^*$.
They introduce a coupling between channels by one eigenphase-preserving supersymmetric transformation (see \S \ref{sec:ept}). 
Solutions $\cU{}_3$, $\cU{}_4$ differ from the canonical form (with vanishing self-Wronskian)
of the eigenphase-preserving transformation solutions
by the normalization factor
\begin{equation}
P(k)=[(\kappa_0-\rmi k)(\kappa_1-\rmi k)]^{-1}\,,
\end{equation}
which is needed to take into account the single-channel transformations performed before the eigenphase-preserving transformation.
These single-channel transformations map $\cU{}_3$, $\cU{}_4$ into transformation solutions with vanishing 
self-Wronskian (see \S \ref{sec:cc1}).
This guarantees the symmetry of the transformed potential at all intermediate steps
and hence of the final potential too.

This eigenphase-preserving supersymmetric transformation results in four
$S$-matrix poles in the complex momentum plane at $\pm \kappa$ and $\pm \kappa^*$.
The mixing parameter of the resulting matrix is given by \eref{et4a.22}, with ${\cal E}_r=0$, ${\cal E}_i=2\chi^2$.
By construction, the resulting $S$ matrix thus reads
\begin{equation}\label{s-matrix-nf}
\fl
S(k)= \frac{1}{k^4+4\chi^4}\left(
\begin{array}{cc}
2\chi^2  & k^2
\\ -k^2 & 2\chi^2
\end{array}
\right) \left(
\begin{array}{cc}
 \dfrac{(k+\rmi \kappa_0)(k+\rmi \kappa_1)}{(k-i
\kappa_0)(k-\rmi \kappa_1)} & 0
\\ 0  & 1
 \end{array}
 \right)\left(
\begin{array}{cc}
2\chi^2 & -k^2
\\ k^2 & 2\chi^2
\end{array}
\right),
\end{equation}
which coincides with the one proposed in \cite{newton:57}.
The free parameters $\kappa_1$ and $\kappa_0$ control the bound-state energy and
the scattering length, as well as the effective range (see \S \ref{sec:npS}),
while parameter $\chi$ controls the coupling between channels.
These parameters are related with the six poles of the scattering matrix.

Using \eref{secSDT-WNij-def} we can present the resulting potential
through determinants constructed from transformation matrices.
To write the $8\times 8$ matrices ${\cal W}$, ${\cal W}_{i,j}$ explicitly we introduce the compact notations
$\varphi_{S,0} \equiv \rmi \varphi_S(\rmi\kappa_0r)$,
$\varphi_{S,\kappa}\equiv -\rmi P(-\kappa) \varphi_S(\kappa r)$,
$f_{S,1} \equiv f_S(-\rmi\kappa_1r)$,
$f_{S,\kappa} \equiv \rmi P(\kappa) f_S(\kappa r)$, 
$\varphi_{D,\kappa}\equiv \varphi_D(\kappa r)$,
$ f_{D,\kappa} \equiv f_D(\kappa r)$.
Then these matrices read
\begin{equation}\label{W-Kohlhoff-v-Geramb}
\fl {\cal W} =
\left(
 \begin{array}{cc:cc:cc:cc}
\varphi_{S,0} & 0 & f_{S,1} & 0 & \varphi_{S,\kappa} & f_{S,\kappa} & \varphi_{S,\kappa}^* & f_{S,\kappa}^* \\
0             & 1 & 0       & 1 & \varphi_{D,\kappa} & f_{D,\kappa} & \varphi_{D,\kappa}^* & f_{D,\kappa}^* \\
\hdashline
\partial\varphi_{S,0} & 0 &\partial f_{S,1} & 0& \partial\varphi_{S,\kappa} & \partial f_{S,\kappa} & \partial\varphi_{S,\kappa}^* & \partial f_{S,\kappa}^* \\
0             & 0 & 0       & 1 & \varphi_{D,\kappa} & f_{D,\kappa} & \varphi_{D,\kappa}^* & f_{D,\kappa}^* \\
\hdashline
\partial^2\varphi_{S,0} & 0 &\partial^2f_{S,1}  & 0&  \partial^2\varphi_{S,\kappa}  &  \partial^2f_{S,\kappa}  & \partial^2\varphi_{S,\kappa}^* & \partial^2f_{S,\kappa}^* \\
0                            &  0 & 0                  &0 & \varphi_{D,\kappa}                   &  f_{D,\kappa}             & \varphi_{D,\kappa}^* & f_{D,\kappa}^* \\%
\hdashline
\partial^3\varphi_{S,0}&  0 & \partial^3f_{S,1} &0 & \partial^3\varphi_{S,\kappa}  & \partial^3f_{S,\kappa}      & \partial^3\varphi_{S,\kappa}^* & \partial^3f_{S,\kappa}^* \\
0                           &  0 & 0                   &0 & \partial\varphi_{D,\kappa}                &\partial f_{D,\kappa}           & \partial\varphi_{D,\kappa}^* & \partial f_{D,\kappa}^*
 \end{array}
%  \begin{array}{rrrrrrrr}
% \varphi_{S,0}         & 0 & f_{S,1}         & 0&    \varphi_{S,\kappa}         &  f_{S,\kappa}         &  \varphi_{S,\kappa}^*     & f_{S,\kappa}^* \\
% 0                            & 1  & 0                  &0&  0                    &  0            & 0                      & 0 \\
% \partial\varphi_{S,0}     &  0 &\partial f_{S,1}     & 0& \partial\varphi_{S,\kappa}        & \partial f_{S,\kappa}       & \partial\varphi_{S,\kappa}^*   & \partial f_{S,\kappa}^* \\
% 0                            &  0 & 0                  & 1& 0                    &  0            & 0                      & 0 \\%
% \partial^2\varphi_{S,0} & 0 &\partial^2f_{S,1}  & 0&  \partial^2\varphi_{S,\kappa}  &  \partial^2f_{S,\kappa}  & \partial^2\varphi_{S,\kappa}^* & \partial^2f_{S,\kappa}^* \\
% 0                            &  0 & 0                  &0 & \varphi_{D,\kappa}                   &  f_{D,\kappa}             & \varphi_{D,\kappa}^* & f_{D,\kappa}^* \\%
% \partial^3\varphi_{S,0}&  0 & \partial^3f_{S,1} &0 & \partial^3\varphi_{S,\kappa}  & \partial^3f_{S,\kappa}      & \partial^3\varphi_{S,\kappa}^* & \partial^3f_{S,\kappa}^* \\
% 0                           &  0 & 0                   &0 & \partial\varphi_{D,\kappa}                &\partial f_{D,\kappa}           & \partial\varphi_{D,\kappa}^* & \partial f_{D,\kappa}^*
%  \end{array}
 \right),
\end{equation}
\begin{equation}\label{W11-Kohlhoff-v-Geramb}
\fl {\cal  W}_{1,1} =
\left(
 \begin{array}{cc:cc:cc:cc}
\ldots & \ldots & \ldots   & \ldots & \ldots &\ldots & \ldots & \ldots \\
\hdashline
\partial^4\varphi_{S,0} & 0 & \partial^4f_{S,1} & 0 & \partial^4\varphi_{S,\kappa} & \partial^4f_{S,\kappa} & \partial^4\varphi_{S,\kappa}^* & \partial^4f_{S,\kappa}^* \\
0                       & 0 & 0                 & 0 & \partial\varphi_{D,\kappa}   & \partial f_{D,\kappa}  & \partial\varphi_{D,\kappa}^*   & \partial f_{D,\kappa}^*
\end{array}
 \right),
\end{equation}
\begin{equation}\label{W12-Kohlhoff-v-Geramb}
\fl {\cal W}_{1,2} =
\left(
 \begin{array}{cc:cc:cc:cc}
\ldots & \ldots & \ldots   & \ldots & \ldots &\ldots & \ldots & \ldots \\
\hdashline
0 & 0 & 0 & 0 & \partial^2\varphi_{D,\kappa} & \partial^2f_{D,\kappa} & \partial^2\varphi_{D,\kappa}^* & \partial^2f_{D,\kappa}^* \\
0 & 0 & 0 & 0 & \partial\varphi_{D,\kappa}   & \partial f_{D,\kappa}  & \partial\varphi_{D,\kappa}^*   & \partial f_{D,\kappa}^*
 \end{array}
 \right),
\end{equation}
\begin{equation}\label{W21-Kohlhoff-v-Geramb}
\fl {\cal W}_{2,1} =
\left(
 \begin{array}{cc:cc:cc:cc}
\ldots & \ldots & \ldots   & \ldots & \ldots &\ldots & \ldots & \ldots \\
\hdashline
\partial^3\varphi_{S,0} & 0 & \partial^3f_{S,1} & 0 & \partial^3\varphi_{S,\kappa} & \partial^3f_{S,\kappa} & \partial^3\varphi_{S,\kappa}^* & \partial^3f_{S,\kappa}^* \\
\partial^4\varphi_{S,0} & 0 & \partial^4f_{S,1} & 0 & \partial^4\varphi_{S,\kappa} & \partial^4f_{S,\kappa} & \partial^4\varphi_{S,\kappa}^* & \partial^4f_{S,\kappa}^*
 \end{array}
 \right),
\end{equation}
\begin{equation}\label{W22-Kohlhoff-v-Geramb}
\fl {\cal W}_{2,2}  =
\left(
 \begin{array}{cc:cc:cc:cc}
\ldots & \ldots & \ldots   & \ldots & \ldots &\ldots & \ldots & \ldots \\
\hdashline
\partial^3\varphi_{S,0} & 0 & \partial^3f_{S,1} & 0 & \partial^3\varphi_{S,\kappa} & \partial^3f_{S,\kappa}  & \partial^3\varphi_{S,\kappa}^* & \partial^3f_{S,\kappa}^* \\
0                       & 0 & 0                 & 0 & \partial^2\varphi_{D,\kappa} & \partial^2 f_{D,\kappa} & \partial^2\varphi_{D,\kappa}^* & \partial^2 f_{D,\kappa}^*
 \end{array}
 \right),
\end{equation}
where we have used the notation $\partial$ for $\frac{\rmd}{\rmd r}$ for compactness,
where the dashed lines stress the block structure
and where the omitted blocks of the last four matrices are identical
to the corresponding blocks of the first matrix.
By linear combination of rows,
the determinants of these matrices can easily be reduced to $6\times 6$ determinants.
Moreover, more simplifications are possible since the second derivative of a factorization solution can be expressed in terms of the
factorization solution itself by using the Schrödinger equation satisfied by this solution.

Having fixed the set of matrices ${\cal W}$, ${\cal W}_{i,j}$, $i,j=1,2$,
one constructs the explicit determinant representation of the potential matrix \eref{VN}
in terms of determinants of these matrices
\begin{equation}\label{secSDT-V4}
V_{4;i,j}(r) = V_{0;i,j}(r) - 2\frac{1}{\det {\cal W}}\frac{\rmd}{\rmd r}\det {\cal W}_{i,j}
+ 2\frac{\det {\cal W}_{i,j}}{(\det {\cal W})^2}\frac{\rmd}{\rmd r} \det {\cal W}\,.
\end{equation}
It is not easy to see directly from \eref{secSDT-V4} that $V_{4;1,2}=V_{4;2,1}$.
Nevertheless,
the symmetry is provided by the equivalence between the chain of transformations
and the determinant representation;
it can also be tested numerically.
Potential \eref{secSDT-V4} gives a determinant representation
of the Kohlhoff-von Geramb potential \cite{kohlhoff:93},
obtained with a Marchenko inversion of the $S$ matrix \eref{s-matrix-nf}.

\begin{figure}[th]
\begin{center}
%\begin{minipage}{15cm}
\includegraphics{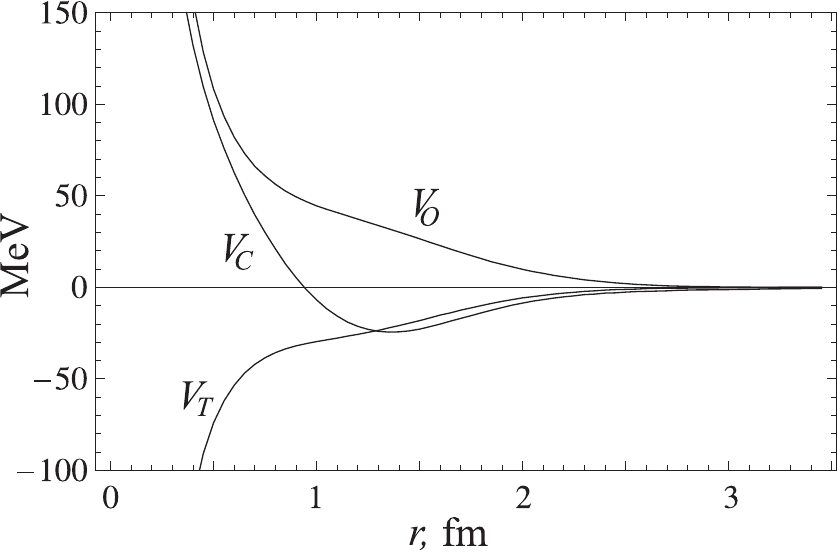}
 \caption{\small
  Exactly-solvable potential
  (central $V_C$, tensor $V_T$ and spin-orbit $V_O$ terms)
 with parameters $\kappa_0=0.944$~fm$^{-1}$, $\kappa_1=0.232$~fm$^{-1}$  and $\chi=1.22$~fm$^{-1}$
 taken from \cite{newton:57}. Reproduced from \cite{pecheritsin:11}.
\label{figSD-hoh}}
%\end{minipage}
\end{center}
\end{figure}

In this example the angle of compensation rotation \eref{et5.3} equals $\pi/2$,
therefore  $V_{4;1,1}$ and $V_{4;2,2}$ correspond to the $D$- and $S$-waves respectively.
From \eref{secSDT-V4} we extract the central, $V_C$, tensor, $V_T$, and spin-orbital, $V_O$,
components of the potential
\begin{equation}
\fl
V_C=V_{2,2}\,,\quad V_T=V_{1,2}/\sqrt{8}\,,\quad
V_O=(V_{2,2}-V_{1,2}/\sqrt{2}-V_{1,1}+6/r^2)/3\,.
\end{equation}
Choosing the same parameters $\kappa_0$, $\kappa_1$ and $\chi$
as in \cite{newton:57,kohlhoff:93},
we show these potential curves in Figure \ref{figSD-hoh}.
They agree with those shown in figure 13 of \cite{kohlhoff:93}.
Note that the potentials \eref{secSDT-V4},
as well as the potentials constructed in \cite{kohlhoff:93},
are singular at the origin and for this reason
differ significantly from the potentials obtained by Newton and Fulton  \cite{newton:57,fulton:56}.
The difference with Ref.\ \cite{newton:57}
can be explained by somewhat different input parameters.
It is well-known
that there is a family of potentials with the same scattering matrix,
when there are bound states.
In our case there is a single bound state and, hence,
there exist two additional parameters,
the ANC's $A_S$ and $A_D$,
which provide a phase-equivalent deformation
of the potential with the given scattering matrix.
%{\bf
The values of
these free parameters are fixed in both our model and the approach of
 Kohlhoff-von Geramb by the following reason.
Physically acceptable components of a matrix potential should not have
long-range tails.
Therefore the corresponding asymptotic normalization
constants can be determined  directly from the $S$-matrix residue at the
bound state pole \cite{stoks:88}.
 In particular, the ratio of the asymptotic normalization
constants reads
\begin{equation}
\eta=\frac{A_D}{A_S}=
\frac{\mathrm{Res}_{k=i \kappa_1} S_{2,1}} {\mathrm{Res}_{k=i \kappa_1} S_{1,1}}\,.
\end{equation}
For our potential \eref{secSDT-V4},
as well as for the Kohlhoff-von Geramb potential,
the parameter $\eta$ is fixed by the value
$\eta=\kappa_1^2/(2\chi^2)=0.018081$
whereas the value is different for the Newton-Fulton potential,
which has a finite value at the origin
but an unphysical slowly-decreasing asymptotic behaviour.

Let us now extend the previous results to fit the scattering matrix on the whole
elastic-scattering energy range, revisiting the results of reference \cite{pupasov:11}.
Combining the single-channel potentials \eref{np-Spot1} and 
\eref{n-p-Dwave-diag-potential} into a diagonal $S$-$D$ potential, 
we obtain the initial Hamiltonian of the neutron-proton interaction 
with realistic phase shifts.
Applying eigenphase-preserving supersymmetric transformations, 
we can fit the mixing angle. 
A three-term expansion is found sufficient to fit realistic data 
on the whole elastic-scattering region and the ratio of ANCs 
for the $S$- and $D$-wave components, $\eta=A_D/A_S$. 
This ratio is taken into account by fixing the tangent of mixing angle 
at the deuteron ground-state energy, $\tan \epsilon_1(-B_d)=-\eta$. 
Three pairs of eigenphase-preserving supersymmetric transformations 
result in the mixing angle 
\begin{equation}
\epsilon(k) = \epsilon_{1}(k)+\epsilon_{2}(k)+\epsilon_{3}(k),
\label{np-mixing}
\end{equation}
where each term corresponds to a pair of complex poles 
$\cE_1 = -0.1590 \pm \rmi\, 0.008153$, 
$\cE_2 = -4.004 \pm \rmi\, 3.107$, 
$\cE_3 = \pm \rmi\, 9.443$. 
The partial mixing angles including the compensation rotation 
of angle \eref{et5.3}, i.e.\ vanishing at $k=0$, are given by 
\begin{eqnarray}
\epsilon_{1}(k) = 
-\arctan \frac{0.008153}{0.1590+k^2}+ 0.051232\,,\\
\epsilon_{2}(k) =
\arctan \frac{3.107}{4.004+k^2}-0.65992\,,\\
\epsilon_{3}(k) =\frac{\pi}{2}-
\arctan \frac{9.443}{k^2}\,.
\label{np-mixing-terms}
\end{eqnarray}

\begin{figure}%[ht]
\begin{center}
%\begin{minipage}{10cm}
\scalebox{0.9}{{\includegraphics{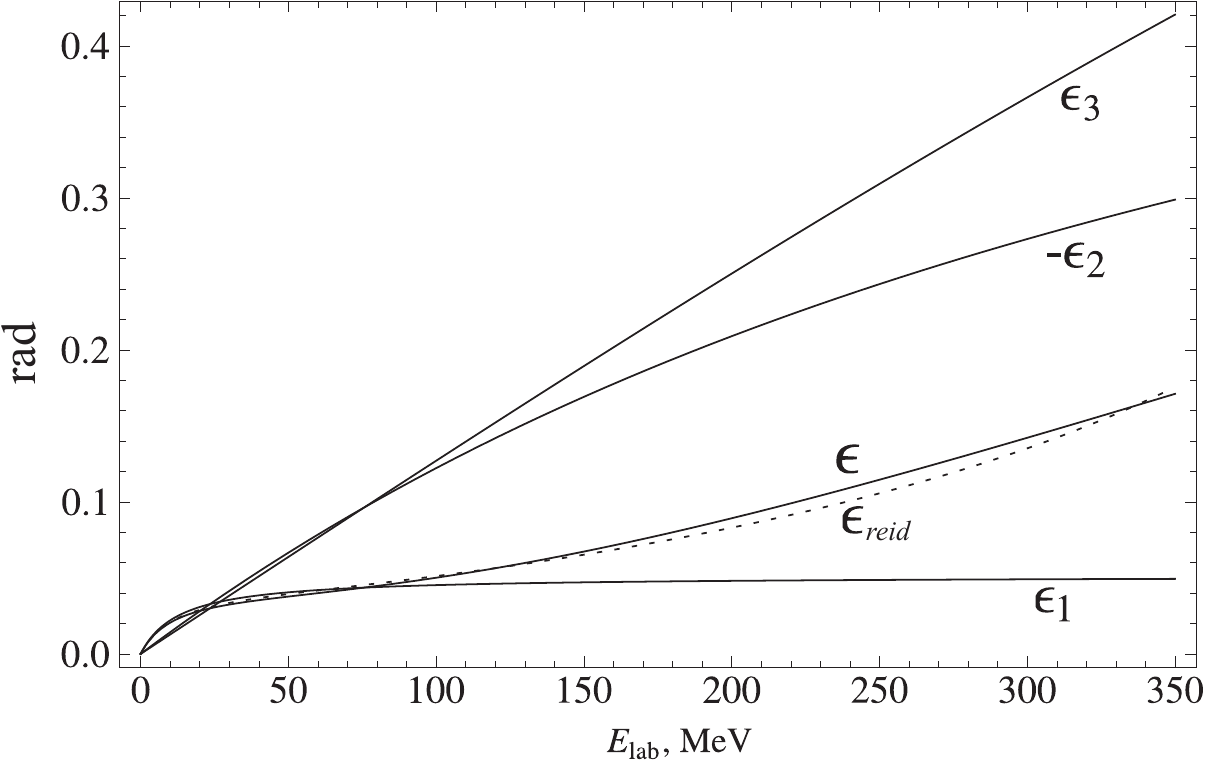}}} \\
\caption{
\small
Components $\epsilon_1$, $\epsilon_2$, $\epsilon_3$ 
of the rational fit $\epsilon$ of the mixing angle $\epsilon_{\rm Reid}$ 
(dotted line) corresponding to the Reid93 $^3S_1$-$^3D_1$ potential \cite{stoks:94} 
as a function of the laboratory energy $E_{\rm lab} = 2E$ \cite{pupasov:11}. 
\label{fig-eps-np}}
%\end{minipage}
\end{center}
\end{figure}

Figure \ref{fig-eps-np} shows the Reid93 mixing parameter 
$\epsilon_{\rm Reid}$, our fit $\epsilon$ 
and contributions from each pair of 
eigenphase-preserving supersymmetric transformations. 
The single eigenphase-preserving transformations with $\cE_1$ 
gives a very accurate fit almost up to 100 MeV. 
Below this energy, the contributions from $\cE_2$ and $\cE_3$ compensate each other. 
They become important at higher energies to reproduce increasing values 
of the mixing angle. 

As shown in \cite{pupasov:11}, the corresponding potential reproduces 
the Reid93 potential and displays a one-pion exchange asymptotic behaviour. 
This example illustrates the effectiveness of 
the iterative supersymmetric inversion approach. 
%
%%%%%%%%%%%%%%%%%%%%%%%%%%%%%%%%%%%%%%%%%%%%%%%%%%%%%%%%%%%%%%%%%%%%%%%%%%%%%%%%%%%%%%

\section{Two-channel inverse problems with different thresholds}
\label{sec:ipwt}

As mentioned above, in the case of threshold differences,
we could not find a way to build coupled potentials
with a non trivial coupling
using conservative transformations only.
Non-conservative transformations, on the other hand, prove very
useful with this respect:
in \cite{sparenberg:06},
they were shown to be able to simplify and generalize the construction
of the exactly-solvable coupled-channel potential obtained by Cox in 1964
by resolution of the inverse problem integral equations \cite{cox:64}.

This potential generalizes the Bargmann potential to the coupled-channel case,
with a Jost function and scattering matrix expressed as first-order rational functions of the wave number.
In reference \cite{pupasov:08},
a detailed study of the spectral properties of this potential was made
for an arbitrary number $N$ of channels.
In particular, a geometrical analysis of the $N 2^{N-1}$ zeros of the Jost-matrix determinant
shows that the number of bound states $n_b$ satisfies $0\le n_b \le N$
and that the maximal number of resonances is $n_r=(N-1) 2^{N-2}$,
the other zeros corresponding to virtual states.
In particular, for two channels, this potential may have 0, 1 or 2 bound states
(contrary to what is stated in \cite{cox:64}),
up to 4 virtual states and 0 or 1 resonance,
depending on the values of its parameters.

In reference \cite{pupasov:08},
a perturbative method is also proposed to calculate the position of these poles in a weakly-coupled case,
for arbitrary $N$.
We do not reproduce these general results here and rather concentrate on the two-channel case,
revisiting the results obtained in \cite{sparenberg:06}
and emphasizing the particular role of non-conservative transformations.
In addition to the potential,
we provide detailed expressions for the scattering matrix
and scattering wave functions,
hence building an ideal test case for checking coupled-channel numerical methods.

\subsection{Derivation of generalized two-channel Cox potentials}

We apply a single non-conservative supersymmetric transformation to the zero potential.
Two useful alternative expressions for the transformation function read
\begin{equation}
u(r)=\cosh(\kappa r) I + \sinh(\kappa r) \kappa^{-1} w(0)
=\exp(\kappa r) C + \exp(-\kappa r) D, 
\label{sigexp}
\end{equation}
with
\begin{equation}
w(0)= \left(
\begin{array}{cc}
\alpha_1 & \beta \\[.5em] \beta & \alpha_2
\end{array}
\right), \quad
C, D= {\textstyle \frac{1}{2}}\left(
\begin{array}{cc}
1\pm\frac{\alpha_1}{\kappa_1} & \pm\frac{\beta}{\kappa_1} \\[.5em]
\pm\frac{\beta}{\kappa_2} & 1\pm\frac{\alpha_2}{\kappa_2}
\end{array}
\right)
\end{equation}
where the plus sign refers to $C$ and the minus sign to $D$.

Comparing \eref{sigexp} with the equivalent expression \eref{Bargsig0} for the single-channel case
shows that this transformation solution can also be seen as resulting from the application
of a first-order differential operator on the regular solution $\sinh(\kappa r)$ of the vanishing potential.
However, in the coupled-channel case
this operator does not display the standard structure \eref{cc.4} of conservative supersymmetric transformations:
the zeroth-order term is a multiplication by a matrix $w(0)$ on the right
and the square of this matrix is not related to a factorization energy.
Nevertheless, this simple underlying structure suggests to explore more general
conservative supersymmetric transformations,
a study we defer to a future work.

The determinant of the factorization solution is necessary to invert it and
to build the superpotential,
the asymptotic value of which also determines the Jost and scattering matrices.
This determinant reads
\begin{eqnarray}
 \det u(r) & = & \textstyle \cosh (\kappa_1 r) \cosh (\kappa_2 r)
+ \frac{\alpha_1\alpha_2-\beta^2}{\kappa_1\kappa_2}
\sinh (\kappa_1 r) \sinh (\kappa_2 r) \nonumber \\[.5em]
& & + \textstyle \frac{\alpha_1}{\kappa_1} \sinh (\kappa_1 r) \cosh (\kappa_2 r)
+ \frac{\alpha_2}{\kappa_2} \cosh (\kappa_1 r)%
 \sinh (\kappa_2 r) 
\label{detsig} 
\end{eqnarray}
with $\det u(0)=1$.
It behaves asymptotically like
\begin{equation}
 \det u(r) \mathop{\to}_{r\to \infty}
 \textstyle \det C \ \rme^{(\kappa_1+\kappa_2) r}
+ \frac{1}{4}\left(1-\frac{\alpha_1}{\kappa_1}+
\frac{\alpha_2}{\kappa_2}
- \frac{\alpha_1\alpha_2-\beta^2}{\kappa_1\kappa_2}\right)
\rme^{(\kappa_2-\kappa_1) r}.
\label{detsigas}
\end{equation}

Equation \eref{detsigas} shows that two cases have to be distinguished,
depending on whether $\det C=0$ or not.
When $\det C \neq 0$, only the $C$ term in equation
\eref{sigexp} determines the asymptotic behaviour of the superpotential,
which reads $w_\infty=\kappa$,
in agreement with \eref{cc.14bis}.
The potential built in \cite{cox:64} (for $q=1$)
corresponds to this case.
From \eref{cc.25b} with  $F_0 = I$ and $G_0 = \rmi k$, its Jost matrix reads 
\begin{equation}
 F_1(k)=[w_\infty-\rmi k]^{-1}[w(0)-\rmi k]
=\left(
\begin{array}{cc}
\frac{\alpha_1-\rmi  k_1}{\kappa_1-\rmi  k_1}  &
\frac{\beta}{\kappa_1-\rmi  k_1}  \\[.5em]
\frac{\beta}{\kappa_2 -\rmi  k_2}  &
\frac{\alpha_2-\rmi  k_2}{\kappa_2-\rmi  k_2}
\end{array}
\right) \quad (\det C \neq 0)
\label{F1Coxbis}
\end{equation}
and depends on five real parameters:
the threshold energy, the factorization energy and the three parameters entering
the real symmetric matrix $w(0)$.
When $\det C=0$, i.e.\
\begin{equation}
\beta = \pm \sqrt{(\kappa_1 + \alpha_1)(\kappa_2 + \alpha_2)}\,,
\end{equation}
one has first to notice that the second term of \eref{detsigas} cannot vanish;
otherwise, the
(super)potential becomes trivial ($\alpha_2=-\kappa_2$, $\beta=0$).
An explicit calculation of the superpotential shows that
$w_\infty=\diag (-\kappa_1,\kappa_2)$ in this case,
in agreement with \eref{cc.14bis},
hence leading to the Jost matrix
\begin{equation}\fl
F_1(k)=[w_\infty-\rmi k]^{-1}[w(0)-\rmi k]
=\left(
\begin{array}{cc}
\frac{\alpha_1- \rmi  k_1}{-\kappa_1 - \rmi  k_1}  &
\frac{\beta}{-\kappa_1 - \rmi  k_1}  \\[.5em]
\frac{\beta}{\kappa_2 - \rmi  k_2}  &
\frac{\alpha_2 - \rmi  k_2}{\kappa_2 - \rmi  k_2}
\end{array}
\right) \quad (\det C = 0)\,.
\label{F1Cox}
\end{equation}
In the following, as in \cite{sparenberg:06}, we concentrate on this case,
which is not covered by the original Cox paper;
however, the obtained expressions can be directly extended
 to Cox' potential by simply changing the sign of $\kappa_1$
(which reveals an error in equation (5.3) of \cite{cox:64}).

\subsection{Scattering-matrix properties}

Let us first establish the explicit expression of the
$2\times 2$ scattering matrix for $\det C = 0$.
Its diagonal elements and its determinant can be directly expressed
in terms of the determinant of the Jost function \cite{newton:82},
which reads
\begin{equation}
{\cal F}_1(k_1,k_2)\equiv\det F_1(k)=
\frac{(k_1+\rmi \alpha_1)(k_2+\rmi \alpha_2)+%
\beta^2}{({\kappa_1+\rmi  k_1})(\kappa_2-\rmi  k_2)}.
\label{det1}
\end{equation}
Its off-diagonal element is explicitly calculated using \eref{mc.21}.
The symmetric $S$ matrix reads
\begin{equation}
S_1(k_1,k_2)= \frac{1}{{\cal F}_1(k_1,k_2)}
\left(\begin{array}{cc}
{\cal F}_1(-k_1,k_2) & \dfrac{-2\rmi \beta\sqrt{k_1k_2}}{k_1^2+\kappa_1^2}
\\[.25em]
\dfrac{-2\rmi \beta\sqrt{k_1k_2}}{k_2^2+\kappa_2^2} & {\cal F}_1(k_1,-k_2)
\end{array}
\right)
\label{sm}
\end{equation}
and it can be checked that
\begin{equation}
 \det S_1(k_1,k_2)= \frac{{\cal F}_1(-k_1,-k_2)}{{\cal F}_1(k_1,k_2)}.
\label{detS}
\end{equation}
The expression \eref{mc.21} of the scattering matrix in terms
of the Jost matrix,
combined with the expression of the Jost matrix \eref{F1Cox}
for the Cox potential,
in particular the diagonal character of $w_\infty-\rmi k$,
suggests that the Blatt-Biedenharn decomposition \eref{mc.22}
of the scattering matrix
is not the most convenient here.
We rather use the Stapp parametrization \cite{stapp:1957},
which reads
\begin{eqnarray}
 S & = & \left( \begin{array}{cc}
           \rme^{\rmi \bar\delta_1} & 0 \\
           0 & \rme^{\rmi \bar\delta_2}
          \end{array} \right)
\left( \begin{array}{cc}
           \cos 2\bar\epsilon & \rmi  \sin2\bar\epsilon \\[.5em]
           \rmi  \sin2\bar\epsilon & \cos 2\bar\epsilon
          \end{array} \right)
\left( \begin{array}{cc}
           \rme^{\rmi \bar\delta_1} & 0 \\[.5em]
           0 & \rme^{\rmi \bar\delta_2}
          \end{array} \right)
\label{Stapp1} \\[.5em]
& = & \rme^{\rmi (\bar\delta_1+\bar\delta_2)}
\left( \begin{array}{cc}
           \rme^{\rmi (\bar\delta_1-\bar\delta_2)} %
           \cos 2\bar\epsilon & \rmi  \sin2\bar\epsilon \\[.5em]
           \rmi  \sin2\bar\epsilon & \rme^{\rmi (\bar\delta_2-\bar\delta_1)}%
            \cos 2\bar\epsilon
          \end{array} \right),
\label{Stapp2}
\end{eqnarray}
where $\bar\delta_1$, $\bar\delta_2$ and $\bar\epsilon$ are known as the bar phase shifts and mixing parameter.

When the second channel is closed, i.e., for energies $0\le E \le \Delta$,
the physical scattering matrix is just the first diagonal
element of $S$ matrix~\eref{sm},
with $k_1$ real positive and $k_2$ imaginary positive.
It reads 
\begin{equation}
S_{1;11}(k_1,\rmi |k_2|)=\frac{\kappa_1+\rmi  k_1}{\kappa_1-\rmi  k_1}
\frac{[(\alpha_1+\rmi  k_1)(|k_2|+\alpha_2)+\beta^2]}
{[(\alpha_1-\rmi  k_1)(|k_2|+\alpha_2)+\beta^2]}
= \rme^{2\rmi  \delta_1} = \rme^{2\rmi  \bar\delta_1},
\label{sf}
\end{equation}
with $0\le k_1\le \sqrt{\Delta}$ and $k_2=\rmi \sqrt{\Delta-k_1^2}$.
The one-channel phase shift,
which has the same value in both parametrizations,
has the rather simple expression 
\begin{eqnarray}
\delta_1(k_1)& = & \bar\delta_1(k_1)
=\arctan \frac{k_1}{\kappa_1} +
\arctan \frac{k_1(|k_2|+\alpha_2)}{\alpha_1 |k_2|+\alpha_1\alpha_2-\beta^2}
\label{d1} \\
& \equiv & \arctan \frac{k_1}{\kappa_1} + \xi_0(k_1),
\label{d0}
\end{eqnarray}
where $\xi_0$ is introduced for further use.

Above threshold, the sum of the two eigenphase shifts also
 has a rather simple expression,
identical in both parametrizations.
It can be established from \eref{detS},
with
\begin{equation}
\det S_1=\rme^{2\rmi (\delta_1+\delta_2)}=
\rme^{2\rmi (\bar\delta_1+\bar\delta_2)}
\end{equation}
the sum reads 
\begin{equation}
 \delta_1+\delta_2 = \bar\delta_1+\bar\delta_2
= \arctan \frac{k_1}{\kappa_1}-\arctan\frac{k_2}{\kappa_2}
-\arctan\frac{\alpha_2 k_1 + \alpha_1 k_2}{k_1 k_2+\beta^2-\alpha_1 \alpha_2}.
\label{d1+d2}
\end{equation}
At threshold ($k_2=0$), this expression agrees with \eref{d1}.
In the Stapp parametrization (and not in the Blatt-Biedenharn parametrization),
the difference of eigenphase shifts also has a simple expression, 
\begin{equation}
 \bar\delta_1-\bar\delta_2
= \arctan \frac{k_1}{\kappa_1}+\arctan\frac{k_2}{\kappa_2}
-\arctan\frac{\alpha_2 k_1- \alpha_1 k_2}{-k_1 k_2+\beta^2-\alpha_1 \alpha_2},
\label{d1-d2}
\end{equation}
which happens to be identical to expression
\eref{d1+d2} after a change of sign for $k_2$.
Another advantage of the Stapp parametrization
is that the bar mixing parameter only depends on $w(0)$, 
\begin{equation}
 \tan 2\,\bar\epsilon=
%{\textstyle\frac{1}{2}} \arctan
2\,\beta \sqrt{\frac{k_1 k_2}{(k_1 k_2+\alpha_1 \alpha_2%
 -\beta^2)^2+(\alpha_2 k_1 - \alpha_1 k_2)^2}},
\end{equation}
while in the Blatt-Biedenharn parametrization,
the mixing parameter also depends on the factorization energy,
i.e.\ on $\kappa_1$ and $\kappa_2$, as 
\begin{eqnarray}
 \tan 2\,\epsilon & = & \frac{2 S_{1;12}}{S_{1;11}-S_{1;22}} =
 \frac{\tan 2\,\bar\epsilon}{\sin (\bar\delta_1-\bar\delta_2)}.
% \\ & = & \frac{2\beta\sqrt{k_1 k_2}}
%{\sin\left(\arctan \frac{k_1}{\kappa_1}+\arctan\frac{k_2}{\kappa_2}\right)
% (k_1 k_2+\alpha_1 \alpha_2 -\beta^2) +
%\cos \left( \arctan \frac{k_1}{\kappa_1}+\arctan\frac{k_2}{\kappa_2}\right)
% (\alpha_2 k_1 - \alpha_1 k_2)}.
\end{eqnarray}
This implies that the Stapp parametrization
is more convenient from an inverse-scattering-problem perspective:
instead of fitting all parameters on the full scattering matrix at once,
$\bar\epsilon$ can be fitted first, using $\alpha_1$,
$\alpha_2$ and $\beta$ only,
then $\bar\delta_1$ and $\bar\delta_2$ can be fitted with
the factorization energy.

If the nonconservative transformation is followed by a conservative transformation with factorization energy $\cal{E}$,
the scattering matrix is modified according to the formal relations established in \cite{amado:88b},
\begin{equation}
 S_2 = (ik+w_\infty)^{-1}S_1(-ik+w_\infty)={\rm diag}(\rme^{\rmi \phi_1}, \rme^{\rmi \phi_2})
 S_1{\rm diag}(\rme^{\rmi \phi_1}, \rme^{\rmi \phi_2})\,,
\end{equation}
where $ik+w_\infty =(E-{\cal E}) {\rm diag}(\rme^{-\rmi \phi_1}, \rme^{-\rmi \phi_2})$
is a diagonal matrix and $\phi_{1,2}$ are phases given by arctangents. 
This implies that the conservative transformation only modifies
the Stapp eigenphase shifts with simple arctangent terms
without affecting the mixing parameter $\bar\epsilon$.
The conservative transformations can then be iterated to fit the bar phase shifts with a sum of arctangents,
as in the single-channel inverse problem.
Combining a few non-conservative transformations 
(to fit the bar mixing parameter) with a chain of conservative transformations 
(to fit the bar phase shifts) could thus provide a full solution to the coupled-channel 
inverse problem with threshold.
However, a difficulty might occur as the second non-conservative transformation 
might mix up all the parameters of the first one.
Therefore chains of non-conservative transformations require a specific study.
\subsection{Explicit expressions for the potential and its solutions}
Let us now construct the superpotential and the potential
corresponding to this scattering matrix.
The general expression for the diagonal elements of $w(r)$ is 
\begin{eqnarray}
 w_{11}(r) & = & \frac{\kappa_1}{\det u(r)}
\left[\textstyle \sinh (\kappa_1 r) \cosh (\kappa_2 r)
+ \frac{\alpha_1\alpha_2-\beta^2}{\kappa_1\kappa_2} \cosh (\kappa_1 r) \sinh (\kappa_2 r) \nonumber\right. \\
& & + \left. \textstyle \frac{\alpha_1}{\kappa_1}%
 \cosh (\kappa_1 r) \cosh (\kappa_2 r)
+ \frac{\alpha_2}{\kappa_2} \sinh (\kappa_1 r) \sinh (\kappa_2 r) \right],
\label{w11}
\end{eqnarray}
with the same expression for $w_{22}(r)$ with subscripts 1 and 2 exchanged. 
The off-diagonal term has the simpler expression 
\begin{equation}
 w_{12}(r) = \frac{\beta}{\det u(r)}.
\end{equation}
Equations \eref{detsig} and \eref{w11} simplify when 
\begin{equation}
\frac{\alpha_1\alpha_2-\beta^2}{\kappa_1 \kappa_2} = \pm 1,
\qquad
\frac{\alpha_1}{\kappa_1} = \pm \frac{\alpha_2}{\kappa_2},
\end{equation}
where the two signs can be chosen arbitrarily and independently from each other.
Choosing twice a minus sign is equivalent
to choosing det $C$ = det $D$ = 0, as in \cite{sparenberg:06},
and implies that 
$\kappa_1 \kappa_2 > \beta^2 >0$ and $\left|\alpha_1/\kappa_1\right|<1$.
In this case, the superpotential can be written in a very compact form 
\begin{equation}
w(r)= \frac{1}{\cosh y}\left(
\begin{array}{cc}
-\kappa_1 \sinh y & \sqrt{\kappa_1\kappa_2} \\[.5em]
\sqrt{\kappa_1\kappa_2} & \kappa_2 \sinh y
\end{array}
\right),
\label{w}
\end{equation}
with 
\begin{equation}
y=(\kappa_2-\kappa_1)r-\mathrm{arcosh} \sqrt{\frac{\kappa_1\kappa_2}{\beta^2}}
= (\kappa_2-\kappa_1)r - \mathrm{arctanh} \frac{\alpha_1}{\kappa_1}.
\end{equation}
The potential then reads
\begin{equation}
V_1(r)= \frac{2(\kappa_2-\kappa_1)}{\cosh^2 y}\left(
\begin{array}{cc}
\kappa_1  & \sqrt{\kappa_1\kappa_2}\sinh y \\[.5em]
\sqrt{\kappa_1\kappa_2}\sinh y & -\kappa_2
\end{array}
\right).
\end{equation}

To calculate the regular solution, we find a solution $\psi_0$ of
the Schrödinger equation with the zero potential which transforms into the
regular solution $\varphi_1$ of the transformed equation 
\begin{eqnarray}
\psi_0(k,r)& \equiv & L^\dagger \varphi_1(k,r) \\
& = & \frac{1}{2\,\rmi }%
\left[\,L^\dagger f_1(k,r)k^{-1}F_1(-k)-L^\dagger f_1(-k,r)k^{-1}F_1(k)
\,\right].
\end{eqnarray}
Indeed, due to the simple expressions 
of the Jost solution \eref{cc.15} and of the Jost function \eref{F1Coxbis}, \eref{F1Cox} for the $V_1$ potential, 
this equation reads 
\begin{eqnarray}
2\rmi \psi_0(k,r) & = &
L^\dagger Lf_0(k,r)[w_\infty-\rmi k]^{-1}k^{-1}[w_\infty+%
\rmi k]^{-1}[w(0)+\rmi k] \nonumber \\[.5em]
&& - L^\dagger Lf_0(-k,r)[w_\infty+\rmi k]^{-1}k^{-1}[w_\infty-%
\rmi k]^{-1}[w(0)-\rmi k] \\[.5em]
& = & \rme^{\rmi kr}k^{-1}[w(0)+\rmi k]-\rme^{-\rmi kr}k^{-1}[w(0)-\rmi k]\,,
\end{eqnarray}
where the last expression has been obtained by using the factorization property 
\begin{eqnarray}
L^\dagger L=H_0+\kappa^2\,,
\end{eqnarray}
with
\begin{eqnarray}
H_0=-I \frac{\rmd^2}{\rmd r^2}\,,\quad
f_0(k,r)=\rme^{\rmi kr}
\end{eqnarray}
for the vanishing initial potential,
together with the value of $w_\infty$.
Finally, we get 
\begin{equation}
\label{psi0}
\psi_0(k,r)=\cos (kr) +\sin(kr) k^{-1} w(0) \equiv
\{\vec\psi_{0;1},\vec\psi_{0;2}\},
%=\left(
% \begin{array}{cc}
% \cos k_1 r + \frac{\alpha_1}{k_1} \sin k_1 r  &
%\frac{\beta}{k_1} \sin k_1 r \\[.5em]
% \frac{\beta}{k_2} \sin k_2 r & \cos k_2 r + \frac{\alpha_2}{k_2} \sin k_2 r
% \end{array}
% \right).
\end{equation}
where $\vec\psi_{0;1}$ and $\vec\psi_{0;2}$ are vector solutions that
do not vanish at the origin.
Like the factorization function \eref{sigexp},
this matrix solution can be seen as resulting from the transformation of the regular matrix solution $\sin(kr)$
of the vanishing potential
by a first-order differential operator containing a matrix multiplication by $w(0)$ from the right.

Using \eref{psi0}, the regular solution of the transformed equation may be
calculated as 
\begin{equation}
\varphi_1(k,r) = L\psi_0(k,r) 
= -\psi_0'(r)+w(r)\psi_0(r)\equiv \{\vec\varphi_{1;1},\vec\varphi_{1;2}\}.
\end{equation}
For the particular superpotential \eref{w}, the vectors read 
\begin{equation}
\vec\varphi_{1;1} = \left( 
\begin{array}{c}
\varphi_{1;11} \\ \varphi_{1;12}
\end{array}
\right), 
\quad
\vec\varphi_{1;2} = \left(
\begin{array}{c}
\varphi_{1;21}
\\
\varphi_{1;22}
\end{array}
\right),
\label{phi11}
\end{equation}
where 
\begin{eqnarray}
\varphi_{1;11} = 
&-&\alpha_1 \cos (k_1 r) + k_1 \sin (k_1 r) + \frac{\beta
\sqrt{\kappa_1\kappa_2}\sin (k_2 r)}{k_2 \cosh y}
\nonumber \\[.5em]
&-&\left(\kappa_1 \cos (k_1 r) + \frac{\alpha_1 \kappa_1}{k_1}%
 \sin (k_1 r)\right) \tanh y, 
\label{2ncomp11}
\end{eqnarray}
\begin{equation}
\fl
\varphi_{1;12}=
\frac{\sqrt{\kappa_1\kappa_2}}{\cosh y}
\left(\cos (k_1 r) +
\frac{\alpha_1}{k_1} \sin (k_1 r)\right)-\beta \left(\cos (k_2 r) - %
\frac{\kappa_2}{k_2} \tanh y \sin (k_2 r)\right)
\label{2ncomp12}
\end{equation}
with symmetrical expressions for $\varphi_{1;21}$ and $\varphi_{1;22}$
obtained by exchanging the roles of subscripts 1 and 2
(leaving the definition of $y$ unaffected) and
by changing the sign of the $\tanh y$ terms in \eref{2ncomp12} and \eref{2ncomp11} respectively.
%
% \begin{equation}
% \fl
% \varphi_{1;21}=
% \frac{\sqrt{\kappa_1\kappa_2}}{\cosh y} \left(\cos (k_2 r) +
% \frac{\alpha_2}{k_2} \sin (k_2 r)\right)-\beta \left(\cos (k_1 r) +
% \frac{\kappa_1}{k_1} \tanh y \sin (k_1 r)\right)
% \label{2ncomp21}
% \end{equation}
% %
% \begin{eqnarray}
% \nonumber
% \varphi_{1;22}=
% &-&\alpha_2 \cos (k_2 r)  + k_2 \sin (k_2 r)+ \frac{\beta
% \sqrt{\kappa_1\kappa_2}\sin (k_1 r)}{k_1 \cosh y}
% \\[.5em]
% &+&\left(\kappa_2 \cos (k_2 r) +
% \frac{\alpha_2 \kappa_2}{k_2} \sin (k_2 r)\right)\tanh y
% \label{2ncomp22}
% \end{eqnarray}
%
Using the values 
$\cosh y(0)=\sqrt{\kappa_1 \kappa_2}/\beta$ and $\tanh y(0)=\alpha_2/\kappa_2$,
one checks that these vectors vanish at the origin, as they should.
The difference of behaviour at the origin for $\psi_0$
and $\varphi_1$ illustrates the non-conservative character
of the supersymmetric transformation.

\begin{figure}
\begin{center}
\scalebox{1}{\includegraphics{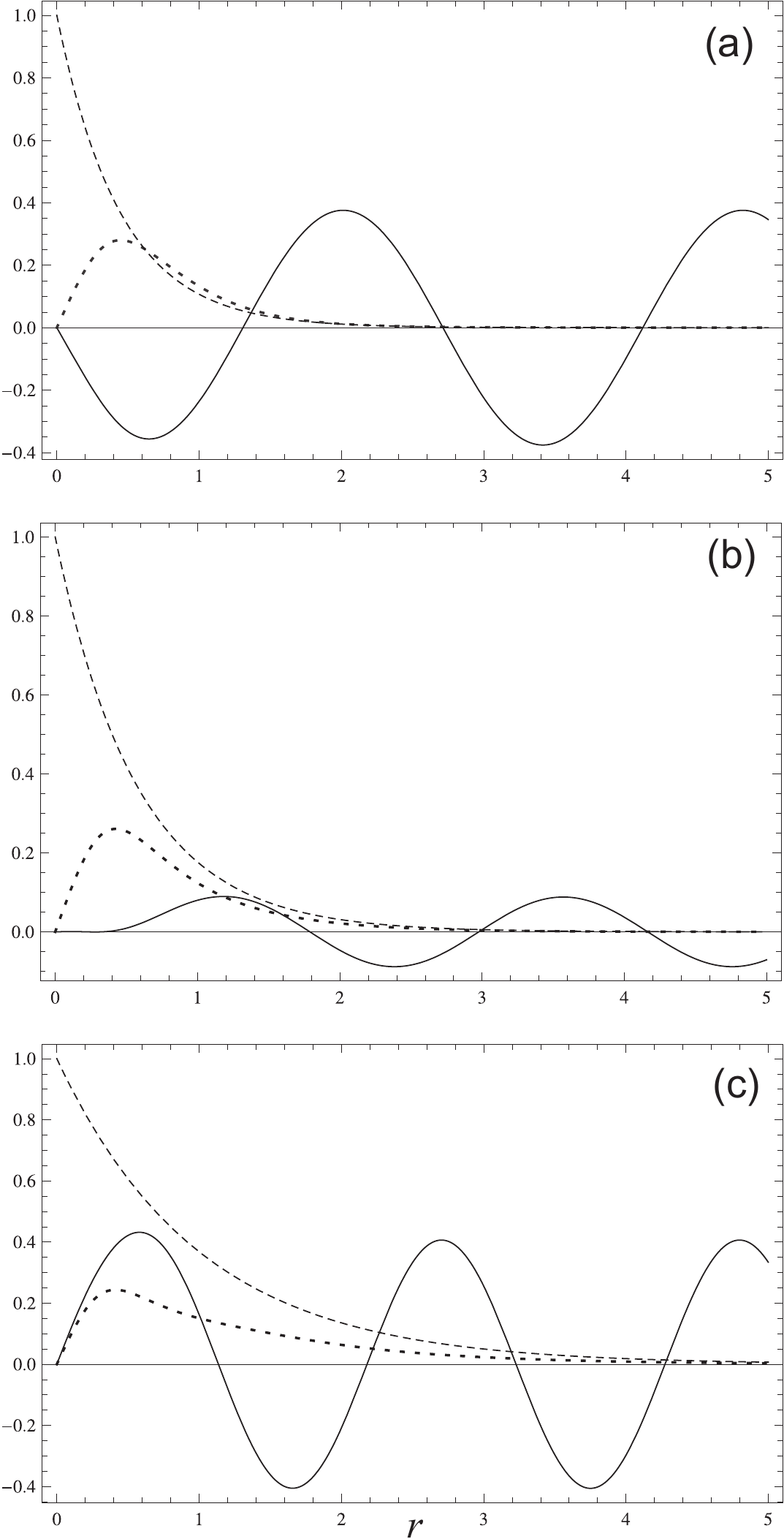}} 
\caption{Examples of solutions \eref{mixsol} (normalized as $\varphi_{1;2}(r\rightarrow0)\sim r$)
when the second channel is closed.
The component $\varphi_{1;1}(r)$ is shown by solid lines,
$\varphi_{1;2}(r)$ is shown by dotted lines.
Besides, $\exp(-|k_2|r)$ is plotted with dashed lines for comparison.
The potential parameters are those chosen in figures 1 and 2 of
\cite{sparenberg:06}
($\Delta=10$, $E_\mathrm{R}=7$, $\Gamma=1$, which correspond
to $\kappa_1\approx0.17207$, $\kappa_2\approx3.16696$,
$\beta\approx0.61710$, $\alpha_1\approx0.094431$, $\alpha_2\approx-1.73801$)
and the energies are 5, 7, 9 (respectively below (a),
at (b) and above (c) the resonance energy).
\label{figvecpsi}}
\end{center}
\end{figure}

Let us now consider the physical solution below the threshold.
In this case, the second components of vectors
\eref{phi11} increase exponentially.
The physical solution is proportional
to a linear combination of these vectors,
the second component of which vanishes at large distances. 
% One can find that the vector solution
% \begin{equation}
% \vec{\tilde\varphi}(r)=\vec{\tilde\varphi}_{1}(r)+q\vec{\tilde\varphi}_{2}(r),
% \qquad q=\frac{\beta}{i k_2-\alpha_2},
% \end{equation}
% has the necessary asymptotic behaviour,
% where $q$ is constant with respect to $r$ but depends on $k_1$ through $k_2$.
The simplest way to obtain this vector is to apply operator $L$,
which conserves the exponentially-decreasing asymptotic behaviour,
to the vector solution of the initial Schrödinger equation,
the second component of which vanishes at infinity.
This vector reads 
\begin{equation}
\vec\psi_{0}(k_1, r)=\vec\psi_{0;1} -
 \frac{\beta}{|k_2|+\alpha_2}\, \vec\psi_{0;2}=
\left(\begin{array}{c}
       \frac{1}{\sin \xi_0(k_1)} \sin \left[k_1 r + \xi_0(k_1)\right]\\
    -\frac{\beta \rme^{\rmi  k_2 r}}{|k_2|+\alpha_2}
      \end{array}\right),
\label{vecpsi0}
\end{equation}
where the first component displays the shift $\xi_0$ defined in \eref{d0}.
Note that solution \eref{vecpsi0} is not regular at the origin.
The regular transformed vector solution is then 
\begin{equation}
\vec{\varphi}_{1}(k_1,r) = L \vec\psi_{0}(k_1, r),
\end{equation}
which, for the particular superpotential \eref{w}, reads 
\begin{equation} \label{mixsol}
\fl \vec{\varphi}_{1}(k_1,r)\!  =\! \left(\!\!\!
\begin{array}{c}
  - \frac{\sqrt{\kappa_1 \kappa_2} \beta}{|k_2|+\alpha_2} \,
   \frac{e^{-|k_2| r}}{\cosh y}
  - \frac{\kappa_1 \tanh y}{\sin \xi_0(k_1)} \sin \left[k_1 r 
   + \xi_0(k_1)\right]
    - \frac{k_1}{\sin \xi_0(k_1)}\cos\left[k_1 r +\xi_0(k_1)\right]
  \\[.5em]
   \frac{\sqrt{\kappa_1 \kappa_2}}{\cosh y \ \sin\xi_0(k_1)}
   \sin\left[k_1 r + \xi_0(k_1)\right] 
   - \frac{|k_2|+\kappa_2 \tanh y}{|k_2|+\alpha_2} \beta e^{-|k_2| r} 
\end{array} \!\!  \right)\!.
\end{equation}
It can be checked that it is regular at the origin, as it should. 
Asymptotically, the second component vanishes exponentially while the first 
component oscillates with the one-channel scattering phase shift \eref{d1}. 
In figure \ref{figvecpsi},
such physical solutions are represented for the potential
presented in figure 1 of \cite{sparenberg:06}.
They illustrate the Feshbach-resonance phenomenon:
for a fixed normalization of the closed-channel component at the origin (dotted lines),
the ANC of this component varies slowly while crossing the resonance.
In contrast, the open-channel component varies very strongly:
it changes sign while crossing the resonance
and is much smaller than the closed-channel component at the resonance energy
while both components have the same order of magnitude for energies below and above the resonance energy.

\section{Conclusion}
\label{sec:conc}

A detailed study of the properties of conservative
and non-conservative supersymmetric transformations
has been performed, both in the single-channel and coupled-channel cases.
In the single-channel case, for a zero initial potential,
a single non-conservative transformation has been shown
to be equivalent to a particular pair of
conservative transformations.
We conjecture that this equivalence can be generalized to chains of transformations.
Hence, in the single-channel case,
non-conservative transformations do not have
a more practical interest than the conservative ones, but they sometimes
lead to more elegant analytical expressions for the obtained potentials.
Similarly, for a zero initial potential too,
chains including zero-energy transformations have been shown
to be replaceable by transformations
performed on purely centrifugal potentials.
Conservative transformations with non-vanishing factorization energies are
thus a self-sufficient tool to solve all practical
inversion problems in the single-channel case.

For coupled-channel cases, the situation strongly depends on the presence or absence of thresholds.
For equal thresholds, conservative transformations are sufficient to solve practical two-channel inversion problems.
In the most convenient algorithm we could find,
these inverse problems are decomposed in single-channel inverse problems that allow one
to fit eigenphase shifts,
followed by eigenphase-preserving transformations that introduce a coupling to the system
without modifying the constructed eigenphase shifts.
The solution of the inverse problem is thus based
on the Blatt-Biedenharn decomposition of the scattering matrix
and has reached a sufficient level of maturity to be applicable to practical problems.
Moreover, compact expressions have been established for the inversion potential in terms of solutions
of the initial Schrödinger equation,
which leads to analytical exactly-solvable potentials when this initial potential is zero or purely centrifugal.
As an important physical illustration,
we could build satisfactory neutron-proton potentials for the triplet spin state.

In contrast, the coupled-channel inverse problem with threshold differences is still in progress.
The present work brings promising first steps but also raises several questions.
Non-conservative transformations appear to be an essential tool
to solve the inverse problems in this case,
as conservative transformations do not seem able to introduce coupling between channels.
However, much as in the single-channel case,
a single non-conservative transformation of the two-channel zero potential
leads to a Jost matrix which can be decomposed into two factors
[see \eref{F1Coxbis} or \eref{F1Cox}],
hence suggesting that a formulation of the non-conservative transformation
in terms of two simpler transformations might be possible.
The second transformation is visibly a conservative transformation,
as the corresponding Jost-matrix factor is diagonal;
the nature of the conjectured first transformation, which would generate a non-diagonal Jost matrix,
is unknown.
A detailed comparison between the single-channel and coupled-channel cases could be useful
to unveil it.

Iterations of non-conservative transformations could also be studied and could provide rather convenient tools
to solve inverse problems, if combined with conservative transformations.
With this respect, the Stapp decomposition of the scattering matrix seems more convenient
as conservative transformations following non-conservative ones only modify the bar eigenphase shifts
without affecting the bar mixing parameter.
Moreover, compact expressions for these iterative potentials could be sought for,
hence extending the results obtained in the case of absence of threshold.
Finally, most of the results presented here for the two-channel case
should be explicitly generalized to an arbitrary number of channels.
We plan to tackle all these questions in future works.

%\section*{Acknowledgments}
\ack
DB, AMPM and JMS remember with emotion the many years of fruitful collaboration and friendship with Boris Samsonov.
The final form of this work, that unfortunately Boris did not know, is dedicated to him.

JMS thanks Bikashkali Midya for his careful reading of several parts of this text.
AMPM thanks the Brazilian foundation CAPES (program PNPD/2011) and
the Russian Foundation for Basic Research (grant 12-02-31552) for the financial support in different stages of this work.
This text presents research results
of the IAP program P7/12 initiated by the Belgian-state
Federal Services for Scientific, Technical, and Cultural Affairs.

\section*{References}
\providecommand{\newblock}{}J.\ Phys.\ A: Math.\ Theor.

\end{document}